\documentclass{JHEP3}

\usepackage{graphicx}
\usepackage{cite}
\usepackage[tbtags]{amsmath}

\newcommand{\new}[1]{#1}

\newcommand{\lf}{\left}
\newcommand{\rg}{\right}

\newcommand{\ii}{i}     
\newcommand{\de}{d}     

\newcommand{\xb}{\hat{x}}
\newcommand{\zd}{\hat{z}}

\newcommand{\tr}{\operatorname*{Tr}\nolimits}

\newcommand{\bm}{\boldsymbol}
\newcommand{\h}{\hat{\bm{h}}}

\newcommand{\ms}{\mskip 1.5mu}
\newcommand{\cdott}{{\mskip -1.5mu} \cdot {\mskip -1.5mu}}

\newcommand{\lsim}{\raisebox{-4pt}{%
    $\,\stackrel{\textstyle <}{\sim}\,$}}

\newcommand{\ff}[1]{\mathcal{F}\bigl[ #1 \bigr]}
\newcommand{\lff}[1]{\mathcal{F}\biggl[\ms #1 \ms\biggr]}

\newcommand{\intobs}[1]{\langle\!\langle\ms #1 \ms\rangle\!\rangle}

\newcommand{\nslash}{\kern 0.2 em n\kern -0.50em /}
\newcommand{\kslash}{\kern 0.2 em k\kern -0.45em /}
\newcommand{\kbarslash}{\kern 0.2 em \bar{k}\kern -0.50em /}
\newcommand{\pslash}{\kern 0.2 em p\kern -0.44em /}
\newcommand{\pbarslash}{\kern 0.2 em \bar{p}\kern -0.44em /}
\newcommand{\lslash}{\kern 0.2 em l\kern -0.39em /}
\newcommand{\lbarslash}{\kern 0.2 em \bar{l}\kern -0.39em /}
\newcommand{\vslash}{\kern 0.2 em v\kern -0.50em /}
\newcommand{\Sslash}{\kern 0.2 em S\kern -0.57em /}
\newcommand{\Pslash}{\kern 0.2 em P\kern -0.55em /}

\allowdisplaybreaks[2]

\preprint{DESY 08-023}

\title{Matches and mismatches in the descriptions of semi-inclusive
  processes at low and high transverse momentum}

\author{Alessandro Bacchetta$^{\ms a}$\footnote{%
present address: Theory Center, Jefferson Lab, 12000
Jefferson Ave., Newport News, VA 23606, USA}\ ,
Dani\"el Boer$^{\ms b}$, Markus Diehl$^{\ms a}$, Piet J. Mulders$^{\ms
  b}$ \\ 
${}^{a}$ Theory Group, Deutsches Elektronen-Synchroton DESY,
22603 Hamburg, Germany \\
${}^{b}$ Dept.\ of Physics and Astronomy, Vrije Universiteit Amsterdam,
1081 HV Amsterdam, The Netherlands \\
}

\abstract{We investigate the transverse-momentum-dependence in
  semi-inclusive deep inelastic leptoproduction of hadrons.  There are
  two different theoretical approaches to study this dependence, one
  for low and one for high transverse momentum of the observed hadron.
  We systematically investigate their connection, paying special
  attention to azimuthal distributions and to polarization dependence.
  In the region of intermediate transverse momentum, where both
  approaches are applicable, we find that their results match for
  certain observables but not for others.  Interpolating expressions
  are discussed for the case where one has no matching.  We then use
  power counting to determine which mechanism is dominant in various
  azimuthal and spin asymmetries that are integrated over the
  transverse momentum.  Our findings have consequences for the
  extension of transverse-momentum-dependent factorization beyond
  leading twist.  They also shed light on the problem of resumming
  logarithms of transverse momentum for azimuthal distributions.  Our
  results can be carried over to the Drell-Yan process and to
  two-hadron production in $e^+e^-$ annihilation.
}


\begin{document}

\section{Introduction}
\label{sec:intro}

Short-distance factorization is a key concept in quantum
chromodynamics, providing much of the predictive power of the theory
in high-energy scattering processes.  Among the simplest processes to
which this concept can be applied are one-particle or single jet
inclusive production in lepton-nucleon scattering, two-particle or
dijet inclusive production in $e^+ e^-$ annihilation, and Drell-Yan
lepton pair production via a photon or electroweak gauge boson in
hadron-hadron collisions.  Crossing of the hard-scattering subprocess
closely relates these reactions, and many results obtained for one of
them carry over to the other ones.
A number of nontrivial issues for factorization arise especially when
one observes the transverse momentum $q_T$ and the angular
distribution of the produced particle with respect to a suitable
reference direction.  The problem then involves three scales, namely
the scale of nonperturbative QCD dynamics, which we represent by the
nucleon mass $M$, the transverse momentum $q_T$, and the photon or
electroweak boson virtuality $Q$, which throughout this paper we
require to be large compared with $M$.  

There are two basic descriptions for the production of a particle with
specified transverse momentum.  One of them is applicable for $q_T \ll
Q$ and involves transverse-momentum-dependent (also called
unintegrated) parton distribution and fragmentation functions.  The
other one requires that $q_T \gg M$ and generates transverse momentum
in the final state by perturbative radiation, using collinear (or
integrated) distribution and fragmentation functions as
nonperturbative input.  In the following we refer to the two momentum
regions and the associated theoretical descriptions as ``low-$q_T$''
and ``high-$q_T$'', respectively.  The low- and high-$q_T$ domains
overlap for $M \ll q_T \ll Q$, where both descriptions can hence be
applied.  An important question is whether in this intermediate $q_T$
region they describe the same dynamics or two competing mechanisms.
Depending on the answer, one can either try to construct a formulation
that smoothly interpolates between the two descriptions, or to add
their results in a consistent manner.

For the cross section depending on $q_T^2$ but integrated over the
angular distribution of the produced particle, the work of Collins,
Soper and Sterman \cite{Collins:1984kg} showed that the descriptions
based on intrinsic transverse momentum and on hard perturbative
radiation indeed match at intermediate $q_T$ and permit a smooth
interpolation at all orders in $\alpha_s$.  A key element of the
derivation was that for sufficiently large transverse momentum one can
express unintegrated parton distributions and fragmentation functions
in terms of their integrated counterparts and of perturbatively
calculable hard-scattering kernels.  The matching of the two
descriptions allowed the authors of \cite{Collins:1984kg} to resum
large logarithms $\ln(Q^2/q_T^2)$ to all orders using renormalization
group techniques---a procedure that remains the cornerstone for
transverse-momentum resummation in a wide range of collider processes.

For the angular distribution, however, the situation is less well
understood.  Both mechanisms just mentioned give rise to nontrivial
angular dependence, as has been pointed out long ago for
semi-inclusive deep inelastic scattering (SIDIS) by Cahn
\cite{Cahn:1978se,Cahn:1989yf} and by Georgi and Politzer
\cite{Georgi:1977tv} for the low- and high-$q_T$ mechanism,
respectively.  To the best of our knowledge, the relation between the
two descriptions for the angular distribution in unpolarized
scattering has not been analyzed so far.  Important progress has
recently been made in the understanding of a particular azimuthal
asymmetry for a transversely polarized target in SIDIS or Drell-Yan
production.  The authors of
\cite{Ji:2006ub,Ji:2006vf,Ji:2006br,Koike:2007dg} have shown that the
description of this asymmetry by the Sivers effect for small $q_T$ and
by the Qiu-Sterman mechanism for large $q_T$ match at order $\alpha_s$
in the intermediate region $M \ll q_T \ll Q$.  It is natural to ask if
one has a similar situation for other observables as well.

In the present work we therefore present a systematic analysis of the
interplay between the low-$q_T$ and the high-$q_T$ mechanisms for
angular distributions, both in unpolarized and in polarized
scattering.  This provides guidance for the theoretical description of
a variety of observables, determining in particular whether or not one
should add different contributions.  For definiteness we will consider
the case of SIDIS, but as remarked above, analogous studies can be
performed for $e^+ e^-$ collisions and for the Drell-Yan process.  Our
results are relevant to the possible extension of transverse-momentum
resummation for specific azimuthal distributions, which was recently
considered for the case of Drell-Yan production in \cite{Boer:2006eq}
and \cite{Berger:2007jw}.

A key finding of our work is that for certain observables, the
\new{leading terms of the} 
low-and high-$q_T$ descriptions match in the region $M \ll q_T \ll Q$ of
intermediate transverse momenta, whereas for others they do \emph{not}
match.  That this may happen can be understood already at the level of
power counting.  The low-$q_T$ description, which uses transverse
momentum dependent parton densities and fragmentation functions, is
based on taking $Q^2$ large compared with $q_T^2$ and all
nonperturbative scales.  
\new{We chose $q_T/Q$ rather than $M/Q$ as parameter for power counting,
  since in the intermediate-$q_T$ region it is the larger of the two.}
Taking for example an observable $F$ with mass
dimension $-2$, we can thus expand
\begin{align}
  \label{low-expand}
F(q_T,Q) & \stackrel{q_T \ll Q}{=} \frac{1}{M^2} \sum_{n}\, 
  \biggl[\frac{q_T}{Q}\biggr]^{n-2}\,
  l_{n} \biggl(\frac{M}{q_T}\biggr) \,,
  \intertext{where $l_n$ are dimensionless functions.  In our
    applications, the term with index $n$ will correspond to twist-$n$
    accuracy in the low-$q_T$ calculation, where $n\ge 2$.  In the
    region of intermediate $q_T$ we can further expand the functions
    $l_n(M/q_T)$ for small $M/q_T$ and then have}
  \label{low-inter}
F(q_T,Q) & \stackrel{M \ll q_T \ll Q}{=}
  \frac{1}{M^2} \sum_{n,k}  l_{n,k}\,
  \biggl[\frac{q_T}{Q}\biggr]^{n-2}\,
  \biggl[\frac{M}{q_T}\biggr]^{k}
  \intertext{with coefficients $l_{n,k}$.  The high-$q_T$ calculation,
    which is based on collinear factorization, treats both $Q$ and
    $q_T$ as large compared with nonperturbative scales like $M$.  The
    relevant parameter for power counting in the intermediate region
    is therefore $M/q_T$, and we have}
  \label{hi-expand}
F(q_T,Q) & \stackrel{M \ll q_T}{=} \frac{1}{M^2} \sum_{n}\,
  \biggl[\frac{M}{q_T}\biggr]^{n}\,
  h_{n} \biggl(\frac{q_T}{Q}\biggr)
  \intertext{with dimensionless functions $h_n$.  In our applications,
    the term with index $n$ will correspond to twist $n$ in the
    high-$q_T$ calculation, where again $n\ge 2$.  In the
    intermediate-$q_T$ region we can then expand $h_n(q_T/Q)$ for
    small $q_T/Q$\,:}
  \label{hi-inter}
F(q_T,Q) & \stackrel{M \ll q_T \ll Q}{=}
  \frac{1}{M^2} \sum_{n,k} h_{n,k}\,
  \biggl[\frac{M}{q_T}\biggr]^{n}\,
  \biggl[\frac{q_T}{Q}\biggr]^{k-2}
\end{align}
with coefficients $h_{n,k}$.  Since both \eqref{low-inter} and
\eqref{hi-inter} are valid in the intermediate region, we can identify
the coefficients $l_{n,k} = h_{k,n}$.
As a consequence, a term of twist $n$ in the low-$q_T$ calculation
will only correspond to a term of the same twist in the high-$q_T$
calculation if $n=k$.  We will for instance encounter observables with
\begin{align}
  \label{low-ex1}
M^2 F(q_T,Q) =
  l_{2,2}\, \biggl[\frac{q_T}{Q}\biggr]^{0}\,
            \biggl[\frac{M}{q_T}\biggr]^{2}
+ l_{4,2}\, \biggl[\frac{q_T}{Q}\biggr]^{2}\,
            \biggl[\frac{M}{q_T}\biggr]^{2} 
+ \ldots \,,
\intertext{where the term with $l_{2,2} = h_{2,2}$ is of leading twist
  in both the low- and high-$q_T$ calculations.  The term with
  $l_{4,2} = h_{2,4}$ is subleading in the low-$q_T$ calculation and
  becomes subleading in the high-$q_T$ when one takes the additional
  limit $q_T \ll Q$.  For other observables, we will find}
  \label{low-ex2}
M^2 F(q_T,Q) =
  l_{2,4}\, \biggl[\frac{q_T}{Q}\biggr]^{0}\,
            \biggl[\frac{M}{q_T}\biggr]^{4}
+ l_{4,2}\, \biggl[\frac{q_T}{Q}\biggr]^{2}\,
            \biggl[\frac{M}{q_T}\biggr]^{2}
+ \ldots \,.
\end{align}
\new{Here the term with $l_{2,4} = h_{4,2}$ is leading in the low-$q_T$
calculation but subleading in the high-$q_T$ one, whereas the reverse
holds for the term with $l_{4,2} = h_{2,4}$.  The respective leading-order
terms in the two calculations will hence not match in the intermediate
region of $q_T$.}
Which term in \eqref{low-ex2} is larger in given kinematics obviously
depends on the relative size of the two small parameters $q_T/Q$ and
$M/q_T$.  We will discuss in section \ref{sec:interpol} how one can
construct interpolating expressions using both terms.

\new{An important question is which terms in the expansions
  \eqref{low-expand} and \eqref{hi-expand}, and hence in \eqref{low-inter}
  and \eqref{hi-inter}, can actually be calculated in practice.  We
  discuss this in some detail in the main body of the paper, but already
  mention here that in the high-$q_T$ framework there is a large number of
  results at twist two and three.  In low-$q_T$ framework factorization is
  rather well understood at twist-two level, whereas its status is less
  clear at twist three.  Little is known about the validity of
  factorization at twist-four accuracy in either framework.  In the
  example \eqref{low-ex2} one can thus envisage
  to compute the terms $l_{2,4}$ and $h_{2,4}$,
  which are leading in their respective power counting scheme.  The
  simultaneous validity of the expansions \eqref{low-inter} and
  \eqref{hi-inter} in the intermediate region requires that $h_{4,2} =
  l_{2,4}$ and $l_{4,2} = h_{2,4}$, but at present one cannot check this
  explicitly because a calculation of the power-suppressed terms $h_{4,2}$
  and $l_{4,2}$ is beyond the state of the art.}

\new{Several investigations have been performed assuming factorization for
  twist-three observables in the low-$q_T$ description.  Detailed
  calculations at tree level
  \cite{Boer:1997nt,Mulders:1995dh,Bacchetta:2006tn,Boer:2003cm} are found
  to be self-consistent and give results with a structure similar to that
  of twist-two observables.  Their extension to higher orders in
  $\alpha_s$, including a proper treatment of soft gluon exchange has not
  been achieved yet, and the study \cite{Gamberg:2006ru} suggests that
  such an extension will not be trivial.  In sections \ref{sec:lowtm} and
  \ref{sec:tailstructure} we will investigate observables where the
  leading terms in the expansions \eqref{low-inter} and \eqref{hi-inter}
  coincide and have the coefficient $l_{3,2} = h_{2,3}$.  The twist-two
  quantity $h_{2,3}$ is readily computed, and its comparison with the
  result for $l_{3,2}$ obtained with a candidate factorization formula
  will shed light on low-$q_T$ factorization at twist three.}

In experimental analyses one often has to integrate over the observed
$q_T$ in order to accumulate statistics.  One may simply integrate an
observable over $q_T^2$ or consider weighted observables like $\int
\de q_T^2\, (q_T^{}/M)^p F(q_T^{},Q)$ with some power $p$.  If in turn
the measurement of the $q_T$-dependence suffers from large
uncertainties, then both a differential observable and weighted
integrals will be affected with large errors, so that the simple
integral $\int \de q_T^2\, F(q_T^{},Q)$ may be the best quantity to
consider from an experimental point of view.  For the theoretical
analysis it is important to identify the relative importance of the
different $q_T$ regions in an integrated observable, and to clarify
their interplay if several regions are important.  Our results will
allow us to address this question at the level of power counting.

Our paper is organized as follows.  In the next section we define the
structure functions for SIDIS, which are the observables we study in
detail in this work.  To set the stage, we recall in section
\ref{sec:example} some important results for SIDIS taken differential
in $q_T$ but integrated over the angular distribution of the observed
hadron, recalling in particular the foundations of $q_T$ resummation
in this context.  In section \ref{sec:large} we collect the well-known
results of the calculation of SIDIS with $q_T \gg M$ in collinear
factorization at leading order in $\alpha_s$, and then approximate
these results for $q_T \ll Q$.  In section \ref{sec:small} we take the
opposite path, recalling the results for SIDIS with $q_T \ll Q$ and
approximating them for $q_T \gg M$.  For this we need the behavior of
distribution and fragmentation functions at high transverse momentum,
and we will derive the corresponding power behavior of these functions
based on general grounds.  In section \ref{sec:comparison} we will see
for which observables the calculations of the two previous sections
match for intermediate $q_T$ and for which ones they do not.  The
consequences for integrated observables are discussed in section
\ref{sec:int-obs}.  Whereas in section \ref{sec:small} we derive the
power behavior for all structure functions introduced in section
\ref{sec:sidis}, we give in section \ref{sec:tails} explicit results
for those observables that appear at twist two in the high-$q_T$
regime.  The comparison of the high-$q_T$ with the low-$q_T$
expressions will allow us to draw some conclusions about the unsolved
problem of $q_T$ resummation for angular distributions, as well as the
possibility of extending low-$q_T$ factorization to twist three.  The
main results of our work are summarized in section \ref{sec:sum}, and
some technical details are given in the appendices.


\section{Structure functions in semi-inclusive deep inelastic
  scattering}
\label{sec:sidis}

The physical process we investigate in this work is semi-inclusive
DIS,
\begin{equation}
  \label{sidis}
\ell(l) + p(P) \to \ell(l') + h(P_h) + X ,
\end{equation}
where $\ell$ denotes the beam lepton, $p$ the proton target, and $h$
the observed hadron, with four-momenta given in parentheses.  We allow
for polarization of beam and target, but restrict ourselves to the
case of an unpolarized final state, i.e.\ to the situation in which
$h$ has spin zero or where its polarization is not observed.  The
corresponding observables cover a variety of situations with different
types of power behavior we wish to discuss.  Many of them have been
measured in experiment, see \cite{Arneodo:1986cf,Breitweg:2000qh,%
Avakian:2003pk,Airapetian:1999tv,Airapetian:2004tw,Alexakhin:2005iw}
and the recent review in \cite{D'Alesio:2007jt}.

Working in the one-photon exchange approximation, we define the photon
momentum $q = l - l'$ and its virtuality $Q^2 = - q^2$.  We use the
conventional variables for SIDIS
\begin{align}
x &= \frac{Q^2}{2\,P\cdot q} \,,
&
y &= \frac{P \cdot q}{P \cdot l} \,,
&
z &= \frac{P \cdot P_h}{P\cdot q} \,,
\end{align} 
and write $M$ and $M_h$ for the respective masses of the proton target
and the produced hadron $h$.  We take the limit of large $Q^2$ at
fixed $x$, $y$, $z$, and throughout this paper we neglect corrections
in the masses of the hadrons or the lepton.  

It is convenient to discuss the experimental observables for SIDIS in
a frame where $P$ and $q$ collinear.  We define the transverse part
$P_{h\perp}^\mu$ of $P_h^\mu$ as orthogonal with respect to the
momenta $P$ and $q$.  Likewise, we define the transverse part
$S_\perp^\mu$ of the spin vector $S^\mu_{\phantom{\perp}}$ of the
target, as well as its longitudinal projection $S_\parallel$ along
$P^\mu$.  We further define the azimuthal angles $\phi_h$ and $\phi_S$
of $P_h^\mu$ and $S^\mu_{\phantom{h}}$ with respect to the lepton
plane in accordance with the Trento conventions
\cite{Bacchetta:2004jz}, as shown in Fig.~\ref{f:angles}.  Covariant
expressions for the quantities just discussed can be found in
\cite{Bacchetta:2006tn}.  Finally, we write $\lambda_e$ for the
longitudinal polarization of the incoming lepton, with $\lambda_e = 1$
corresponding to a purely right-handed beam.

\FIGURE[ht]{
\includegraphics[width=10cm]{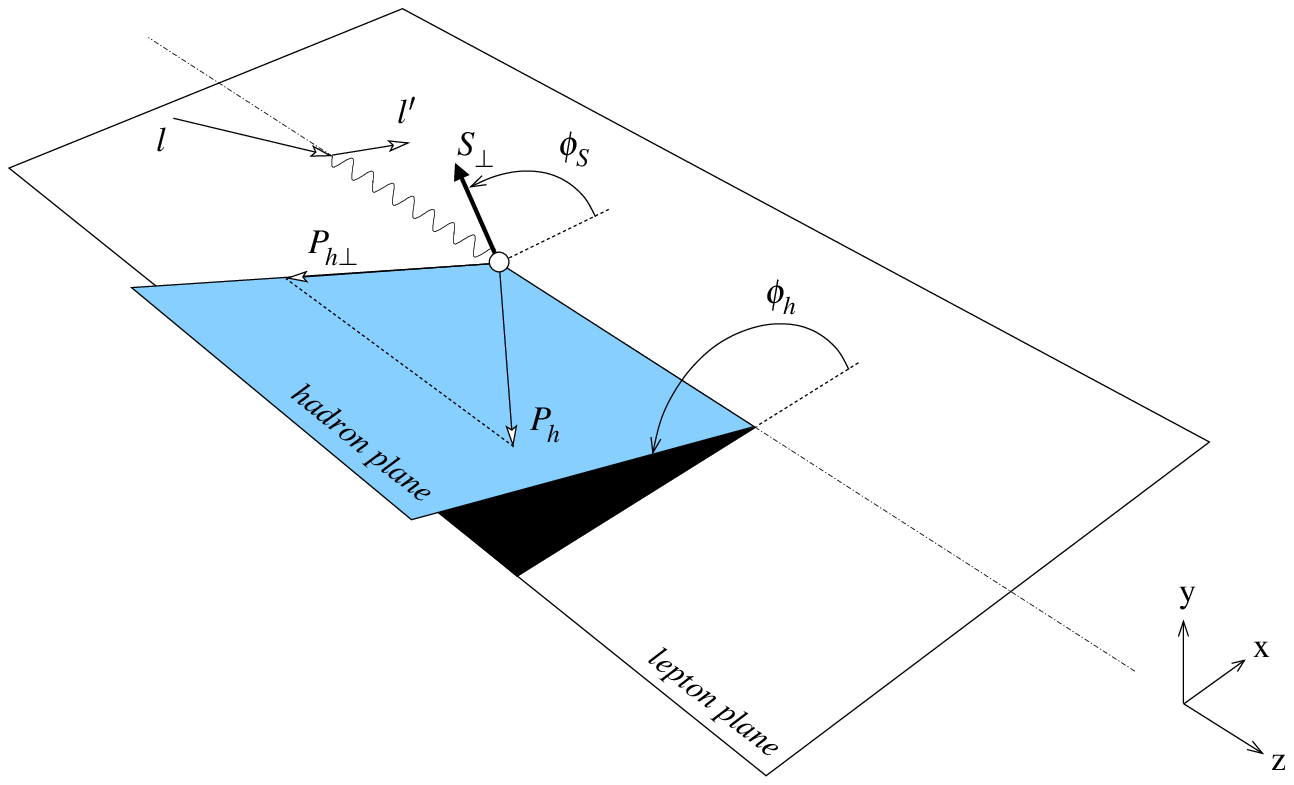}
\caption{\label{f:angles} Definition of azimuthal angles for
  semi-inclusive deep inelastic scattering in the target rest
  frame~\cite{Bacchetta:2004jz}.  $P_{h\perp}$ and $S_\perp$ are the
  transverse parts of $P_h$ and $S$ with respect to the photon
  momentum.}
}

The lepton-hadron cross section can then be parameterized as
\cite{Bacchetta:2006tn}
\begin{align}
& \frac{d\sigma}{dx \, dy \,dz \,
         d\phi_S \, d\phi_h\, d P_{h\perp}^2} =
\frac{\alpha^2}{x \ms Q^2}\,
\frac{y}{2\,(1-\varepsilon)}\,
\nonumber \\[0.2em]
 & \quad \times \biggl\{
F_{UU ,T}
+ 
\varepsilon\ms
F_{UU ,L}
+ \sqrt{2\,\varepsilon (1+\varepsilon)}\,\cos\phi_h\,
F_{UU}^{\cos\phi_h}
+ \varepsilon \cos(2\phi_h)\, 
F_{UU}^{\cos 2\phi_h}
\nonumber \\
  & \quad \qquad
+ \lambda_e\, \sqrt{2\,\varepsilon (1-\varepsilon)}\, 
           \sin\phi_h\, 
F_{LU}^{\sin\phi_h}
\phantom{\bigg[ \bigg] }
\nonumber \\
 & \quad \qquad
+ S_\parallel\, \bigg[ 
 \sqrt{2\, \varepsilon (1+\varepsilon)}\,
  \sin\phi_h\, 
F_{UL}^{\sin\phi_h}
+  \varepsilon \sin(2\phi_h)\, 
F_{UL}^{\sin 2\phi_h}
\bigg]
\nonumber \\
 & \quad \qquad
+ S_\parallel\ms \lambda_e\, \bigg[ \,
  \sqrt{1-\varepsilon^2}\; 
F_{LL}
+\sqrt{2\,\varepsilon (1-\varepsilon)}\,
  \cos\phi_h\, 
F_{LL}^{\cos \phi_h}
\bigg]
\nonumber \\
 & \quad \qquad
+ |\bm{S}_\perp|\; \bigg[
  \sin(\phi_h-\phi_S)\,
\Bigl(F_{UT ,T}^{\sin\lf(\phi_h -\phi_S\rg)}
+ \varepsilon\, F_{UT ,L}^{\sin\lf(\phi_h -\phi_S\rg)}\Bigr)
\nonumber \\ 
 & \qquad  \qquad \qquad
+ \varepsilon\, \sin(\phi_h+\phi_S)\, 
F_{UT}^{\sin\lf(\phi_h +\phi_S\rg)}
+ \varepsilon\, \sin(3\phi_h-\phi_S)\,
F_{UT}^{\sin\lf(3\phi_h -\phi_S\rg)}
\phantom{\bigg[ \bigg] }
\nonumber \\ 
 & \qquad \qquad \qquad
+ \sqrt{2\,\varepsilon (1+\varepsilon)}\, 
  \sin\phi_S\, 
F_{UT}^{\sin \phi_S }
+ \sqrt{2\,\varepsilon (1+\varepsilon)}\, 
  \sin(2\phi_h-\phi_S)\,  
F_{UT}^{\sin\lf(2\phi_h -\phi_S\rg)}
\bigg]
\nonumber \\ 
 & \quad \qquad 
+ |\bm{S}_\perp|\, \lambda_e\, \bigg[
  \sqrt{1-\varepsilon^2}\, \cos(\phi_h-\phi_S)\, 
F_{LT}^{\cos(\phi_h -\phi_S)}
+\sqrt{2\,\varepsilon (1-\varepsilon)}\, 
  \cos\phi_S\, 
F_{LT}^{\cos \phi_S}
\nonumber \\ 
 & \qquad \qquad \qquad
+\sqrt{2\,\varepsilon (1-\varepsilon)}\, 
  \cos(2\phi_h-\phi_S)\,  
F_{LT}^{\cos(2\phi_h - \phi_S)}
\bigg] \biggr\} \,,
\label{e:crossmaster}
\end{align}
where $\alpha$ is the fine structure constant and $\varepsilon$ the
ratio of longitudinal and transverse photon flux,
\begin{equation}
\varepsilon = \frac{1-y}{1-y+ y^2/2} \,.
\end{equation}  
The 18 structure functions $F$ on the r.h.s.\ depend on $x$, $Q^2$,
$z$ and $P_{h\perp}^2$ and encode the strong-interaction dynamics of
the hadronic subprocess $\gamma^* + p\to h + X$.  Their first and
second subscript respectively specifies the polarization of the beam
and the target.  In the structure functions $F^{}_{UU,T}$,
$F^{}_{UU,L}$ and $F_{UT,T}^{\sin\lf(\phi_h -\phi_S\rg)}$,
$F_{UT,L}^{\sin\lf(\phi_h -\phi_S\rg)}$, the third subscript refers to
the transverse and longitudinal polarization of the photon.

To calculate the SIDIS structure functions it is convenient to use
light-cone coordinates with respect to the directions of the relevant
hadron momenta.  We introduce light-like vectors $n_+$ and $n_-$ with
$n_+\cdot n_- = 1$ such that, up to mass corrections, $n_+$ is
proportional to $P$ and $n_-$ proportional to $P_h$.  A rescaling
\begin{align}
  \label{boost}
n_+ &\to \kappa\ms n_+ \,, & n_- &\to \kappa^{-1}\ms n_-
\end{align}
corresponds to boosts in the collinear direction.  The
off-collinearity of the process is determined by the vector
\begin{equation}
  \label{qT-def}
q_{T}^\mu =
  q^\mu_{\phantom{T}} + (1-r)\ms x P^\mu_{\phantom{T}} - P_h^\mu /z
\end{equation}
with $r = q_T^2/Q^2$.  For ease of notation we denote the length of
this vector by
\begin{equation}
q_T^{} = 
  \bigl( - q_T^{\ms\mu}\, q_{T \mu}^{\phantom{\mu}} \bigr)^{1/2} \,,
\end{equation}
so that $q_T^2$ is positive.
There is a simple relation between the transverse momentum $q_T^\mu$
of the photon with respect to the hadrons and the transverse momentum
$P_{h\perp}^\mu$ of the produced hadron with respect to the photon and
proton: $\smash{P_{h\perp}^{\mu} = -z q_T^{\mu} - 2r z x
  P^{\mu}_{\phantom{T}}}$.  The SIDIS cross section differential in
$q_T^2$ instead of $P_{h\perp}^2$ is hence equal to $z^2$ times the
r.h.s.\ of~\eqref{e:crossmaster}.


\section{Factorization and $q_T$ resummation}
\label{sec:example}

In this section we recall some important results for the description
of hard processes with measured $q_T$, in particular the factorization
for low $q_T$ formulated by Collins and Soper \cite{Collins:1981uk}
and its connection to the procedure of transverse momentum resummation
by Collins, Soper and Sterman \cite{Collins:1984kg}.  In the following
we refer to these authors as CS and CSS, respectively.  In the next
two subsections we focus on the unpolarized SIDIS cross section
differential in $q_T$ but integrated over the azimuthal angle
$\phi_h$.

\subsection{Collins-Soper factorization}
\label{sec:CS-fact}

In the work of CS, factorization was derived for the production of
back-to-back jets in electron-positron annihilation, or more
specifically for $e^+ e^- \to A + B + X$, where $A$ and $B$ are two
hadrons belonging to opposite-side jets in the $e^+e^-$ c.m.  In
general the momenta $P_A$ and $P_B$ of the two hadrons are not exactly
back-to-back because of their recoil against the additional particles
$X$ produced in the process.  The cross section, or equivalently the
hadron tensor, depends on the transverse momentum $q_T^\mu$ of the
virtual photon w.r.t.\ the hadrons in the c.m.\ of $A$ and $B$, which
is the analog of $q_T^\mu$ introduced for SIDIS in \eqref{qT-def}.
For $q_T \ll Q$ the CS paper derived a factorized expression of the
hadron tensor, which is a convolution in transverse momentum of a soft
factor $U$ and two fragmentation functions $D_1^{A\smash{/}a}$ and
$D_1^{B\smash{/}\bar{a}}$ for the fragmentation of a quark or
antiquark into $A$ or $B$.  In addition there is a hard-scattering
factor $H$, which does not depend on any transverse momentum.  To be
specific, Eq.~(7.14) in \cite{Collins:1981uk} gives the following
expression for the hadron tensor:
\begin{align}
 \hspace{-0.5em}
W_{e^+ e^-}^{\mu\nu}  &\propto
  \tr\bigl\{\!\Pslash_A \gamma^\mu \Pslash_B \gamma^\nu \bigr\} \,
\bigl| H_{e^+ e^-}^{}\bigl(
  z_A^{-1\phantom{/}\!\!} \zeta_A^{1/2},
  z_B^{-1\phantom{/}\!\!} \zeta_B^{1/2} \ms\bigr)
\bigr|^2 \,
\sum_a e_a^2 
  \int \de^2 \bm{p}_T^{}\, \de^2 \bm{k}_T^{}\, \de^2 \bm{l}_T^{}\,
\nonumber \\
& \quad\times
\delta^{(2)}(\bm{p}_T^{}+\bm{k}_T^{}+\bm{l}_T^{}-\bm{q}_T^{})\,
D_1^{A\smash{/}a} (z_A, {p}_T^{2}; \zeta_A)\,
D_1^{B\smash{/}\bar a} (z_B, {k}_T^{2}; \zeta_B)\,
U({l}_T^{2})
+ \ldots \,,
\label{Wepem}
\end{align}
where $\ldots$ stands for terms that either vanish after integration
over the azimuthal angle of $\bm{q}_T$ or are power suppressed in
$1/Q$.  The index $a$ runs over flavors of quarks and of antiquarks
with fractional charge $e_a$.  For consistency within the present
paper we have slightly changed notation compared with CS.\footnote{%
  We write $D_1$ instead of $\mathcal{P}$ for the fragmentation
  functions, $v$ instead of $n$ for the gauge fixing vector, and
  $n_+$, $n_-$ instead of $v_A$, $v_B$ for the light-cone directions.
  Our normalization condition for $U$ differs from the one of CS by a
  factor $(2\pi)^2$.}
The individual factors $H$, $D_1^{A\smash{/}a}$,
$D_1^{B\smash{/}\bar{a}}$, and $U$ depend on an ultraviolet
factorization scale $\mu$, which we have not displayed for brevity.
The derivation by CS is done in an axial gauge specified by a
spacelike vector $v$, and the dependence of individual factors on this
vector is through the parameters $\zeta_{A} = - (2P_{A}\cdot v)^2
/v^2$ and $\zeta_{B} = - (2P_{B}\cdot v)^2 /v^2$.  More generally,
$\zeta_{A}$ and $\zeta_B$ serve as cut-offs for rapidity divergences
and in a gauge-invariant definition of the fragmentation functions
arise from path-ordered exponentials involving the vector $v$, see
e.g.\ \cite{Collins:2003fm,Ji:2004wu,Collins:2007ph}.  The hadron
tensor as a whole is of course independent of $\zeta_A$ and $\zeta_B$,
so that the dependence on these parameters has to cancel between the
fragmentation functions and the hard-scattering factor.  The fact that
the soft factor defined by CS does not depend on them is less
obvious.\footnote{\label{U-footnote}%
  The four-vectors entering the construction of $U$ are $v$, $n_+$,
  $n_-$ and $l_T$.  The gauge vector $v$ used by CS has a zero
  transverse component, so that the only scalar products involving $v$
  are $v\cdot n_+$, $v\cdot n_-$ and $v^2= 2\ms (v\cdot n_+) (v\cdot
  n_-)$.  Gauges related by scaling $v\to \kappa v$ are equivalent,
  which leaves only a possible dependence on $v\cdot n_+ /v\cdot n_-$.
  This is however excluded by boost invariance, which requires that
  $U$ must not change under the rescaling \protect\eqref{boost}.}

In this work we will assume that the factorization of Collins and
Soper also holds for the hadron tensor in SIDIS at low $q_T^{}$.  Such
an expression, albeit with some differences, has been obtained by Ji,
Ma and Yuan \cite{Ji:2004wu}, and another relevant investigation has
recently been made by Collins, Rogers and Sta{\'s}to
\cite{Collins:2007ph}, which gives us confidence that our assumption
can be justified.  The analog of \eqref{Wepem} then reads
\begin{align}
W_{\text{SIDIS}}^{\mu\nu}  & \propto
  \tr\bigl\{\!\Pslash\ms \gamma^\mu \Pslash_h \gamma^\nu \bigr\} \,
\bigl| H_{\text{SIDIS}}^{}
  \bigl(x\ms \zeta^{1/2}, z^{-1}\zeta_{\smash{h}}^{1/2} \ms\bigr)
\bigr|^2 \,
\sum_a e_a^2 
\int \de^2 \bm{p}_T^{}\, \de^2 \bm{k}_T^{}\, \de^2 \bm{l}_T^{}\,
\nonumber \\
& \quad \times
\delta^{(2)}(\bm{p}_T^{}-\bm{k}_T^{}+\bm{l}_T^{}+\bm{q}_T^{}) \,
f_1^a(x,p_T^2;\zeta)\, D_1^a(z,k_T^2;\zeta_h)\,
U({l}_T^{2}) + \ldots \,,
\label{Wsidis}
\end{align}
where one of the fragmentation functions has been replaced by the
distribution function $f_1^a$ for quarks or antiquarks in the target.
In the following we refer to the factorization expressed in
\eqref{Wsidis} as ``CS factorization''.  The result \eqref{Wsidis}
gives rise to just one structure function,
\begin{align}
F_{UU,T} &=
\bigl| H\bigl(x\ms \zeta^{1/2}, z^{-1}\zeta_{\smash{h}}^{1/2} \ms\bigr)
\bigr|^2 \,
\sum_a x\ms e_a^2
\int \de^2 \bm{p}_T\, \de^2 \bm{k}_T^{}\, \de^2 \bm{l}_T^{}\,
\nonumber \\
& \quad \times
  \delta^{(2)}\bigl(\bm{p}_T - \bm{k}_T^{} +\bm{l}_T^{} + \bm{q}_T^{}
  \bigr)\,
f_1^a(x,p_T^2;\zeta)\, D_1^a(z,k_T^2;\zeta_h)\, U(l_T^2) \,,
\label{FUUTconv}
\end{align}
where we recall that $H$, $f_1$, $D_1$, and $U$ depend on a
renormalization scale $\mu$.  For brevity we omit the subscript
``SIDIS'' in $H$ from now on.

In the intermediate region $M \ll q_T \ll Q$, one can go further since
at least one of the momenta $\bm{p}{}_T$, $\bm{k}_T$, $\bm{l}_T$ in
\eqref{FUUTconv} is of order $\bm{q}{}_T$ and hence large compared
with the nonperturbative scale $M$.  For large transverse momentum the
soft factor $U(l_T^2)$ can be calculated order by order in $\alpha_s$,
whereas $f_1^a(x,p_T^2;\zeta)$ and $D_1^a(z,k_T^2;\zeta_h)$ are
respectively given as convolutions of perturbatively calculable
kernels with the collinear distribution and fragmentation functions
$f_1^a(x)$ and $D_1^a(z)$.  We will follow this path in sections
\ref{sec:small} and~\ref{sec:tails}.  The accuracy of this procedure
is however limited: up to mass corrections one has $(\zeta
\zeta_h)^{1/2} = x^{-1} z\ms Q^2$, and the power counting in the CS
derivation requires both $\zeta$ and $\zeta_h$ to be of order $Q^2$,
so that the perturbative expressions for $f_1^a(x,p_T^2;\zeta)$ and
$D_1^a(z,k_T^2;\zeta_h)$ involve large logarithms
$\ln(Q/q_T)$.  Let us sketch how these logarithms are resummed in the
work of CS.  The variation of $D_1$ with $\zeta_h$ is described by the
Collins-Soper equation, which gives $\partial D_1/\partial \ln\zeta_h$
as a convolution in transverse momentum of an evolution kernel with
$D_1$.  Analogous considerations apply to the distribution function
$f_1^a(x,p_T^2;\zeta)$.  For the Fourier transformed functions
\begin{align}
  \label{Dfb}
\tilde{f}_1^a(x,b^2;\zeta) &=
 \int \de^2 \bm{p}_T\, e^{\ii \bm{b}\cdot \bm{p}_T}\,
   f_1^a(x,p_T^2;\zeta) \,,
\nonumber \\
\widetilde{D}_1^a(z,b^2;\zeta_h) &=
 \int \de^2 \bm{k}_T\, e^{\ii \bm{b}\cdot \bm{k}_T}\,
   D_1^a(z,k_T^2;\zeta_h)
\end{align}
one obtains ordinary differential equations, whose solutions can be
written as
\begin{align}
  \label{CS-solve}
\tilde{f}_1^a(x,b^2;\zeta) &= \hat{f}_1^a(x,b^2)\,
  \exp\bigl[- \widehat{S}\ms'(x\ms \zeta^{1/2}, b) \bigr] \,,
\nonumber \\[0.3em]
\widetilde{D}_1^a(z,b^2;\zeta_h) &= \widehat{D}_1^a(z,b^2)\,
  \exp\bigl[- \widehat{S}\ms'(z^{-1} \zeta_{\smash{h}}^{1/2}, b) \bigr] \,,
\end{align}
where the Sudakov factor $\widehat{S}$ is constructed from the
evolution kernel.  The structure function in \eqref{FUUTconv} can then
be rewritten as
\begin{align}
F_{UU,T} &=
\bigl| H(Q,Q;\mu) \bigr|^2 \,
\sum_a x\ms e_a^2
\int \frac{\de^2 \bm{b}}{(2\pi)^2}\, e^{-\ii \bm{b}\cdot \bm{q}_T}\,
\exp\bigl[- 2\widehat{S}\ms'(Q,b) \bigr] \,
\widetilde{U}(b^2;\mu)
\nonumber \\
& \quad\times
\hat{f}_1^a(x,b^2;\mu)\, \widehat{D}_1^a(z,b^2;\mu)\,
\label{FUUTb}
\end{align}
with
\begin{equation}
  \label{Ub}
\widetilde{U}(b^2;\mu) = \int \de^2 \bm{l}_T\,
  e^{\ii \bm{b}\cdot \bm{l}_T}\,
  U(l_T^2;\mu) \,.
\end{equation}
Here we have fixed the gauge parameters as $x^2 \zeta = z^{-2} \zeta_h
= Q^2$ and restored the dependence on the renormalization scale $\mu$.
The Sudakov factor $\widehat{S}\ms'$ resums large logarithms of $Q b$,
which corresponds to large logarithms of $Q/q_T$ in $F_{UU,T}$ since
the typical values of $b$ in the integral \eqref{FUUTb} are of order
$1/q_T$.

For $b \ll 1/M$ the factors $\widehat{S}\ms'$, $\widetilde{U}$,
$\hat{f}_1^a$ and $\widehat{D}_1^a$ can be expanded in perturbation
theory.  To avoid large logarithms of $\mu b$ in this expansion, one
should take the renormalization scale of order $1/b$.  A common choice
in the $\overline{\text{MS}}$ scheme is $\mu= b_0/b$ with $b_0=2
e^{-\gamma_E} \approx 1.1$, where $\gamma_E$ is the Euler constant.
This simplifies a number of perturbative coefficients: in particular
the $O(\alpha_s)$ term in the soft factor is then zero, and one has
$\widetilde{U}(b^2,\mu=b_0/b) = 1 + O(\alpha_s^2)$ up to power
corrections in $Mb$.  The small-$b$ expansion for the distribution and
fragmentation functions reads
\begin{align}
  \label{small-b-exp}
\hat{f}_1^a(x,b^2;\mu=b_0/b) &=
  \sum_{i} \bigl( \widehat{C}^{\ms\text{in}}_{ai}
  \otimes f_1^i \bigr)(x;\mu=b_0/b) \,,
\nonumber \\
z^2\ms \widehat{D}_1^a(z,b^2;\mu=b_0/b) &=
  \sum_j \bigl( D_1^j
  \otimes \widehat{C}^{\ms\text{out}}_{ja}\bigr)(z;\mu=b_0/b) \,,
\end{align}
where the indices $i$ and $j$ run over quarks, antiquarks and the
gluon.  $f_1^i(x;\mu)$ and $D_1^j(z;\mu)$ are the usual collinear
distribution and fragmentation functions, and $\otimes$ denotes the
familiar convolution in longitudinal momentum fractions,
\begin{align}
  \label{con-def}
\bigl(C \otimes f \bigr)(x;\mu) &= \int_{x}^1
  \frac{\de\xb}{\xb}\; C(\xb;\mu)\, f\Bigl(\frac{x}{\xb};\mu\Bigr) \,,
\nonumber \\
\bigl(D \otimes C \bigr)(z;\mu) &= \int_{z}^1
  \frac{\de\zd}{\zd}\; D\Bigl(\frac{z}{\zd};\mu\Bigr)\, C(\zd;\mu) \,.
\end{align}
With the scale choice $\mu=b_0/b$ we find large logarithms of $Qb$ in
$|H|^2 = 1 + O(\alpha_s)$.  These can readily be resummed using the
renormalization group equation for this factor, which allows one to
write $|H(Q,Q;\mu=b_0/b)|^2 = |H(Q,Q;\mu=Q)|^2\;
e^{-\widehat{R}(Q,b)}$.  In the intermediate region $M\ll q_T \ll Q$
one therefore has
\begin{align}
F_{UU,T} &=
\bigl| H(\mu=Q) \bigr|^2\;
\frac{1}{z^2}\ms \sum_a x\ms e_a^2
\int \frac{\de^2 \bm{b}}{(2\pi)^2}\, e^{-\ii \bm{b}\cdot \bm{q}_T}\,
\exp\bigl[- \widehat{S}(Q,b) \bigr]\,
\widetilde{U}(\mu=b_0/b)\,
\nonumber \\
& \quad \times
\sum_i \bigl(
   \widehat{C}^{\ms\text{in}}_{ai} \otimes f_1^i \bigr)(x;\mu=b_0/b)\,
\sum_j \bigl( 
   D_1^j \otimes \widehat{C}^{\ms\text{out}}_{ja} \bigr)(z;\mu=b_0/b)
  \label{FUUTb-expand}
\end{align}
with $\widehat{S} = 2\widehat{S}\ms' + \widehat{R}$.  Here we have
used that for dimensional reasons $H(Q,Q;\mu)$ depends on $Q$ only in
the combination $Q/\mu$, so that the only $Q$-dependence in
$H(Q,Q;\mu=Q)$ is through the argument of the running coupling.  An
analogous statement holds for the $b$-dependence in
$\widetilde{U}(b^2;\mu=b_0/b)$ at small $b$.  The result
\eqref{FUUTb-expand} only involves the usual collinear distribution
and fragmentation functions, together with factors $H$, $\widehat{S}$,
$\widetilde{U}$, $\widehat{C}^{\ms\text{in}}$,
$\widehat{C}^{\ms\text{out}}$ whose perturbative expansions are free
of large logarithms.


\subsection{Collins-Soper-Sterman resummation}
\label{sec:CSS-resum}

We now turn to the region of large $q_T \gg M$, where one can evaluate
the hadron tensor in standard collinear factorization.  To leading
order in $\alpha_s$ we have
\begin{equation} \begin{split} 
& F_{UU, T} =
\frac{1}{Q^2}\,
\frac{\alpha_s}{(2\pi z)^2}
\sum_a x\ms e_a^2
\int_{x}^1 \frac{\de\xb}{\xb} \int_{z}^1 \frac{\de\zd}{\zd} \;
\delta\biggl(\frac{q_T^2}{Q^2} - 
  \frac{(1-\xb)(1-\zd)}{\xb \ms \zd}\biggr)
\\
& \quad \times \biggl[
  f_1^a \Bigl(\frac{x}{\xb}\Bigr)\,
  D_1^a \Bigl(\frac{z}{\zd}\Bigr)\,
  C_{UU, T}^{(\gamma^* q \to qg)}
+ f_1^a \Bigl(\frac{x}{\xb}\Bigr)\,
  D_1^g \Bigl(\frac{z}{\zd}\Bigr)\,
  C_{UU, T}^{(\gamma^* q \to gq)}
+ f_1^g \Bigl(\frac{x}{\xb}\Bigr)\,
  D_1^a \Bigl(\frac{z}{\zd}\Bigr)\,
  C_{UU, T}^{(\gamma^* g \to q\bar{q})}
\biggr]
\raisetag{-0.2em}\label{e:FUUThigh0}
\end{split} 
\end{equation}
with power corrections in $M/q_T$.  The hard-scattering coefficients
$C_{UU,T}$ for the indicated partonic subprocesses are functions of
$\hat x$, $\hat z$, and $q_T^{}/Q$, and will be given in
section~\ref{sec:large}.  Approximating \eqref{e:FUUThigh0} for $q_T
\ll Q$ one obtains
\begin{align}
F_{UU, T}
&= \frac{1}{q_T^2}\, \frac{\alpha_s}{2\pi^2 z^2}
\sum_a x\ms e_a^2\,
\biggl[f_1^a(x)\,D_1^a(z)\,L\biggl( \frac{Q^2}{q_T^2} \biggr)
+ f_1^a(x)\, \bigl( D_1^a \otimes P_{qq}
                     + D_1^g \otimes P_{gq} \bigr)(z)
\nonumber \\
& \qquad\hspace{7em}
+ \bigl( P_{qq} \otimes f_1^a 
+ P_{qg} \otimes f_1^g \bigr)(x)\, D_1^a(z)
\biggr]
\label{FUUTlog}
\end{align}
with power corrections in $q_T/Q$ and in $M/q_T$.  The factor $L$ is
defined as
\begin{equation}
L\biggl( \frac{Q^2}{q_T^2}\biggr) =
2 C_F \ln \frac{Q^2}{q_T^2} - 3 C_F \,,
\label{e:sudakovleading}
\end{equation} 
and $P_{qq}$, $P_{gq}$, $P_{qg}$ are the DGLAP splitting functions at
lowest order in $\alpha_s$, given in \eqref{splitting-fcts} below.
We see that a large logarithm of $Q^2/q_T^2$ appears in the
fixed-order calculation when $q_T \ll Q$.  Corresponding logarithms at
higher orders in $\alpha_s$ spoil the convergence of the perturbative
series.  Collins, Soper and Sterman \cite{Collins:1984kg} have shown
that these logarithms exponentiate and that their resummation results
in a factorized expression, which we will refer to as ``CSS
factorization''.  The discussion in the CSS paper is given for the
cross section of Drell-Yan production differential in $q_T^2$ but
integrated over the azimuthal angle of $\bm{q}_T$.  The corresponding
result for SIDIS is given in \cite{Nadolsky:1999kb} and can be written
as
\begin{align}
F_{UU,T} &=
\frac{1}{z^2}\ms \sum_a x\ms e_a^2
\int \frac{\de^2 \bm{b}}{(2\pi)^2}\, e^{-\ii \bm{b}\cdot \bm{q}_T}\,
\exp\bigl[- S(Q,b) \bigr]\,
\nonumber \\
& \quad \times
\sum_i \bigl( {C}^{\text{in}}_{ai}
   \otimes f_1^i \bigr)(x;\mu=b_0/b)\,
\sum_j \bigl( D_1^j
   \otimes {C}^{\text{out}}_{ja} \bigr)(z;\mu=b_0/b) \,.
\label{FCSS85}
\end{align}
This form is valid for $M \ll q_T \ll Q$.  It can be extended to the
full large-$q_T$ region, $q_T \gg M$, by adding the difference between
\eqref{e:FUUThigh0} and its approximated form \eqref{FUUTlog}.  For
further discussion of this matching of resummed and fixed-order terms
we refer to \cite{Kulesza:2002rh}.
As an aside, we remark that at $q_T \sim Q$ the longitudinal structure
function $F_{UU,L}$ is parametrically of the same order as $F_{UU,T}$,
whereas at $q_T \ll Q$ it is suppressed by a relative factor
$q_T^2/Q^2$.  One may hence also apply the CSS prescription to
$F_{UU,T} + \varepsilon F_{UU,L}$ or to $F_{UU,T} + F_{UU,L}$ instead
of $F_{UU,T}$.  The term to which resummation is applied is the same
in all cases, and only the unresummed part of the fixed-order
calculation is different.

Let us see how \eqref{FCSS85} reduces to \eqref{FUUTlog} at leading
order in $\alpha_s$.  The Sudakov factor $S(Q,b)$ reads
\begin{equation}
S(Q,b) = \int_{b_0^2/b^2}^{Q^2} \frac{\de\mu^2}{\mu^2}
\left[ A\bigl(\alpha_s(\mu)\bigr)\ln\frac{Q^2}{\mu^2}
     + B\bigl(\alpha_s(\mu)\bigr) \right]
\label{Sudakov}
\end{equation}
with
\begin{align}
A\left(\alpha_s\right) &=
\sum_{k=1}^\infty A_k\left(\frac{\alpha_s}{\pi}\right)^k \,,
&
B\left(\alpha_s\right) &=
\sum_{k=1}^\infty B_k\left(\frac{\alpha_s}{\pi}\right)^k \,,
\label{AandB}
\end{align}
where $A_1=C_F$ and $B_1= -3C_F/2$.  The coefficient functions
$C^{\text{in}}$ can be written as
\begin{equation}
C_{ai}^{\text{in}}(x;\mu=b_0/b) =
\delta_{ai}^{}\ms \delta(1-x)
+ \sum_{k=1}^\infty C_{ai}^{\text{in}\,\smash{(k)}}(x)\,
  \left(\frac{\alpha_s}{\pi}\right)^k \,,
\end{equation}
and an analogous expansion holds for $C^{\text{out}}$.  Using the
DGLAP equation we can evolve $f_1$ from the scale $\mu=b_0/b$ to
$\mu=Q$ and obtain
\begin{equation}
f_1^a(x;b_0/b) = f_1^a(x;Q)
 - \frac{\alpha_s}{2\pi}\,
   \bigl( P_{qq} \otimes f_1^a + P_{qg} \otimes f_1^g \bigr)(x)\,
   \ln\frac{b^2Q^2}{b_0^2} 
 + O(\alpha_s^2) \,.
\end{equation}
Evolving $D_1$ in the same way and putting everything together, we
obtain
%
%
\begin{align}
 F_{UU,T} &=
\frac{1}{z^2}\ms \sum_{a} x\ms e_a^2
\int \frac{\de^2 \bm{b}}{(2\pi)^2}\, e^{-\ii \bm{b}\cdot \bm{q}_T^{}}
\Biggl[ 1 - \frac{\alpha_s}{2\pi}\, C_F 
\left(\ln^2\frac{b^2Q^2}{b_0^2} - 3 \ln\frac{b^2Q^2}{b_0^2} \right)
\Biggr]
\nonumber\\
& \times
\Biggl[ f_1^a(x;Q) - \frac{\alpha_s}{2\pi}\,
  \bigl( P_{qq} \otimes f_1^a + P_{qg} \otimes f_1^g \bigr)(x)\,
    \ln\frac{b^2Q^2}{b_0^2} 
+ \frac{\alpha_s}{\pi} \sum_i
    \bigl( {C}^{\text{in}\,\smash{(1)}}_{ai} \otimes f_1^i \bigr)(x)
\Biggr]
\nonumber\\
& \times 
\Biggl[ D_1^a(z;Q) - \frac{\alpha_s}{2\pi}\,
   \bigl( D_1^a \otimes P_{qq} + D_1^g \otimes P_{gq} \bigr)(z)\,
   \ln\frac{b^2Q^2}{b_0^2}
+ \frac{\alpha_s}{\pi} \sum_j
    \bigl( D_1^j \otimes {C}^{\text{out}\,\smash{(1)}}_{ja} \bigr)(z)
 \Biggr]
\nonumber \\[0.3em]
&  + O(\alpha_s^2) \,.
\label{FCSS85c}
\end{align}
The running of $\alpha_s$ is irrelevant at the accuracy of this
expression.  Expanding the square brackets one obtains a term
$f_1^a(x;Q)\, D_1^a(z;Q)$ of order $\alpha_s^0$, which is independent
of $b$ and hence gives a contribution proportional to
$\delta^{(2)}(\bm{q}_T)$ to $F_{UU,T}$.  Since we require $q_T \gg M$,
this term must be discarded.  For the same reason, the first-order
coefficients ${C}^{\text{in}\,\smash{(1)}}_{ai}$ and
${C}^{\text{out}\,\smash{(1)}}_{ja}$ do not contribute to $F_{UU,T}$
at order $\alpha_s$.  With the integrals \cite{Nadolsky:2001sf}
\begin{align}
\int \de^2 \bm{b}\, e^{-\ii \bm{b}\cdot \bm{q}_T^{}}
  \ln^2\frac{b^2Q^2}{b_0^2} &=
  - \frac{8\pi}{q_T^2}\, \ln\frac{Q^2}{q_T^2} \,,
&
\int \de^2 \bm{b}\, e^{-\ii \bm{b}\cdot \bm{q}_T^{}}
  \ln\frac{b^2Q^2}{b_0^2} &=
  - \frac{4\pi}{q_T^2}
\label{NadolskyIds}
\end{align}
we recover the lowest-order result \eqref{FUUTlog} from
\eqref{FCSS85c}.  Going to higher orders in $\alpha_s$, one finds that
the term with $A_1$ in the Sudakov factor produces the leading
logarithms $\alpha_s^k\,\ln^{2k-1}\left(Q^2 /q_T^2\right)$ in
$F_{UU,T}$, whereas the next-to-leading logarithms
$\alpha_s^k\,\ln^{2k-2}\left(Q^2 /q_T^2\right)$ also receive
contributions from the coefficient $B_1$, from the one-loop running of
$\alpha_s$, and from the leading-order evolution of $f_1(x)$ and
$D_1(z)$.

The $\alpha_s$ expansion of the CSS factorization formula
\eqref{FCSS85}, which we have just performed to leading order, allows
one to determine the functions $S$, ${C}^{\text{in}}$ and
${C}^{\text{out}}$ at a given order in perturbation theory by
comparing with the collinear fixed-order calculation in the high-$q_T$
region.  The \emph{functional form} of \eqref{FCSS85} was however
derived by CSS using the result of CS factorization in the
intermediate region $M \ll q_T \ll Q$.  This is not immediately
obvious by comparing \eqref{FCSS85} with \eqref{FUUTb-expand}, because
in the former expression there is no hard and no soft factor.  As
pointed out in \cite{Catani:2000vq}, one can however introduce the
hard factor into the CSS expression.  $|H(\mu=Q)|^2$ depends on $Q$
only through the argument of $\alpha_s$, so that one can use the
renormalization group equation for the running coupling to rewrite it
as $|H(\mu=Q)|^2 = |H(\mu=b_0/b)|^2\; e^{R(Q,b)}$.  Since
${C}^{\text{in}}$ and ${C}^{\text{out}}$ in \eqref{FCSS85} are also
evaluated at $\mu= b_0/b$, we can combine factors into
${C}^{\ms\prime\,\text{in}} = |H|^{-1}\, {C}^{\text{in}}$ and
${C}^{\ms\prime\,\text{out}} = |H|^{-1}\, {C}^{\text{out}}$ at that
scale, which gives
\begin{align}
F_{UU,T} &=
\bigl| H(\mu=Q) \bigr|^2\;
\frac{1}{z^2}\ms \sum_a x\ms e_a^2
\int \frac{\de^2 \bm{b}}{(2\pi)^2}\, e^{-\ii \bm{b}\cdot \bm{q}_T}\,
\exp\bigl[- {S}^{\ms\prime}(Q,b) \bigr]\,
\nonumber \\
& \quad \times
\sum_i \bigl(
  C^{\ms\prime\,\text{in}}_{ai} \otimes f_1^i \bigr)(x;\mu=b_0/b)\,
\sum_j \bigl( 
  D_1^j \otimes C^{\ms\prime\,\text{out}}_{ja} \bigr)(z;\mu=b_0/b) \,.
  \label{FCSS85-hard}
\end{align}
with ${S}^{\ms\prime} = S + R$.  Identifying
${C}^{\ms\prime\,\text{in}} = \widetilde{U}^{1/2}\,
\widehat{C}^{\ms\text{in}}$, ${C}^{\ms\prime\,\text{out}} =
\widetilde{U}^{1/2}\, \widehat{C}^{\ms\text{out}}$, and
${S}^{\ms\prime} = \widehat{S}$, we finally recognize the CS result
\eqref{FUUTb-expand}.

The upshot of this discussion is essential in our context: the CSS
derivation of transverse momentum resummation for $F_{UU,T}$ and its
analogs in $e^+e^-$ annihilation and Drell-Yan production makes use of
two facts:
\begin{enumerate}
\item CS factorization is valid for these observables at $q_T \ll Q$,
  and
\item in the intermediate region $M \ll q_T \ll Q$ its results match
  those obtained from collinear factorization in the high-$q_T$
  region.
\end{enumerate}
In such a situation one can go further and construct expressions that
interpolate between the factorization formulae for low and for high
$q_T$ and are thus valid for all $q_T$, from $q_T=0$ to $q_T \sim Q$.
A prescription for this had already been given by CSS, and a number of
different methods have been proposed later; see \cite{Koike:2006fn}
for a discussion and references.  We shall not dwell on this issue
here.


\subsection{Azimuthal dependence and polarization}
\label{sec:polar}

So far we have discussed only $F^{}_{UU,T}$.  The situation for the
other unpolarized structure functions, $F^{}_{UU,L}$, $F_{UU}^{\cos
  \phi_h}$, $F_{UU}^{\cos 2\phi_h}$, cannot readily be inferred from
the results of CSS.  In section \ref{sec:large} we will see that the
splitting functions appearing in the analogs of \eqref{FUUTlog} for
these three structure functions differ from the usual ones.  Since the
splitting functions are relevant at next-to-leading logarithmic
accuracy, it is not clear if and how resummation beyond the leading
logarithmic approximation can be performed in this case.  An analogous
observation for the angular distribution in Drell-Yan production has
been made in \cite{Boer:2006eq}.  (Resummation at leading logarithmic
accuracy has recently been considered in \cite{Berger:2007jw}.)

The extension of CSS factorization for polarized scattering is
relatively straightforward as long as one integrates over the
azimuthal angle of $\bm{q}_T$.  For SIDIS this concerns the structure
function $F_{LL}$, and a corresponding calculation in this framework
has been presented in \cite{Koike:2006fn}.  CSS resummation for
Drell-Yan production with longitudinal beam polarization has been
investigated in \cite{Weber:1991wd,Nadolsky:2003fz}.  A detailed
discussion of polarization in the context of collinear factorization
can be found in \cite{Collins:1992xw}.

In the present work we will apply CS factorization to polarized
scattering and to the SIDIS cross section depending on the azimuth
$\phi_h$, as has been done in \cite{Ji:2004xq}.  The factors
$\Pslash\ms f_1^a(x,p_T^2; \zeta)$ and $\Pslash_h^{}\ms
D_1^{a}(z,{k}_T^{2}; \zeta_h)$ in the factorization formula
\eqref{Wsidis} are then replaced by the quark-quark correlator
$\Phi^a(x,\bm{p}_T^{}; \zeta)$ and the fragmentation correlator
$\Delta^a(z,\bm{k}_T^{};\zeta_h)$, which will be defined in section
\ref{sec:corr-def}.  The result reads
\begin{align}
W_{\text{SIDIS}}^{\mu\nu}  & \propto
\bigl| H\bigl(x\ms \zeta^{1/2}, z^{-1}\zeta_{\smash{h}}^{1/2} \ms\bigr)
\bigr|^2 \,
\sum_a e_a^2 
\int \de^2 \bm{p}_T^{}\, \de^2 \bm{k}_T^{}\, \de^2 \bm{l}_T^{}\,
\delta^{(2)}(\bm{p}_T^{}-\bm{k}_T^{}+\bm{l}_T^{}+\bm{q}_T^{}) \,
\nonumber \\
& \quad \times
\tr\left\{ \Phi^a(x,\bm{p}_T^{}; \zeta)\ms
  \gamma^\mu \Delta^a(z,\bm{k}_T^{};\zeta_h)\ms \gamma^\nu\right\} \,
U({l}_T^{2})
\label{Wsidis-full}
\end{align}
and gives rise to a number of spin and azimuthal asymmetries at
leading order in $1/Q$.  It is an open question if and how CS
factorization can be extended to power suppressed observables, at
least to those coming with one factor of $1/Q$.  The calculations of
our work are relevant to this question, as we shall see in section
\ref{sec:tailstructure}.

A structure function that has received much attention in the
literature is $F_{UT,T}^{\sin(\phi_h-\phi_S)}$, which arises when the
initial hadron is transversely polarized.  At low $q_T$ this
observable is nonzero due to the Sivers effect \cite{Sivers:1990cc}:
the CS factorization formula \eqref{Wsidis-full} gives a leading
contribution in $1/Q$ proportional to the Sivers function
$f_{1T}^\perp(x,p_T^2)$ \cite{Boer:1997nt}.  At high $q_T$ one can
describe the same observable in terms of the Qiu-Sterman mechanism
\cite{Qiu:1991pp}.  The corresponding calculation uses collinear
factorization at twist-three level, i.e.,\
$F_{UT,T}^{\sin(\phi_h-\phi_S)}$ is suppressed by $1/q_T$ compared
with $F_{UU,T}$.  Calculating the behavior of the Sivers function at
high transverse momentum, the analysis in
\cite{Ji:2006br,Koike:2007dg} has shown that at order $\alpha_s$ the
two descriptions exactly match in the intermediate region $M \ll q_T
\ll Q$.  The situation is the same for the corresponding asymmetry in
Drell-Yan production \cite{Ji:2006ub,Ji:2006vf,Koike:2007dg}.  This
suggests that for $F_{UT,T}^{\sin(\phi_h-\phi_S)}$ and its Drell-Yan
analog it should be possible to use the CSS resummation procedure for
large logarithms of $Q/q_T$, as discussed in \cite{Idilbi:2004vb}.

In the following sections we will derive the power behavior for the
full set of SIDIS structure functions, both in the low-$q_T$ and in
the high-$q_T$ description.  This will in particular determine whether
or not one can envisage to use CSS resummation for these observables.
To determine the power behavior we can restrict our calculations to
the leading order in $\alpha_s$.  Since we will not attempt to
actually perform a resummation of large logarithms, we need not go to
$b$-space as in \eqref{FUUTb-expand} or \eqref{FCSS85}.  We will
instead directly work with the momentum-space version
\eqref{Wsidis-full} of CS factorization.  In particular, we shall
recover the fixed-order result \eqref{FUUTlog} for the intermediate
region when expanding the CS expression \eqref{FUUTconv} of $F_{UU,T}$
in the limit $q_T \gg M$.


\section{From high to intermediate $q_T$}
\label{sec:large}

In this section we present the calculation of SIDIS structure
functions at high transverse momentum in terms of collinear
distribution and fragmentation functions, and then take the limit $q_T
\ll Q$.  We give explicit results for the six structure functions
appearing at leading twist and order $\alpha_s$.  For seven of the
remaining structure functions, which are of higher twist or higher
order in $\alpha_s$, there exist studies in the literature, which we
will briefly discuss.

\FIGURE[b]{
\includegraphics[width=14.5cm]{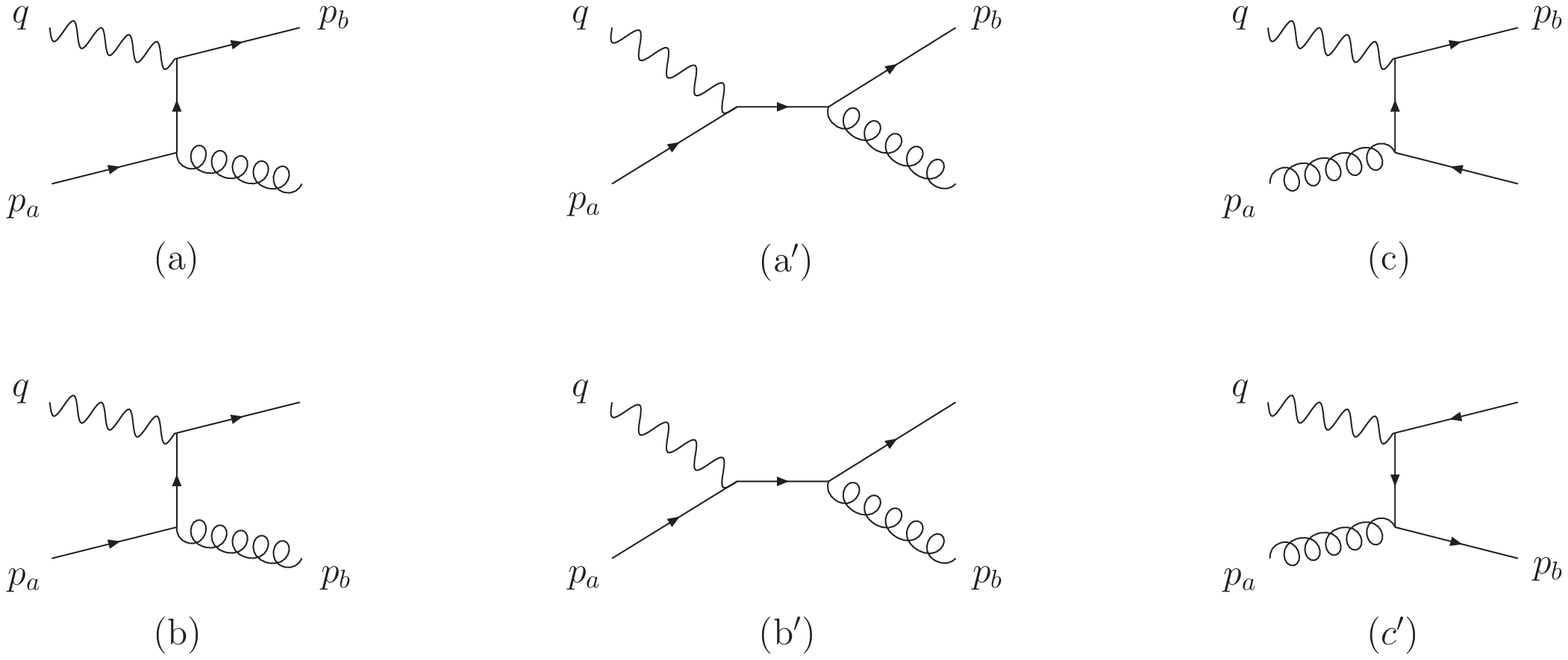}
\caption{\label{f:gqqG} Feynman diagrams for the processes $\gamma^* q
  \to qg$, $\gamma^* q \to gq$, and $\gamma^* g \to q\bar{q}$.}  
}

In the high-$q_T$ calculation, the generation of large transverse
momentum is described by hard-scattering processes at parton level.
The diagrams for the contributions at first order in $\alpha_s$ are
shown in Fig.~\ref{f:gqqG}.
We introduce the scaling variables
\begin{align}
\xb &= \frac{Q^2}{2 p_a \cdot q} \,,
&
\zd &= \frac{p_a\cdot p_b}{p_a \cdot q}
\end{align} 
for the partonic subprocess, where $p_a$ is the momentum of the
incoming parton and $p_b$ the momentum of the parton which fragments
into the observed hadron $h$.  We furthermore use the transverse
momentum $q_T^\mu$ introduced in \eqref{qT-def}.  Neglecting mass
corrections we have $\xb\ms p_a = x P$ and $p_b /\zd = P_h /z$, so
that
\begin{equation}
q_T^\mu = q^\mu_{\phantom{T}}
   + (1-r)\ms \hat{x}\ms p_{a\phantom{b}\!\!}^\mu - p_b^\mu /\zd
\end{equation}
with $r = q_T^2/Q^2$.  The partonic Mandelstam variables are then
given by
\begin{alignat}{4}
\hat{s} &= (q+p_a)^2 & &= \frac{1-\xb}{\xb}\, Q^2 \,,
\qquad
\hat{t} = (q-p_b)^2 = - \frac{1-\zd}{\xb}\, Q^2 
        = - \frac{\zd}{1-\xb}\, q_T^2 \,,
\nonumber \\[0.2em]
\hat{u} &= (p_a-p_b)^2 & &= - \frac{\zd}{\xb}\, Q^2 \,.
\end{alignat}

The structure functions defined in section \ref{sec:sidis} can be
written as convolutions of hard-scattering coefficients with collinear
parton distribution and fragmentation functions,
\begin{equation} \begin{split} 
& F_{UU, T} =
\frac{1}{Q^2}\,
\frac{\alpha_s}{(2\pi z)^2}
\sum_a x\ms e_a^2
\int_{x}^1 \frac{\de\xb}{\xb} \int_{z}^1 \frac{\de\zd}{\zd} \;
\delta\biggl(\frac{q_T^2}{Q^2} - 
  \frac{(1-\xb)(1-\zd)}{\xb \ms \zd}\biggr)
\\
& \quad \times \biggl[
  f_1^a \Bigl(\frac{x}{\xb}\Bigr)\,
  D_1^a \Bigl(\frac{z}{\zd}\Bigr)\,
  C_{UU, T}^{(\gamma^* q \to qg)}
+ f_1^a \Bigl(\frac{x}{\xb}\Bigr)\,
  D_1^g \Bigl(\frac{z}{\zd}\Bigr)\,
  C_{UU, T}^{(\gamma^* q \to gq)}
+ f_1^g \Bigl(\frac{x}{\xb}\Bigr)\,
  D_1^a \Bigl(\frac{z}{\zd}\Bigr)\,
  C_{UU, T}^{(\gamma^* g \to q\bar{q})}
\biggr] \,,
\raisetag{-0.2em}\label{e:FUUThigh}
\end{split} 
\end{equation}
as we have already seen in section \ref{sec:CSS-resum}.  We recall
that $a$ runs over flavors of quarks and of antiquarks.  Analogous
expressions with different kernels $C$ give the structure functions
$F_{UU,L}^{}$, $F_{UU}^{\cos\phi_h}$, and $F_{UU}^{\cos 2\phi_h}$.  At
order $\alpha_s$ (but not at higher order) one finds the relation
\begin{equation}
  \label{FL-vs-cos}
F_{UU,L}^{\phantom{\phi_h}} = 2 F_{UU}^{\cos 2\phi_h} \,.
\end{equation}
The structure functions $F_{LL}^{}$ and $F_{LL}^{\cos\phi_h}$ for
longitudinal target and beam polarization are also given by
expressions analogous to \eqref{e:FUUThigh}, with different kernels
$C$ and with the unpolarized parton densities $f_1^{a\phantom{g}\!\!}$
and $f_1^{g}$ replaced by their polarized counterparts
$g_1^{a\phantom{g}\!\!}$ and $g_1^{g}$.
The hard-scattering coefficients for the partonic processes $\gamma^*
q \to qg$, $\gamma^* q \to gq$, $\gamma^* g \to q\bar{q}$ can be
computed from the respective diagrams $(a,a')$, $(b,b')$, $(c,c')$ in
Fig.~\ref{f:gqqG}, and those for $\gamma^* \bar{q} \to \bar{q}g$,
$\gamma^* \bar{q} \to g\bar{q}$, $\gamma^* g \to \bar{q}q$ are
identical to their counterparts obtained by charge conjugation.
Process by process we have
\begin{itemize}
\item{$\gamma^* q \to qg$
\begin{align}
  \label{first-coeff}
C_{UU, T}
&= 2 C_F\,\biggl((1-\xb)(1-\zd)
   + \frac{1+\xb^2 \zd^2}{\xb \zd}\,\frac{Q^2}{q_T^2} \biggr) ,
\\
C_{UU}^{\cos\phi_h}
&= -4 C_F\ms \bigl[\ms \xb \zd + (1-\xb)(1-\zd) \ms\bigr]
   \,\frac{Q}{q_T} ,
\phantom{\biggl( \biggr)}
\\
C_{UU}^{\cos 2\phi_h}
&= 4 C_F\ms \xb \zd ,
\phantom{\biggl( \biggr)}
\\ 
C_{LL}
&= 2 C_F\,
\biggl( 2\ms(\xb+\zd)
  + \frac{\xb^2 + \zd^2}{\xb \zd}\,\frac{Q^2}{q_T^2}\biggr) ,
\\ 
C_{LL}^{\cos \phi_h}
&= -4 C_F\, (\xb+\zd-1) \,\frac{Q}{q_T} ,
\end{align} 
}
\item{$\gamma^* q \to gq$
\begin{align}
C_{UU, T}
&= 2 C_F\,\biggl((1-\xb)\,\zd
   + \frac{1+\xb^2 (1-\zd)^2}{\xb \zd}\,
  \frac{1-\zd}{\zd}\,\frac{Q^2}{q_T^2} \biggr) ,
\\
C_{UU}^{\cos\phi_h}
&= 4 C_F\ms \bigl[\ms \xb \,(1-\zd) + (1-\xb)\,\zd \ms\bigr]
   \,\frac{1-\zd}{\zd}\, \,\frac{Q}{q_T} ,
\\
C_{UU}^{\cos 2\phi_h}
&= 4 C_F\ms \xb \,(1- \zd) ,
\phantom{\biggl( \biggr)}
\\ 
C_{LL}
&= 2 C_F\, \biggl(2\xb + 2 (1-\zd)
   + \frac{\xb^2 + (1-\zd)^2}{\xb \zd}
     \,\frac{1-\zd}{\zd}\,\frac{Q^2}{q_T^2}\biggr) ,
\\
C_{LL}^{\cos \phi_h}
&= 4 C_F\, (\xb-\zd)
   \,\frac{1-\zd}{\zd}\, \frac{Q}{q_T} ,
\end{align} 
}
\item{$\gamma^* g \to q\bar{q}$
\begin{align} 
C_{UU, T}
&= 2 T_R\ms
  \bigl[\ms \xb^2+(1-\xb)^2 \ms\bigr]\ms
  \bigl[\ms \zd^2+(1-\zd)^2 \ms\bigr]\,
  \frac{1-\xb}{\xb \zd^2}\,\frac{Q^2}{q_T^2} ,
\\
C_{UU}^{\cos\phi_h}
&= -4 T_R\, (2\xb-1)\,(2\zd-1)\, \frac{1-\xb}{\zd}
   \,\frac{Q}{q_T} ,
\\
C_{UU}^{\cos 2\phi_h}
&= 8 T_R\, \xb\,(1-\xb) ,
   \phantom{\biggl( \biggr)}
\\ 
C_{LL}
&= 2 T_R\,
  (2\xb-1)\, \bigl[\ms \zd^2+(1-\zd)^2 \ms\bigr]\,
  \frac{1-\xb}{\xb \zd^2}\,\frac{Q^2}{q_T^2} ,
\\ 
  \label{last-coeff}
C_{LL}^{\cos \phi_h}
&= -4 T_R\, (2\zd-1) \frac{1-\xb}{\zd} \,\frac{Q}{q_T}
\end{align} 
}
\end{itemize}
with $C_F = 4/3$ and $T_R = 1/2$.  The relation $C_{UU,L}^{} = 2
C_{UU}^{\cos 2\phi_h}$ holds for each individual subprocess.  Our
results agree with those in~\cite{Mendez:1978zx,Koike:2006fn}.

The behavior of the above results in the region $q_T^2 \ll Q^2$ can be
obtained by rewriting the $\delta$ function in Eq.~\eqref{e:FUUThigh}
as~\cite{Meng:1995yn}
\begin{equation} \begin{split} 
 \delta\biggl(\frac{q_T^2}{Q^2}
     -\frac{(1-\xb)(1-\zd)}{\xb \ms \zd}\biggr)
&= \delta(1-\xb)\,\delta(1-\zd)\, \ln \frac{Q^2}{q_T^2}
+ \frac{\xb}{(1-\xb)_{+}} \; \delta(1-\zd)
\\
&\quad
+ \frac{\zd}{(1-\zd)_{+}} \; \delta(1-\xb)
+ \mathcal{O}\biggl(\frac{q_T^2}{Q^2}\ln\frac{Q^2}{q_T^2}\biggr) \,,
\label{e:deltaexpan}
\end{split} \end{equation} 
where the plus-distribution is as usual defined by
\begin{equation}
  \label{plus-def}
\int_z^1 \de y \, \frac{G(y)}{(1-y)_{+}} 
= \int_z^1 \de y \, \frac{G(y)-G(1)}{1-y} 
  - G(1)\ms \ln\frac{1}{1-z} \,.
\end{equation} 
We have written the hard-scattering coefficients in
\eqref{first-coeff} to \eqref{last-coeff} in a way that allows for an
easy extraction of the leading power behavior at small $q_T/Q$.  The
result is
\begin{align}
\label{e:high_FUU}
F_{UU, T}
&= \frac{1}{q_T^2}\, \frac{\alpha_s}{2\pi^2 z^2}
\sum_a x\ms e_a^2\,
\biggl[f_1^a(x)\,D_1^a(z)\,L\biggl( \frac{Q^2}{q_T^2} \biggr)
+ f_1^a(x)\, \bigl( D_1^a \otimes P_{qq}
                     + D_1^g \otimes P_{gq} \bigr)(z)
\nonumber \\
& \qquad\hspace{7em}
+ \bigl( P_{qq} \otimes f_1^a 
+ P_{qg} \otimes f_1^g \bigr)(x)\, D_1^a(z)
\biggr],
\\
F_{UU, L}^{} &= 2 F_{UU}^{\cos 2\phi_h} ,
\phantom{\biggl[ \biggr]}
\label{e:high_FUUL}
\\[0.2em]
\label{e:high_FUUcosphi}
F_{UU}^{\cos\phi_h}
&= - \frac{1}{Q\ms q_T}\, \frac{\alpha_s}{2\pi^2 z^2}
\sum_a x\ms e_a^2\,
\biggl[f_1^a(x)\,D_1^a(z)\,L\biggl( \frac{Q^2}{q_T^2} \biggr)
+ f_1^a(x)\, \bigl( D_1^a \otimes P_{qq}'
                     + D_1^g \otimes P_{gq}'\bigr)(z)
\nonumber \\
& \qquad\hspace{7em}
+ \bigl( P_{qq}' \otimes f_1^a 
+ P_{qg}' \otimes f_1^g\bigr)(x)\, D_1^a(z)
\biggr],
\\
\label{e:high_FUUcos2phi}
F_{UU}^{\cos 2\phi_h}
&= \frac{1}{Q^2}\, \frac{\alpha_s}{2\pi^2 z^2}
\sum_a x\ms e_a^2\,
\biggl[f_1^a(x)\,D_1^a(z)\,L\biggl( \frac{Q^2}{q_T^2} \biggr)
+ f_1^a(x)\, \bigl( D_1^a \otimes P_{qq}''
                     + D_1^g \otimes P_{gq}'' \bigr)(z)
\nonumber \\
& \qquad\hspace{7em}
+ \bigl( P_{qq}'' \otimes f_1^a 
+ P_{qg}'' \otimes f_1^g \bigr)(x)\, D_1^a(z)
\biggr],
\intertext{and}
\label{e:high_FLL}
F_{LL}
&= \frac{1}{q_T^2}\, \frac{\alpha_s}{2\pi^2 z^2}
\sum_a x\ms e_a^2\,
\biggl[g_1^a(x)\,D_1^a(z)\,L\biggl( \frac{Q^2}{q_T^2} \biggr)
+ g_1^a(x)\, \bigl( D_1^a \otimes P_{qq}
                     + D_1^g \otimes P_{gq} \bigr)(z)
\nonumber \\
& \qquad\hspace{7em}
+ \bigl( \Delta P_{qq} \otimes g_1^a 
+ \Delta P_{qg} \otimes g_1^g \bigr)(x)\, D_1^a(z)
\biggr],
\\
\label{e:high_LLcosphi}
F_{LL}^{\cos \phi_h}
&= - \frac{1}{Q\ms q_T}\, \frac{\alpha_s}{2\pi^2 z^2}
\sum_a x\ms e_a^2\,
\biggl[g_1^a(x)\,D_1^a(z)\,L\biggl( \frac{Q^2}{q_T^2} \biggr)
+ g_1^a(x)\, \bigl( D_1^a \otimes P'_{qq}
                     + D_1^g \otimes P'_{gq} \bigr)(z)
\nonumber \\
& \qquad\hspace{7em}
+ \bigl( \Delta P'_{qq} \otimes g_1^a
+ \Delta P'_{qg} \otimes g_1^g \bigr)(x)\, D_1^a(z)
\biggr] .
\end{align}
The factor $L$ contains a logarithm of $Q^2/q_T^2$ and is given in
\eqref{e:sudakovleading}, and the convolutions are defined in
\eqref{con-def}.  The splitting functions are given by
\begin{align} 
  \label{splitting-fcts}
P_{qq}(\xb) &= C_F \biggl[\frac{1+\xb^2}{(1-\xb)_{+}}
               + \frac{3}{2}\,\delta(1-\xb) \biggr] \,,
&
P_{qg}(\xb) &= T_R\, \bigl[\xb^2+(1-\xb)^2\bigr] \,,
\nonumber \\[0.2em]
P_{gq}(\zd) &=C_F\,\frac{1+(1-\zd)^2}{\zd} \,,
\\[10pt]
P_{qq}'(\xb) &= C_F \biggl[\frac{2\xb^2}{(1-\xb)_{+}}
                + \frac{3}{2}\,\delta(1-\xb) \biggr] \,,
&
P_{qg}'(\xb) &= 2T_R\, \xb\,(2\xb-1) \,,
\nonumber \\[0.3em]
P_{gq}'(\zd) &= -2 C_F\,(1-\zd) \,,
\\[10pt]
P_{qq}''(\xb) &= P_{qq}'(\xb) \,,
&
P_{qg}''(\xb) &= 4T_R\, \xb^2 \,,
\nonumber \\[0.4em]
P_{gq}''(\zd) &= 2C_F\ms \zd
\intertext{for convolutions with unpolarized distribution or
  fragmentation functions, and by}
\Delta P_{qq}(\xb) &= P_{qq}(\xb) \,,
&
\Delta P_{qg}(\xb) &= T_R\, (2\xb -1) \,,
\\[0.4em]
\Delta P_{qq}'(\xb) &= P_{qq}'(\xb) \,,
&
\Delta P_{qg}'(\xb) &= 2T_R\, \xb
\label{last-splitting-fct}
\end{align} 
for convolutions with polarized distributions.
Similar results for the Drell-Yan process have been obtained by Boer
and Vogelsang in Ref.~\cite{Boer:2006eq}.  We note that our $P_{qq}'$
and $P_{qq}''$ correspond to $P_{qq}^-$ in Eq.~(38) of
Ref.~\cite{Boer:2006eq}, while our $P_{qg}'$ corresponds to their
$\tilde{P}_{qg}^-$ and our $P_{qg}''$ to their $P_{qg}^{\prime -}/2$.

Let us remark that the $1/q_T^{}$ power behavior of $F_{UU}^{\cos
  \phi_h}$ and $F_{LL}^{\cos \phi_h}$ arises as $q_T^{} /q_T^2$, where
$1/q_T^2$ comes from the $t$-channel propagators in the
hard-scattering graphs of Fig.~\ref{f:gqqG}.  Likewise, the constant
$q_T^{}$ behavior (up to $\ln q_T^2$ terms) of $F_{UU,L}^{}$ and
$F_{UU}^{\cos 2\phi_h}$ arises as $q_T^2 /q_T^2$, where the $1/q_T^2$
from the hard propagators is fully canceled by numerator factors.  We
emphasize that, although they do not have a power-law divergence, the
above expressions for $F_{UU,L}^{}$ and $F_{UU}^{\cos 2\phi_h}$ cannot
be used for small $q_T$, because the approximations giving $1/q_T^2$
for the hard propagators break down when $q_T \lsim M$.  In fact,
angular momentum conservation requires $F_{UU}^{\cos \phi_h}$ and
$F_{LL}^{\cos \phi_h}$ to vanish like $q_T^{}$ and $F_{LL}^{\cos
  2\phi_h}$ to vanish like $q_T^2$ for $q_T^{} \to 0$, as shown e.g.\
in \cite{Diehl:2005pc}.

The six $F_{UT}$ structure functions for transverse target
polarization vanish in the leading-twist approximation at high
transverse momentum, because they would require the combination of the
transversity distribution function $h_1$ with a chiral-odd collinear
fragmentation function of twist two, which does not exist for an
unpolarized hadron.  They are however nonzero at twist three, where
collinear quark-gluon-quark and three-gluon correlation functions
appear, so that many more diagrams than the ones in Fig.~\ref{f:gqqG}
need to be computed.  Such a computation has been performed by Eguchi,
Koike and Tanaka in Refs.~\cite{Eguchi:2006qz,Eguchi:2006mc}.  The
results for the different structure functions involve the product of
$D_1$ with $G_F$ and $\widetilde{G}_F$, which are chiral-even
functions appearing in the decomposition of the quark-gluon-quark
distribution correlator given in \eqref{phi-G-def}.  Some observables
involve in addition the product of $h_1$ with $\widehat{E}_F$, which
is a chiral-odd function appearing in the decomposition of the
quark-gluon-quark fragmentation correlator.  The structure function
$F_{UT,T}^{\sin(\phi_h-\phi_S)}$ has also been computed in
\cite{Ji:2006br}.  The result differs from the one in
\cite{Eguchi:2006qz,Eguchi:2006mc} because both calculations were
missing certain terms.  With the corrections discussed in
\cite{Koike:2007dg}, agreement between the two groups has been
achieved.
We note, however, that the twist-three calculation of the $F_{UT}$
structure functions is presently not complete.  Terms involving
$\widehat{E}_F$ are only considered in \cite{Eguchi:2006qz}, where
they are found to contribute to
$\smash{F_{UT,T}^{\sin(\phi_h-\phi_S)}}$,
$\smash{F_{UT}^{\sin(\phi_h+\phi_S)}}$, and $F_{UT}^{\sin\phi_S}$.
The calculation of that work is restricted to so-called derivative
terms due to soft gluon pole contributions.  The remaining soft gluon
pole contributions, as well as contributions from soft fermion poles
and hard poles are evaluated in \cite{Eguchi:2006mc}, but only for
$G_F$ and $\widetilde{G}_F$ .  In \cite{Koike:2007dg} it is shown that
soft fermion pole contributions from further diagrams must be added to
those results.  Finally, all calculations in the literature are
restricted to quark-gluon-quark functions of twist three, so that
three-gluon correlators do not appear.

Using the soft gluon pole and the hard pole contributions computed in
\cite{Eguchi:2006qz,Eguchi:2006mc} we have extracted the leading
behavior of all $F_{UT}$ structure functions in the limit $q_T \ll Q$.
The structure of the results is listed in Eqs.~\eqref{e:high_FUTT} to
\eqref{e:high_FUTsin2phi} of section~\ref{sec:comparison}, both for
the power law and for the distribution and fragmentation functions
appearing in each observable.  We have verified that this structure is
not changed by the soft fermion pole contributions given in
\cite{Eguchi:2006mc}.  Since the corrections to \cite{Eguchi:2006mc}
discussed in \cite{Koike:2007dg} concern only soft fermion pole
contributions in $\smash{F_{UT}^{\sin(\phi_h+\phi_S)}}$ they do not
affect the structure of \eqref{e:high_FUTcollins} either.  The same
should hold for the remaining five $F_{UT}$ structure functions, but
this has not been checked explicitly.

The structure function $F_{LU}^{\sin\phi_h}$ is nonzero at twist two
and order $\alpha_s^2$.  Depending on a single polarization, it is a
$T$-odd observable and hence requires an absorptive part in the
amplitude, which in this case is provided by a loop in the
hard-scattering subprocess.  The relevant graphs have been calculated
in \cite{Hagiwara:1983cq}, and numerical estimates for specific
kinematics have been given in
\cite{Hagiwara:1983cq,Gehrmann:1995be}.
For the structure functions $F_{UL}^{\sin\phi_h}$ and $F_{UL}^{\sin
  2\phi_h}$ the situation is similar, but no explicit calculation
exists in the literature.  There is no contribution to
$F_{LU}^{\sin\phi_h}$, $F_{UL}^{\sin\phi_h}$, $F_{UL}^{\sin 2\phi_h}$
at twist three and order $\alpha_s$, because the necessary $T$-odd
terms would need to come from twist-three quark-gluon-quark
correlators.  For an unpolarized or longitudinally polarized hadron,
these are chiral-odd \cite{Eguchi:2006qz} and have no twist-two
chiral-odd partners in the other correlator.  {}From the calculation
in \cite{Hagiwara:1983cq} we can extract the power behavior of
$F_{LU}^{\sin\phi_h}$ for $q_T^{} \ll Q$, which we will give in
Eq.~\eqref{e:high_FLUsinphi_power}.


\section{From low to intermediate $q_T\ms$: power counting}
\label{sec:small}

In this section we derive the behavior of distribution and
fragmentation functions at high transverse momentum.  Plugging the
results into the known low-$q_T$ expressions of the SIDIS structure
functions, we will obtain their power behavior in the intermediate
region $M\ll q_T \ll Q$.  To begin with, we specify in the next two
subsections the distribution and fragmentation functions that will
appear in our calculation.


\subsection{Transverse-momentum-dependent distribution and
  fragmentation functions}  
\label{sec:corr-def}

For the discussion of distribution and fragmentation functions we use
light-cone coordinates defined with respect to the momenta $P$ and
$P_h$, which we already introduced at the end of section
\ref{sec:sidis}.  For any four-vector $a$ we then have the
plus-component $a^+ = a\cdot n_-$, the minus-component $a^- = a\cdot
n_+$, and the transverse part $a_T^\mu = a^\mu_{\phantom{+}\!} - a^+
n_+^\mu - a^- n_-^\mu$.  The hadron momenta read
\begin{align}
P^\mu   &= P^+ n_+^\mu + \frac{M^2}{2P^+}\, n_-^\mu \,,
&
P_h^\mu &= P_h^- n_-^\mu + \frac{M_h^2}{2P_h^-}\, n_+^\mu \,,
\end{align}
and the spin vector of the target can be decomposed into longitudinal
and transverse components as
\begin{equation}
S^\mu = S_L \, \biggl( \frac{P^+}{M}\, n_+^\mu
                 - \frac{M}{2P^+}\, n_-^\mu \biggr) + S_T^\mu \,.
\end{equation}
The transverse-momentum-dependent quark distributions appearing in the
description of SIDIS are defined from the quark-quark correlation
function
\begin{equation}  
  \label{e:phi}
\Phi_{ij}^{[{\mathcal U}]}(x,p_T)
= \int \frac{\de\xi^- \de^2 \bm{\xi}_T}{(2\pi)^{3}}\; 
  e^{\ii p \cdot \xi}\,
       \langle P|\bar{\psi}_j(0)\, \mathcal{U}_{(0,\xi)}\,
  \psi_i(\xi)|P \rangle \,\bigg|_{\xi^+=0} \,,
\end{equation}
where $p^+ = x P^+$ and summation over color indices is understood.
The corresponding correlation function for antiquarks is obtained by
replacing the quark fields by their transforms under charge
conjugation, see Ref.~\cite{Mulders:1995dh}.  The quark fields in
\eqref{e:phi} are renormalized fields, and the corresponding scale
dependence of the correlation function is given by a renormalization
group equation involving the quark anomalous dimension
\cite{Collins:1981uk}.

The gauge link $\mathcal{U}_{(0,\xi)}$ in \eqref{e:phi} is a Wilson
line that connects the quark fields and thus makes the correlation
function color gauge invariant.  In the factorization theorems for
scattering processes, the gauge link incorporates the exchange of
gluons between partons that move in the opposite light-cone directions
$n_+$ and $n_-$.  Consideration of gluons collinear to the target
yields Wilson lines with paths that point along $n_-$ and lead to
light-cone infinity, $a^- = \pm\infty$, where they are closed by
transverse segments from $\bm{0}_T$ to $\bm{\xi}{}_T$
\cite{Belitsky:2002sm,Boer:2003cm}.  Different processes require
different gauge links.  In particular, the simplest links closed at
$a^- = \pm\infty$, which we denote by $\mathcal{U}^{\pm}$, give rise
to the correlators $\Phi^{[+]}(x,p_T)$ and $\Phi^{[-]}(x,p_T)$
appearing in SIDIS and Drell-Yan production, respectively.  More
complicated gauge links show up in other processes
\cite{Bomhof:2004aw,Bomhof:2006dp}.

When defined with strictly lightlike Wilson lines, the correlator
\eqref{e:phi} contains divergences in gluon rapidity (sometimes
referred to as ``endpoint singularities'') and hence must be modified
\cite{Collins:2003fm}.  Different schemes have been discussed in the
literature.  One possibility is to use paths that point in a
non-lightlike direction $v$ instead of $n_-$
\cite{Ji:2004wu,Collins:2007ph}.  Up to subtle issues we will mention
in Appendix~\ref{app:FG}, this is equivalent to working in axial
gauge, $A\cdot v=0$, which was used in the original scheme of Collins
and Soper \cite{Collins:1981uk}.  In a number of different schemes,
the proton matrix element in \eqref{e:phi} is divided by vacuum
expectation values of suitably chosen Wilson lines
\cite{Collins:2007ph,Collins:2004nx,Hautmann:2007uw,Cherednikov:2007tw}.
The arguments in the present section use Lorentz invariance and power
counting, so that we need not specify the detailed choice of scheme.
As long as $v$ is a linear superposition of $n_+$ and $n_-$, no new
four-vector is introduced in $\Phi(x,p_T)$, which therefore depends on
$v$ only via the scalar parameter $\zeta = - (2P\cdot v)^2 /v^2$ we
already encountered in section \ref{sec:CS-fact}.

The correlation function \eqref{e:phi} can be parameterized in terms
of distributions functions depending on $x$ and $p_T^2$ as
\cite{Bacchetta:2006tn}
\begin{align}
  \label{phi-dec}
 & \Phi(x,p_T) 
   \phantom{\biggl[ \biggr]}
\nonumber \\
 & = \frac{1}{2}\, \biggl\{ 
f_1^{} \nslash_+ 
+ f_{1T}^\perp\, \frac{S_{T\rho}^{}\ms \epsilon_T^{\rho\sigma}
                       p_{T\sigma}^{}}{M} \, \nslash_+ 
+ g_{1L}^{} S_L^{} \gamma_5\nslash_+ 
- g_{1T}^{}\, \frac{S_T^{} \cdott p_T^{}}{M} \, \gamma_5\nslash_+ 
\nonumber \\[0.2em]
 & \qquad
+ h_{1}^{} \frac{\gamma_5 \bigl[\Sslash_T, \nslash_+ \bigr]}{2}
- h_{1T}^\perp\,
    \frac{S_{T\rho}^{}\, p_T^{(\rho}\ms p_T^{\sigma)}}{M^2}\,
    \frac{\gamma_5 \bigl[\gamma_\sigma, \nslash_+ \bigr]}{2}
+ h_{1L}^\perp S_L^{}\ms
         \frac{\gamma_5 \bigl[\pslash_T, \nslash_+ \bigr]}{2 M}
+ h_1^\perp\, \frac{i \bigl[\pslash_T, \nslash_+ \bigr]}{2M}
\biggr\} 
\nonumber \\[0.2em]
 & \quad + \frac{M}{2 P^+}\,\biggl\{ 
  e 
- e_L^{} S_L^{}\ms i \gamma_5
+ e_T^{}\ms \frac{S_T^{} \cdott p_T^{}}{M}\; i \gamma_5
+ e_T^\perp\, \frac{S_{T\rho}^{}\ms \epsilon_T^{\rho\sigma}
                    p_{T\sigma}^{}\ms }{M} 
\nonumber \\
 & \qquad
+ f^\perp \frac{\pslash_T}{M}
+ f_L^\perp S_L^{}\ms
      \frac{p_{T\rho}^{}\ms \epsilon_T^{\rho\sigma}}{M}\,
      \gamma_\sigma
- f_T^\perp\, \frac{S_{T\rho}^{}\, p_T^{(\rho}\ms p_T^{\sigma)}\,
                    \epsilon_{T \sigma\tau}^{}}{M^2}\, \gamma^\tau
+ f_T^{}\, S_{T\rho}^{}\ms \epsilon_T^{\rho\sigma} \gamma_\sigma^{}
\nonumber \\
 & \qquad
+ g_L^\perp S_L^{}\ms \frac{\gamma_5 \pslash_T}{M}
+ g^\perp\ms \frac{p_{T\rho}^{}\ms \epsilon_{T}^{\rho\sigma}}{M}\,
             \gamma_5 \gamma_\sigma
- g_T^\perp\, \frac{S_{T\rho}^{}\, p_T^{(\rho}\ms p_T^{\sigma)}}{M^2}\,
              \gamma_5 \gamma_\sigma
+ g_T^{}\ms \gamma_5 \Sslash_T
\nonumber \\[0.2em]
 & \qquad
+ h_T^\perp\, \frac{\gamma_5 \bigl[\Sslash_T, \pslash_T \bigr]}{2 M}
- h_T^{}\ms \frac{S_T^{} \cdott p_T^{}}{M}\,
      \frac{\gamma_5 [\nslash_+, \nslash_-]}{2}
+ h_L^{} S_L^{}\ms \frac{\gamma_5[\nslash_+, \nslash_-]}{2}
+ h\, \frac{i \bigl[\nslash_+, \nslash_- \bigr]}{2} 
\biggr\}
\nonumber \\[0.2em]
 & \quad + \frac{M^2}{2\ms (P^+)^2}\, \bigl\{ \, \ldots \, \bigr\} \,,
\end{align}
where the two-dimensional antisymmetric tensor is given by
\begin{equation}
\epsilon_T^{\alpha\beta} 
= \epsilon_{\phantom{T}}^{\alpha\beta\rho\sigma}\ms
  n_{+\rho}\ms n_{-\sigma}
\end{equation}
with $\epsilon^{0123} = 1$.  Index pairs in parentheses indicate that
the trace is subtracted in the two transverse dimensions,
\begin{equation}
p_T^{(\rho}\ms p_T^{\sigma)} =
p_T^\rho\ms p_T^{\sigma\phantom{\rho\!\!}} 
  - \frac{1}{2}\, (p_T^{} \cdott p_T^{})\, g_T^{\rho\sigma} \,,
\end{equation}
where the transverse metric tensor is
$g_T^{\alpha\beta} = g^{\alpha\beta}_{\phantom{T}} -
n_{+}^{\alpha\phantom{\beta}\!\!} n_{-}^{\beta} -
n_{-}^{\alpha\phantom{\beta}\!\!}  n_{+}^{\beta}$.
The first eight distributions in \eqref{phi-dec} are referred to as
twist two, and the next sixteen distributions as twist three.  The
$\ldots$ stand for the remaining eight distributions of twist four,
which are given in \cite{Goeke:2005hb}.  We will not need them in the
following and tacitly omit them in further parameterizations.
Corresponding to the Dirac matrix structure in the decomposition
\eqref{phi-dec}, functions denoted with letters $f$, $g$ or $e$, $h$
are respectively referred to as chiral-even or chiral-odd.  Functions
with subscripts $L$ or $T$ appear in the parts of $\Phi(x,p_T)$ that
depend on the longitudinal or transverse component of the spin vector.
(An exception to this rule of notation is the transversity
distribution $h_1$.)

It is understood that the correlator and each of the functions in
\eqref{phi-dec} should carry a label specifying the gauge link, as
well as a label for the quark flavor.  Time reversal connects
$\Phi^{[\mathcal{U}]}$ with $\Phi^{[\mathcal{U}^T]}$, and in
particular $\Phi^{[+]}$ with $\Phi^{[-]}$.  This provides
relations~\cite{Collins:2002kn}
\begin{equation}
  \label{T-reverse}
f^{[+]}(x,p_T^2) = \eta_f \ms f^{[-]}(x,p_T^2) \,,
\end{equation}
where $f$ stands for any of the distributions in \eqref{phi-dec}.  We
call a distribution $T$-even if $\eta_f = +1$ and $T$-odd if $\eta_f =
-1$.  The values of $\eta_f$ are given in table~\ref{tab:T-even-odd}.
We also anticipate in the table the power behavior
\begin{equation}
  \label{p-def}
f^{[\pm]}(x,p_T^2) \sim 1 /p_T^{n}
\end{equation}
of the distributions for $p_T \gg M$, which we shall derive in
section~\ref{sec:power}. 

\TABLE[t]{
\renewcommand{\arraystretch}{1.2}
\caption{\label{tab:T-even-odd} Behavior of distribution functions
  under time reversal and in the high-$p_T$ limit.  The time reversal
  factor $\eta_f$ is defined in \protect\eqref{T-reverse} and the
  exponent $n$ for the high-$p_T$ behavior in \protect\eqref{p-def}.}
\begin{tabular}{|c|ccc|ccc|cccccc|} \hline
 & \multicolumn{12}{|c|}{target polarization} \\ \cline{2-13}
 & \multicolumn{3}{|c|}{unpolarized}
 & \multicolumn{3}{c|}{longitudinal}
 & \multicolumn{6}{c|}{transverse} \\ \hline
 & $f_1^{}$ & $f^\perp$ & $g^\perp$ & 
   $g_{1L}^{}$ & $g_L^\perp$ &$ f_L^\perp$ &
   $f_{1T}^\perp$ & $f_T^\perp$ & $f_T^{}$ & 
   $g_{1T}^{}$ & $g_T^\perp$ & $g_T^{}$ \\
$\eta_f$ & $+$ & $+$ & $-$ & $+$ & $+$ & $-$ & 
           $-$ & $-$ & $-$ & $+$ & $+$ & $+$ \\
$n$ & 2 & 2 & 2 & 2 & 2 & 2 &
      4 & 4 & 2 & 4 & 4 & 2 \\ \hline
 & $h_1^\perp$ & $h$ & $e$ & $h_{1L}^\perp$ & $h_L^{}$ & $e_L^{}$ &
   $h_{1}^{}$ & $h_{1T}^\perp$ & $h_T^{}$ & $h_T^\perp$ &
   $e_T^{}$ & $e_T^\perp$ \\
$\eta_f$ & $-$ & $-$ & $+$ & $+$ & $+$ & $-$ &
           $+$ & $+$ & $+$ & $+$ & $-$ & $-$ \\
$n$ & 4 & 2 & 2 & 4 & 2 & 2 &
      2 & 4 & 2 & 2 & 2 & 2 \\ \hline
\end{tabular}
}

Fragmentation functions are defined from the correlator
\begin{align}
  \label{e:delta}
\Delta_{ij}(z,k_T) &= \frac{1}{2N_c\ms z}\, \sum_X \, \int
  \frac{\de\xi^+ \de^2\bm{\xi}_T}{(2\pi)^{3}}\; e^{\ii k \cdot \xi}\,
\nonumber \\
 & \quad\times
\langle 0|\, \mathcal{U}_{(\infty,\xi)} 
  \,\psi_i(\xi)| h, X\rangle_{\text{out}} \;
  {}_{\text{out}}\langle h, X| \bar{\psi}_j(0)\,
  \mathcal{U}_{(0,\infty)} |0\rangle \,\bigg|_{\xi^-=0} \,,
\end{align}
where $k^- = P_h^-/z$ and $N_c=3$.  The prefactor $1/(2N_c)$ comes
from averaging over the polarization and color of the fragmenting
quark.  The subscript $\infty$ in the gauge links indicates a
space-time point with plus-coordinate $a^+ =\infty$.  The precise
choice of Wilson lines involves the same issues we mentioned for the
distribution correlator.  Aspects specific to the case of
fragmentation functions are discussed in
\cite{Collins:2004nx,Bomhof:2006dp}.
Notice that fragmentation functions with different gauge links are
\emph{not} related by time reversal, because time reversal transforms
``out'' states $|h, X\rangle_{\text{out}}$ into ``in'' states $|h,
X\rangle_{\text{in}}$, and the difference between these states amounts
in general to more than just a phase.

For an unpolarized hadron $h$ the decomposition of the fragmentation
correlator reads
\begin{align} 
  \label{eq:delta} 
\Delta(z, k_T) &= 
\frac{1}{2} \, \biggl\{ D_1 \nslash_- 
+ H_1^\perp\ms
  \frac{i \bigl[ \kslash_T, \nslash_- \bigr]}{2M_h} \biggr\} 
\nonumber \\[0.2em]
 & \quad + \frac{M_h}{2 P_h^-}\,
\biggl\{ E + D^\perp \frac{\kslash_T}{M_h}
+ H \frac{ i \bigl[\nslash_-, \nslash_+ \bigr]}{2} 
- G^\perp \frac{k_{T\rho}^{}\ms \epsilon_{T}^{\rho\sigma}}{M_h}\,
          \gamma_5 \gamma_\sigma
\biggr\} \,,
\end{align}
where the functions on the r.h.s.\ depend on $z$ and $k_T^2$.  In a
more explicit notation they should also carry a flavor index.


\subsection{Collinear distribution and fragmentation functions}
\label{sec:collinear}

In applications of collinear factorization, the structure of incoming
hadrons is represented by the light-cone distribution correlator
\begin{equation}
  \label{e:phi-coll}
\Phi_{ij}(x)= \int \frac{\de \xi^-}{2\pi}\; 
 e^{\ii p \cdot \xi}\,
       \langle P|\bar{\psi}_j(0)\,
\mathcal{U}^{n_-}_{(0,\xi)}\,
\psi_i(\xi)|P \rangle \bigg|_{\xi^+=0,\, \bm{\xi}_T = \bm{0}_T} \,,
\end{equation}
where the gauge link $\mathcal{U}^{n-}_{(0,\xi)}$ connects the quark
fields along a path in the $n_-$ direction.
This would simply be the integral of the $p_T$-dependent correlator
introduced in the previous subsection,
\begin{equation}
  \label{pt-int}
  \Phi(x) = \int \de^2 \bm{p}_T\, \Phi^{[\cal U]}(x,p_T) \,,
\end{equation}
were it not for two complications.  On the one hand, the correlator
\eqref{e:phi-coll} has ultraviolet divergences due to the fact that
all field operators are taken at the same transverse position.  Their
regularization and subtraction gives rise to a scale dependence
described by the DGLAP equations.  Correspondingly, the integrand on
the r.h.s.\ of \eqref{pt-int} diverges like $1 /p_T^{2}$ at high
$p_T^{}$, as we will see in the next subsection, so that the
$p_T^{}$-integral must be regularized.
On the other hand, the rapidity divergences of the $p_T$-dependent
correlator, which we discussed in the previous subsection, cancel
under the integral over $p_T$ \cite{Collins:2003fm,Hautmann:2007uw}.
They require no regularization in the collinear correlation function
\eqref{e:phi-coll}, which hence is independent of the parameter
$\zeta$.  The different regularization procedures in the correlators
$\Phi(x,p_T)$ and $\Phi(x)$ reflect the different types of
subtractions required when constructing transverse-momentum-dependent
or collinear factorization theorems.  We will shortly discuss how the
relation \eqref{pt-int} should be understood.

The correlation function \eqref{e:phi-coll} can be parameterized as
\begin{align}
  \label{eq:phix}
\Phi(x) & = \frac{1}{2}\, \biggl\{ 
f_1^{} \nslash_+ 
+ g_{1}^{} S_L^{} \gamma_5\nslash_+
+ h_{1}^{} \frac{\gamma_5 \bigl[\Sslash_T, \nslash_+ \bigr]}{2}
\biggr\} 
+ \frac{M}{2 P^+}\,\biggl\{ 
e 
- e_L^{} S_L^{}\ms i \gamma_5
\nonumber \\[0.2em]
 & \quad
+ f_T^{}\ms S_{T\rho}^{}\ms \epsilon_T^{\rho\sigma} \gamma_\sigma^{}
+ g_T^{}\ms \gamma_5 \Sslash_T
+ h_L^{} S_L^{}\ms \frac{\gamma_5[\nslash_+, \nslash_-]}{2}
+ h\, \frac{i \bigl[\nslash_+, \nslash_- \bigr]}{2} 
\biggr\} \,,
\end{align}
where the distributions on the r.h.s.\ depend only on $x$.  They are
given by
\begin{equation}
  \label{f1-int}
f_1(x) =  \int \de^2 \bm{p}_T\; f_{1}^{}(x,p_T^2)
\end{equation} 
and similarly for the other functions, with one common exception of
notation,
\begin{equation}
  \label{g1-int}
g_1(x) = \int \de^2 \bm{p}_T\; g_{1L}^{}(x,p_T^2) \,.
\end{equation}
Since the gauge link in \eqref{e:phi-coll} is along a finite
light-like path from $0$ to $\xi$, time reversal relates the collinear
correlator with itself, and as a consequence $f_T(x) = e_L(x) = h(x) =
0$.  This ensures that \eqref{pt-int} can simultaneously hold for
different links $\mathcal{U}$, in particular for $\mathcal{U}^{+}$ and
$\mathcal{U}^{-}$, which according to \eqref{T-reverse} give
$p_T$-dependent distributions $f_T^{[\pm]}$, $e_L^{[\pm]}$, and
$h_{\phantom{T}}^{[\pm]}$ of opposite sign.

To make the meaning of \eqref{pt-int} more precise, we observe that
the combination of Eqs.~\eqref{CS-solve} and \eqref{small-b-exp} gives
\begin{align}
  \label{small-b-relation}
\tilde{f}_1(x,b; \zeta,\mu)
&= \int \de^2 \bm{p}_T\, e^{\ii \bm{b}\cdot \bm{p}_T}\,
   f_1(x,p_T^2;\zeta,\mu)
 = f_1(x;\mu)
   + O\bigl( \alpha_s \ln^2(\zeta b^2) \bigr)
\end{align}
for small enough $b$, where we have set $\mu = b_0/b$ and used the
perturbative expansions $\widehat{S}\ms' = {O}\bigl( \alpha_s
\ln^2(\zeta b^2) \bigr)$ and $\widehat{C}^{\ms\text{in}}_{ai} =
\delta_{ai}^{}\ms \delta(1-x) + {O}(\alpha_s)$.  The factor $e^{\ii
  \bm{b}\cdot \bm{p}_T}$ in \eqref{small-b-relation} regulates the
logarithmic divergence of $\int \de^2 \bm{p}_T\, f_1(x,p_T^2)$ by
damping the integrand for large $p_T > 1/b$.
Alternatively one can cut off the integral at $p_T = b_0/b$, since
\begin{align}
  \label{cutoff}
\int \de^2 \bm{p}_T\; e^{\ii \bm{b}\cdot \bm{p}_T}\,
  f_1(x,p_T^2;\zeta,\mu)
= \int \de^2 \bm{p}_T\; \theta\bigl( \mu^2-\bm{p}_T^2 \bigr)\,
  f_1(x,p_T^2;\zeta,\mu)
\end{align}
up to corrections of order $b^2$, as we will show in
appendix~\ref{app:cutoff}.
We thus see that the relation \eqref{pt-int} should be understood with
a suitable regulator of the integral on the r.h.s.\ and as up to
corrections of order $\alpha_s$.  The same holds for \eqref{f1-int},
\eqref{g1-int}, and for similar integral relations in the following.
Let us remark that an extension of \eqref{small-b-relation} to the
full correlation function $\Phi$ has not been given in the literature.

As we will see in the next subsection, calculations at subleading
power or those involving azimuthal asymmetries lead in the collinear
expansion to $p_T$-weighted correlation functions
\begin{equation}
  \label{pT-mom-def}
\Phi_\partial^{\alpha [\pm]}(x)
  = \int \de^2 \bm{p}_T\, p_T^\alpha\, \Phi^{[\pm]}(x,p_T) \,,
\end{equation}
where the Lorentz index $\alpha$ is restricted to be transverse, and
where the same remarks about regularization apply as for
\eqref{pt-int}.  In contrast to $\Phi(x)$, the correlator
$\Phi_\partial^{\alpha [\mathcal{U}]}(x)$ does depend on the choice of
Wilson lines in $\Phi^{[\mathcal{U}]}(x,p_T)$ and hence contains both
$T$-even and $T$-odd distributions.  Omitting the superscript
$[\mathcal{U}]$ for the sake of legibility, we have the decomposition
\begin{align}
  \label{pT-moment}
\Phi_\partial^{\alpha}(x)
& = - \frac{M}{2}\, \biggl\{
  f_{1T}^{\perp (1)}\ms S_{T\rho}^{}\ms \epsilon_T^{\rho\alpha}
               \, \nslash_+ 
- g_{1T}^{(1)}\, S_T^{\alpha} \, \gamma_5\nslash_+ 
\nonumber \\[0.2em]
 & \qquad\quad
+ h_{1L}^{\perp (1)} S_L^{}\ms
         \frac{\gamma_5 \bigl[\gamma_T^\alpha, \nslash_+ \bigr]}{2}
+ h_1^{\perp (1)}\;
         \frac{i \bigl[\gamma_T^\alpha, \nslash_+ \bigr]}{2}
\biggr\}
+ \frac{M^2}{2 P^+}\, \bigl\{ \ldots \bigr\} \,,
\end{align}
where we have only displayed the terms of leading twist and defined
$\bm{p}_T^2$ moments
\begin{equation}
  \label{mom-notation}
f_{1T}^{\perp (n)}(x) = \int \de^2 \bm{p}_T\,
   \biggl( \frac{\bm{p}_T^2}{2M^2} \biggr)^n f_{1T}^{\perp}(x,p_T^2)
   \,, 
\end{equation}
and similarly for the other functions.  The functions $f_{1T}^{\perp
  (n)}$ and $h_{1\phantom{T}}^{\perp (n)}$ are $T$-odd and thus change
sign when going from $\Phi_\partial^{\alpha\ms \smash{[+]}}(x)$ to
$\Phi_\partial^{\alpha\ms \smash{[-]}}(x)$.

The factor $p_T^{\alpha}$ in \eqref{pT-mom-def} can be converted into
a derivative $\partial^\alpha$ acting on the matrix element that
appears in the definition \eqref{e:phi} of $\Phi^{}(x,p_T)$.  One can
then express $\Phi_\partial^\alpha(x)$ in terms of correlators with
either a gluon field or a covariant derivative between the antiquark
and quark fields.  The former is a collinear quark-antiquark-gluon
correlation function, whereas the latter can be rewritten in terms of
the quark-quark correlator $\Phi(x)$ using the equation of motion for
the quark field.  In this way, the $\bm{p}_T^2$ moments given in
\eqref{pT-moment} can be traded for functions of twist three, up to
twist-two distributions multiplied by the quark mass
\cite{Bacchetta:2006tn}.

The $k_T$-integrated fragmentation correlator for an unpolarized
hadron has the decomposition
\begin{align} 
  \label{eq:deltaz} 
\Delta(z) &= z^2 \int \de^2 \bm{k}_T\, \Delta(z,k_T)
= \frac{1}{2}\ms D_1 \nslash_- 
+ \frac{M_h}{2 P_h^-}\, \biggl\{ 
E
+ H\ms \frac{ i \bigl[\nslash_-, \nslash_+ \bigr]}{2} 
\biggr\} \,,
\end{align}
where the fragmentation functions on the r.h.s.\ depend on $z$.  They
are given by
\begin{equation}
  \label{D1-int}
  D_1 (z) = z^2 \int \de^2 \bm{k}_T\,D_1 (z, k_T^2) \,,
\end{equation} 
and similarly for the other functions.  Notice the factor $z^2$, which
appears because $D_1(z,k_T^2)$ is a probability density w.r.t.\ the
transverse momentum $k'_T = -z k_T^{}$ of the final-state hadron
relative to the fragmenting quark
\cite{Collins:1982uw,Mulders:1995dh}.  As already discussed, time
reversal invariance does not constrain fragmentation correlators, so
that $H(z)$ can be nonzero unlike its distribution counterpart $h(x)$.
For the $k_T$-weighted correlation function needed in calculations at
twist three and higher, we have
\begin{equation}
\Delta_\partial^\alpha(z) 
 = z^2 \int \de^2 \bm{k}_T^{}\, k_T^\alpha\, \Delta(z,k_T)
 = - \frac{M_h}{2}\ms H_1^{\perp (1)}\,
             \frac{i \bigl[ \gamma_T^\alpha, \nslash_- \bigr]}{2}
   + \frac{M_h^2}{2 P_h^-}\, \bigl\{ \ldots \bigr\}
\end{equation}
with
\begin{equation}
H_1^{\perp (n)}(z) = z^2 \int \de^2 \bm{k}_T\,
   \biggl( \frac{\bm{k}_T^2}{2 M_h^2} \biggr)^n 
   H_1^{\perp}(z,k_T^2) \,,
\end{equation}
where again $\alpha$ is restricted to be transverse.


\subsection{Distribution and fragmentation
  functions at high transverse momentum} 
\label{sec:power}

We are now ready to derive the behavior of correlation functions at
high transverse momentum.  We consider the distribution correlator
$\Phi(x,p_T)$ for transverse momentum $p_T$ much larger than the scale
of nonperturbative interactions.  The generation of the large
transverse momentum can be described in perturbation theory.
Technically, we approximate $\Phi(x,p_T)$ in powers of $1/p_T$ using a
collinear expansion that leads to the factorization of the transverse
momentum dependence.  To derive a formal proof of factorization, one
would use the same techniques as for, say, the production of a
high-$p_T$ jet in deep inelastic scattering.  We shall not attempt
this here, but limit ourselves to determining the power-law behavior
of the distribution functions that parameterize $\Phi(x,p_T)$, using
Lorentz invariance and dimensional analysis as our main tools.  The
explicit calculation in section \ref{sec:tails} and the one for the
Sivers function $f_{1T}^\perp$ in \cite{Ji:2006vf,Koike:2007dg}
provide examples for the consistency of the collinear factorization
formalism at leading order in $\alpha_s$ and to leading and first
subleading power in $1/p_T$.  One should, however, be aware that
factorization might break down at some higher-order or higher-power
accuracy.

\FIGURE[t]{
\includegraphics[width=11cm]{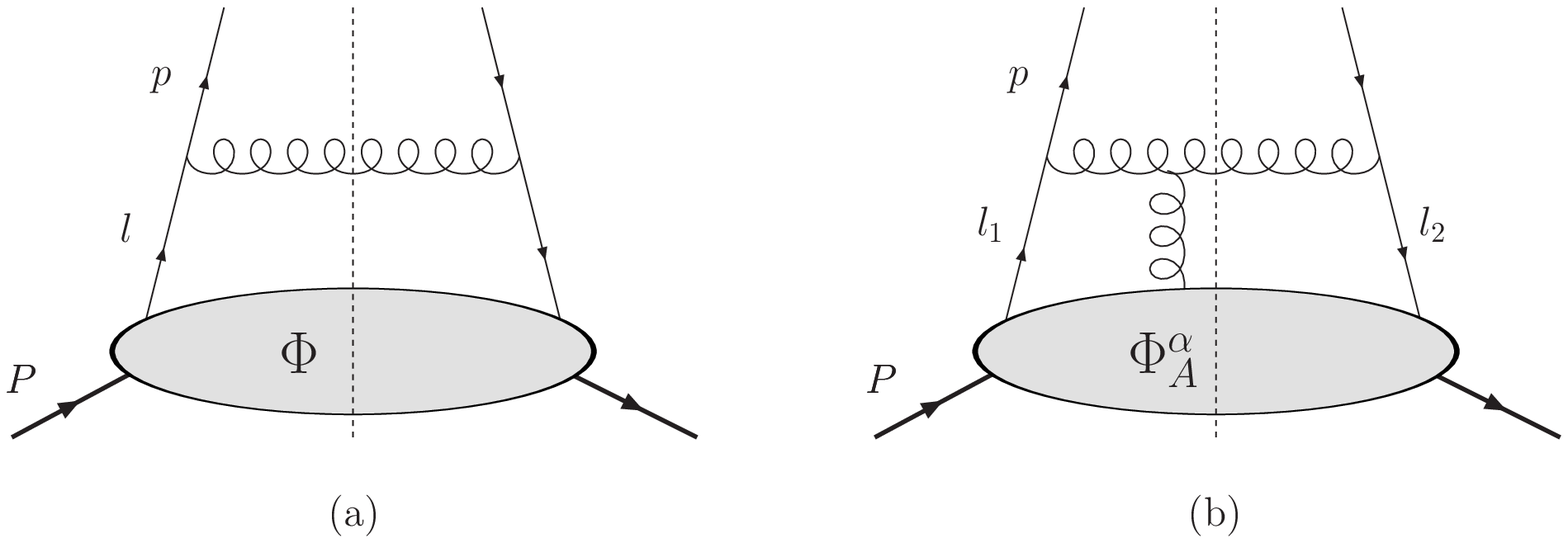}
\caption{\label{f:rungdistA} Example diagrams for the calculation of
  the high-$p_T$ behavior of the quark-quark correlator
  $\Phi(x,p_T)$.  The dashed lines represent the final-state cut.}
}

The evaluation of $\Phi(x,p_T)$ at lowest order in $1/p_T$ involves
diagrams as in Fig.~\ref{f:rungdistA}a, whereas at higher orders
correlators with three or more partons appear at the bottom of the
graphs, as shown in Fig.~\ref{f:rungdistA}b.  To set up the power
counting, we generically write $p$ for the hard scale and take
\begin{alignat}{3}
  \label{pow-ct-0}
p^+, p^-, \bm{p}_T \sim \;&\; P^+ \sim \;& l^+, l_1^+, l_2^+
\;\sim \;&\; p \,,
\\
  \label{pow-ct-1}
 & & \bm{l}_T, \bm{l}_{1T}, \bm{l}_{2T} \;\sim \;&\; M \,, 
\\
  \label{pow-ct-2}
 &\; P^-  \sim \;& l^-, l_1^-, l_2^- \;\sim \;&\; M^2/p \,,
\end{alignat}
where the mass $M$ represents the soft scale.  In \eqref{pow-ct-2} we
have used that the loop momenta $l$, $l_1$ and $l_2$ are attached to a
soft function at the bottom of the graphs and thus have virtualities
of order $M^2$.
Starting point of the calculation is the correlation function
depending on the full four-momentum $p$,
\begin{equation}
\Phi_{ij}^{[\pm]}(p) = \int \frac{\de^4 \xi}{(2\pi)^4}\;
  e^{\ii p \cdot \xi}\, \langle P|\bar{\psi}_j(0)\,
     \mathcal{U}_{(0,\xi)}^{\pm}\, \psi_i(\xi)|P \rangle \,,
\end{equation}
from which we obtain $\Phi^{[\pm]}(x,p_T) = \int \de p^-\,
\Phi^{[\pm]}(p)$.  We omit the superscript $[\pm]$ for clarity of
notation in the next few steps, and will restore it when required
later on.  We restrict ourselves to the leading and first subleading
order of $\Phi(x,p_T)$ in the $1/p$ expansion.  To this order, the
relevant factorized graphs can be written as the convolution of
hard-scattering kernels and correlation functions in the form
\begin{align}
  \label{hi-pT-start}
\Phi(x, p_T) &=
\int \de p^-\ms d^4 l\; H(p,l)\, \Phi(l)
+ \int \de p^-\ms d^4 l_1\, d^4 l_2\;
  H_{A}^\alpha(p,l_1,l_2)\, \Phi^{}_{A\ms \alpha}(l_1,l_2)
\nonumber \\
& \quad + \{ \text{terms with two- and three-gluon correlators} \}
  + \ldots
\phantom{\int}
\end{align}
with $\ldots$ representing terms of higher order in $1/p$.  It is
understood that the hard-scattering kernels $H(p,l)$ and
$H_A^\alpha(p,l_1,l_2)$ include $\delta$ functions putting the cut
lines on shell---this can readily be used to perform the integration
over $p^-$.
The lower blob in Fig.~\ref{f:rungdistA}b is parameterized by the
quark-gluon-quark correlator $\Phi_{A}^\alpha$, which contains the
gluon potential $A^\alpha$ between the quark and antiquark fields.
The gluon polarization index $\alpha$ in \eqref{hi-pT-start} is
restricted to be transverse: the contribution from $A^-$ gluons is
power suppressed by at least $1/p^2$, whereas the corresponding
contribution of $A^+$ gluons ends up in the gauge link of the
quark-quark correlator when the terms in the factorization formula are
arranged in a gauge-invariant manner.

We now expand the hard-scattering kernels in the small momentum
components \eqref{pow-ct-1} and \eqref{pow-ct-2}.  To the order we are
considering, we can neglect $l_1^-$, $l_2^-$ and $l_{1T}^{}$,
$l_{2T}^{}$ in $H_A^\alpha(p,l_1,l_2)$, whereas in $H(p,l)$ we can
neglect $l^-$ but must expand the $l_T$-dependence up to first order.
This gives
\begin{align}
  \label{H23-def}
\int \de p^-\, H(p,l)
&= \frac{1}{p^+}\, \hat{x}\ms H_2(\hat{x}, p^+, p_T) 
  + \frac{l_{T \alpha}}{p^+}\, \hat{x}\ms H_3^\alpha(\hat{x}, p^+, p_T)
  + \ldots \,,
\\
\int \de p^-\, H_{A}^\alpha(p,l_1,l_2)
&= \frac{1}{p^+}\, \hat{x}_1^{} \hat{x}_2^{} \,
  H_{A,3}^\alpha(\hat{x}_1^{},\hat{x}_2^{}, p^+, p_T)
  + \ldots \,,
\end{align}
where we have introduced the plus-momentum fractions
\begin{equation}
  \label{x-hat-def}
  \hat{x} = p^+ /\, l^+ \,, 
\qquad 
  \hat{x}_1^{} = p^+ /\, l^+_1 \,, 
\qquad
  \hat{x}_2^{} = p^+ /\, l^+_2
\end{equation}
and chosen prefactors such that $H_2^{}$, $H_3^\alpha$ and
$H_{A,3}^\alpha$ are invariant under longitudinal boosts, i.e.\ under
the rescaling \eqref{boost} of $n_\pm$ and the corresponding change of
plus- and minus components.
The convolution \eqref{hi-pT-start} now takes the form
\begin{align}
  \label{hi-pT-next}
& \Phi(x, p_T)
\nonumber \\[0.2em]
 &\quad
= \int \frac{\de\hat{x}}{\hat{x}}\, H_2(\hat{x}, p^+, p_T) 
  \int \de\ms l^-\, \de^2 \bm{l}_T\, \Phi(l)
+ \int \frac{\de\hat{x}}{\hat{x}}\, H_3^\alpha(\hat{x}, p^+, p_T) 
  \int \de\ms l^-\, \de^2 \bm{l}_T\, l_{T\alpha}\, \Phi(l)
\nonumber \\[0.2em]
 &\qquad + p^+
  \int \frac{\de\hat{x}_1}{\hat{x}_1}\,
       \frac{\de\hat{x}_2}{\hat{x}_2}\,
    H_{A,3}^\alpha(\hat{x}_1^{},\hat{x}_2^{}, p^+, p_T)
  \int \de\ms l^-_1\, \de\ms l^-_2\,
       \de^2 \bm{l}_{1T}^{}\, \de^2 \bm{l}_{2T}^{}\,
    \Phi_{A\ms \alpha}(l_1,l_2)
\nonumber \\
& \qquad + \{ \text{terms with two- and three-gluon correlators} \}
  + \ldots \,.
\phantom{\int}
\end{align}
In the first two terms we recognize the collinear quark-quark
correlators from \eqref{pt-int} and \eqref{pT-mom-def},
\begin{align}
  \label{Phi-coll}
\Phi(y) &= \int \de\ms l^-\ms \de^2 \bm{l}_T\, \Phi(l) \,,
&
\Phi_\partial^\alpha(y)
  &= \int \de\ms l^-\ms \de^2 \bm{l}_T\; l_T^\alpha\, \Phi(l)
\end{align}
with $y = l^+/P^+$, whereas the third term involves a collinear
quark-gluon-quark correlation function
\begin{align}
  \label{PhiA-coll}
\Phi_A^\alpha(y_1,y_2) &=
\int \de\ms l^-_1\, \de\ms l^-_2\,
     \de^2 \bm{l}_{1T}^{}\, \de^2 \bm{l}_{2T}^{}\;
     \Phi_{A}^\alpha(l_1,l_2)
\end{align}
with $y_1^{} = l_1^+/P^+$ and $y_2^{} = l_2^+/P^+$.  In order for
these correlators to be gauge invariant one must reshuffle certain
pieces among the different terms in \eqref{hi-pT-next}, as shown for
instance in \cite{Boer:2003cm}.  On the r.h.s.\ of \eqref{Phi-coll}
and \eqref{PhiA-coll} this implies subtraction of terms with the gluon
potential $A^\alpha$ at light-cone infinity, which we have not
displayed.  One also finds that taking the gauge link
$\mathcal{U}^\pm$ in $\Phi^{[\pm]}(x,p_T)$ leads to the corresponding
path-dependent correlators $\Phi_\partial^{\alpha [\pm]}(y)$ and
\begin{equation}
\Phi_A^{\alpha [\pm]}(y_1,y_2) =
\frac{1}{\ii P^+}\,
  \frac{\Phi_G^\alpha(y_1,y_2)}{y_1-y_2 \pm i\epsilon}
\end{equation}
in the factorization formula, where
\begin{align}
  \label{phi-G-def}
\Phi_{G\, ij}^\alpha(y_1,y_2) &=
  \int \frac{\de\xi_1^-}{2\pi}\, \frac{\de\xi_2^-}{2\pi}\;
  e^{\ii l_1\cdot \xi_1}\, e^{\ii (l_2-l_1)\cdot \xi_2}\,
\nonumber \\
 & \quad\times
  \langle P| \bar{\psi}_j(0)\, \mathcal{U}^{n_-}_{(0,\xi_2)}\,
  gG^{+\alpha}(\xi_2)\, \mathcal{U}^{n_-}_{(\xi_2,\xi_1)}\,
  \psi_i(\xi_1) |P \rangle\,
  \biggr|_{\xi_1^+ = \xi_2^+ =0,\;
           \bm{\xi}{}_{1T} = \bm{\xi}{}_{2T} = \bm{0}_T}
\end{align}
does not carry a superscript $[\pm]$ because, like $\Phi(y)$, it
involves only Wilson lines of finite length along $n_-$.
We now decompose the correlation functions into terms of definite
twist,
\begin{align}
  \label{coll-twist-dec}
\Phi(y) &= \Phi_2(y) + \frac{M}{P^+}\, \Phi_3(y) + \ldots \,,
&
\Phi_\partial^\alpha(y) 
  &= M\, \Phi_{\partial, 3}^\alpha(y) + \ldots \,,
\nonumber \\
\Phi_A^\alpha(y_1,y_2) 
  &= \frac{M}{P^+}\,
     \Phi_{A,3}^\alpha(y_1, y_2) + \ldots \,,
\end{align}
where the prefactors are chosen such that $\Phi_2$, $\Phi_3$ and
$\Phi_{\partial,3}$, $\Phi_{A,3}$ are dimensionless and independent of
$P^+$.  Under a longitudinal boost $\Phi_3$ is invariant, whereas the
other correlators transform like $n_+$.  Dimensional counting readily
gives $H_2 \sim 1/p^2$ and $H_3, H_{A,3} \sim 1/p^3$.  Using $p^+ =
xP^+$ and \eqref{x-hat-def} we can then rewrite \eqref{hi-pT-next} as
\begin{align}
  \label{hi-pt-23}
\Phi(x,p_T) &=
   \int \frac{\de\hat{x}}{\hat{x}}\, H_2(\hat{x}, p^+, p_T)\,
     \Phi_2\Bigl( \frac{x}{\hat{x}} \Bigr)
\nonumber \\[0.2em]
& \quad + M\; \biggl\{ \,
   \int \frac{\de\hat{x}}{\hat{x}}\,
     \frac{H_2(\hat{x}, p^+, p_T)}{p^+} \,
     x\ms \Phi_3\Bigl( \frac{x}{\hat{x}} \Bigr)
+ \int \frac{\de\hat{x}}{\hat{x}}\, H_3^\alpha(\hat{x}, p^+, p_T)\,
     \Phi_{\partial,3\ms \alpha}\Bigl( \frac{x}{\hat{x}} \Bigr)
\nonumber \\[0.2em]
& \qquad\qquad 
+ \int \frac{\de\hat{x}_1}{\hat{x}_1}\,
       \frac{\de\hat{x}_2}{\hat{x}_2}\,
    H_{A,3}^\alpha(\hat{x}_1^{},\hat{x}_2^{}, p^+, p_T)\;
    x\ms \Phi_{A,3 \ms\alpha}\Bigl( \frac{x}{\hat{x}_1},
                                    \frac{x}{\hat{x}_2} \Bigr)
\biggr\}
\nonumber \\[0.4em]
& \quad
+ \{ \text{terms with two- and three-gluon correlators} \}
+ \mathcal{O}\bigl( 1/p^4 \bigr) \,,
\phantom{\int}
\end{align}
where the first term is of order $1/p^2$ and the terms with prefactor
$M$ are of order $1/p^3$.  To obtain the high-$p_T$ behavior of the
individual distribution functions parameterizing $\Phi(x,p_T)$, we
need to analyze the dependence of the hard-scattering kernels on $p^+$
and $p_T$.  The kernels carry four Dirac indices, so that
\eqref{hi-pt-23} explicitly reads
\begin{equation}
\Phi_{ij}(x,p_T) =
   \int \frac{\de\hat{x}}{\hat{x}}\, 
   H_{2,ijkl}(\hat{x}, p^+, p_T)\,
     \Phi_{2,kl} \Bigl( \frac{x}{\hat{x}} \Bigr) + \ldots \,,
\end{equation}
and similarly for the terms of order $1/p^3$.  We can decompose
$H_{2}$ as
\begin{align}
  \label{H2-dec}
H_{2}(\hat{x}, p^+, p_T) =
  \frac{1}{\bm{p}_T^2}\, & \biggl[
    \sum_{mn} \Gamma^{}_m\otimes \Gamma^{}_n\,
      t_{mn}^{}(\hat{x}, p^+, p_T)
  + \sum_{mn} \Gamma^{}_{m,\mu}\otimes \Gamma^{}_{n,\nu}\;
      t_{mn}^{\smash{\mu}\nu}(\hat{x}, p^+, p_T)
\nonumber \\
& + \sigma_{\lambda\mu}^{}\ms \sigma_{\nu\rho}^{}\,
      t^{\lambda\mu\nu\rho}_{}(\hat{x}, p^+, p_T)
  \biggr]
\end{align}
with $\Gamma_m \in \{1,\gamma_5\}$ and $\Gamma_{m,\mu} \in
\{\gamma_\mu, \gamma_\mu\gamma_5\}$, where the first matrix in the
tensor products carries Dirac indices $ik$ and the second one indices
$jl$.  Only an even number of $\gamma$ matrices appears in this
decomposition, i.e., no structures like $\Gamma_{m,\mu} \otimes
\Gamma_n$ or $\Gamma_{m,\lambda} \otimes \Gamma_{n,\mu\nu}$, which
reflects that chirality is conserved in the hard scattering kernel.
The scalars $t_{mn}^{}$ and the tensors $t_{mn}^{\smash{\mu}\nu}$,
$t^{\lambda\mu\nu\rho}_{}$ are dimensionless and invariant under
longitudinal boosts, and therefore they can be constructed from
$g_{\mu\nu}$, $\epsilon_{\lambda\mu\nu\rho}$, and the vectors
\begin{equation}
  \label{basis-vectors}
\frac{p_T^\mu}{|\bm{p}_T|} \,, \qquad
\frac{p^+ n_+^\mu}{|\bm{p}_T|} \,, \qquad
\frac{|\bm{p}_T|\, n_-^\mu}{p^+} \,.
\end{equation}
Since the tensors have an even number of indices, the factors
$|\bm{p}_T|$ combine such that $H_2(\hat{x},p^+,p_T)$ depends only on
integer powers of $1/ \bm{p}_T^2$.  The same is readily shown for the
kernels $H_3^\alpha(\hat{x}, p^+, p_T)$ and
$H_{A,3}^\alpha(\hat{x}_1^{},\hat{x}_2^{}, p^+, p_T)$, which go like
$1/p^3$ instead of $1/p^2$ but involve one more Lorentz index in the
analog of the decomposition \eqref{H2-dec}.  Analogous arguments can
be given for the kernels connected with two- or three-gluon
correlation functions at the bottom of the graphs (which have two
instead of four Dirac indices and two additional Lorentz indices for
the exchanged gluons) and for the kernels that appear when
$\Phi(x,p_T)$ is evaluated to order $1/p^4$ or higher.

The upshot of this argument is that the distributions parameterizing
the correlator $\Phi(x,p_T)$ behave like integer powers of $1/
\bm{p}_T^2$ for $\bm{p}_T^2 \gg M^2$.  Together with the constraints
from dimensional counting and Lorentz invariance, this allows us to
determine the leading power behavior of each distribution.
Matching the dependence on $p^+$, we find for instance
that terms on the r.h.s.\ of \eqref{hi-pt-23} contribute to the
twist-two and twist-three parts of $\Phi(x,p_T) = \Phi_2(x,p_T) +
(M/P^+)\, \Phi_3(x,p_T) + \ldots$ as
\begin{align}
  \label{hight-pT-cont}
\frac{1}{\bm{p}_T^2} &\;\to\; \Phi_2(x,p_T) \,,
&
\frac{p_T^\rho}{p^+\, \bm{p}_T^2} &\;\to\; x\ms \Phi_3(x,p_T) \,,
\nonumber\\[0.2em]
\frac{M}{p^+ \bm{p}_T^2} &\;\to\; x\ms \Phi_3(x,p_T) \,,
&
\frac{M p_T^\rho}{\bm{p}_T^4} &\;\to\; \Phi_2(x,p_T) \,,
\end{align}
where we have used that $1/P^+ = x/p^+$.  Comparing with the
parameterization \eqref{phi-dec} of $\Phi(x,p_T)$ we see e.g.\ that
$f_{1T}^\perp(x,p_T^2)$ and $g_{1T}^{}(x,p_T^2)$ decrease as $M^2
/\bm{p}_T^4$.  If there had been terms going like $p_T^\rho
/|\bm{p}_T^{}|^3$ on the r.h.s.\ of \eqref{hi-pt-23}, they would instead
decrease as $M /|\bm{p}_T|^3$.

A number of selection rules specify which collinear distributions can
contribute to the high-$p_T$ behavior of a given $p_T$-dependent
distribution.  Clearly, the dependence on the target polarization must
match.  Because the hard scattering conserves chirality, the chirality
of distributions must match as well.  Finally, we recall that the
correlator $\Phi(x,p_T)$ depends on the gauge link and contains terms
which are even or odd under the exchange $\Phi^{[+]}(x,p_T)
\leftrightarrow \Phi^{[-]}(x,p_T)$.  The $T$-odd terms in
$\Phi_\partial^{[\pm]}(y)$ and $\Phi_A^{[\pm]}(y_1,y_2)$ thus
contribute to the $T$-odd distributions in $\Phi^{[\pm]}(x,p_T)$, and
vice versa.  The collinear correlator $\Phi(y)$ only contains $T$-even
terms, but it can contribute to $T$-odd distributions in
$\Phi^{[\pm]}(x,p_T)$ through graphs with absorptive parts in the
hard-scattering subprocess, starting at order $\alpha_s^2$.  An
example is shown in Fig.~\ref{f:alphas2}.  Graphs involving the
quark-gluon-quark correlator (such as the one in
Fig.~\ref{f:rungdistA}b) have absorptive parts already at order
$\alpha_s$, which provides further contributions to the $T$-odd part
of $\Phi^{[\pm]}(x,p_T)$.

\FIGURE[ht]{
\parbox{\textwidth}{\centering
\includegraphics[height=3.2cm]{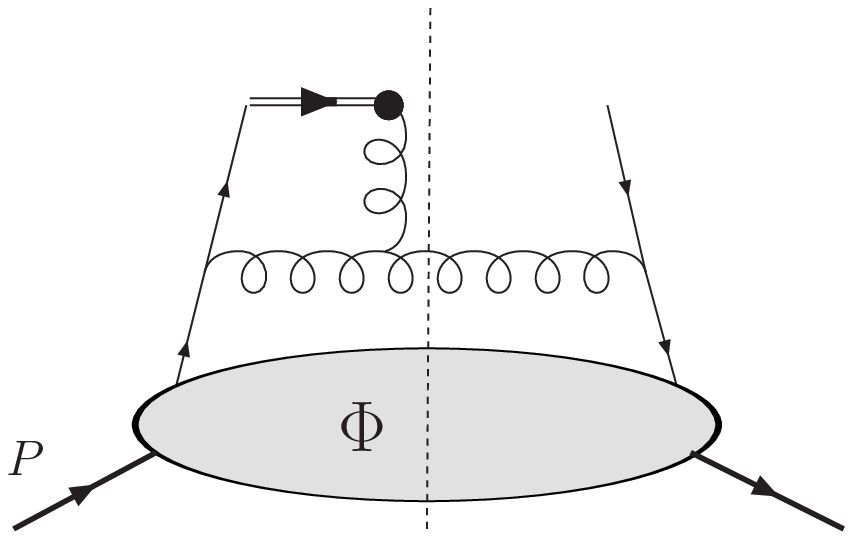}
\caption{\label{f:alphas2} Example of a diagram at order $\alpha_s^2$,
  whose absorptive part gives rise to $T$-odd terms in the quark-quark
  correlator.  The double line denotes an eikonal line originating
  from the gauge link in $\Phi(x,p_T)$, as specified in
  appendix~\protect\ref{app:FG}.}
}}

With the time reversal properties given in table~\ref{tab:T-even-odd},
we find that at order $\alpha_s$ collinear functions denoted by
letters $f$, $g$, $e$, $h$ can only contribute to the high-$p_T$
behavior of distributions denoted by the same letter.  Putting
everything together we find
\begin{align}
f_1 &\sim \dfrac{1}{\bm{p}_T^2} \, \alpha_s\, \ff{f_1} \,,
& 
g_{1L}^{} &\sim \dfrac{1}{\bm{p}_T^2} \, \alpha_s\, \ff{g_1} \,,
&
h_1 &\sim \dfrac{1}{\bm{p}_T^2} \, \alpha_s\, \ff{h_1} \,,
\nonumber \\
x f_{}^\perp &\sim \dfrac{1}{\bm{p}_T^2} \, \alpha_s\, \ff{f_1} \,,
&
x g_L^\perp &\sim \dfrac{1}{\bm{p}_T^2} \, \alpha_s\, \ff{g_1} \,,
\nonumber \\
x h_T^{} &\sim \dfrac{1}{\bm{p}_T^2} \, \alpha_s\, \ff{h_1} \,,
&
x h_T^\perp &\sim \dfrac{1}{\bm{p}_T^2} \, \alpha_s\, \ff{h_1}
  \label{tails-2}
\\
\intertext{from the $1/p^2$ part of $\Phi(x,p_T)$, and}
f_{1T}^\perp &\sim \dfrac{M^2}{\bm{p}_T^4} \, \alpha_s \,
                   \ff{f_{1T}^{\perp (1)}, \ldots} \,,
&
g_{1T}^{} &\sim \dfrac{M^2}{\bm{p}_T^4} \, \alpha_s\,
                \ff{g_{1T}^{(1)}, \ldots} \,,
\nonumber \\
h_{1L}^\perp &\sim \dfrac{M^2}{\bm{p}_T^4} \, \alpha_s\,
                   \ff{h_{1L}^{\perp (1)}, \ldots} \,,
&
h_1^\perp &\sim \dfrac{M^2}{\bm{p}_T^4} \, \alpha_s\,
                \ff{h_{1}^{\perp (1)}, \ldots} \,,
\nonumber \\
x f_T^\perp &\sim \dfrac{M^2}{\bm{p}_T^4} \, \alpha_s\,
                  \ff{f_{1T}^{\perp (1)}, \ldots} \,,
&
x g_T^\perp &\sim \dfrac{M^2}{\bm{p}_T^4} \, \alpha_s\,
                  \ff{g_{1T}^{(1)}, \ldots} \,,
\nonumber \\
x f_T^{} &\sim \dfrac{1}{\bm{p}_T^2} \, \alpha_s\,
               \ff{f_{1T}^{\perp (1)}, \ldots} \,,
&
x g_T^{} &\sim \dfrac{1}{\bm{p}_T^2} \, \alpha_s\,
               \ff{g_{1T}^{(1)}, \ldots} \,,
\nonumber \\
x h_L^{} &\sim \dfrac{1}{\bm{p}_T^2} \, \alpha_s\,
               \ff{h_{1L}^{\perp (1)}, \ldots} \,,
&
x h &\sim \dfrac{1}{\bm{p}_T^2} \, \alpha_s\,
          \ff{h_1^{\perp (1)}, \ldots} \,,
\nonumber \\
x e_L^{} &\sim \dfrac{1}{\bm{p}_T^2} \, \alpha_s\, \ff{\ldots} \,,
&
x e &\sim  \dfrac{1}{\bm{p}_T^2} \, \alpha_s\, \ff{x e, \ldots}
  \label{tails-3}
\end{align}
from the $1/p^3$ part of $\Phi(x,p_T)$.  The distributions on the
l.h.s.\ depend on $x$ and $p_T$, and on the right-hand side we have
convolutions of the form
\begin{align}
  \label{F-def}
\ff{f} &= K_q \otimes f^a + K_g \otimes f^g \,,
\end{align}
where $a$ is the quark or antiquark flavor of the $p_T$-dependent
functions on the l.h.s.\ of \eqref{tails-2} or \eqref{tails-3}.  We
note that at higher orders in $\alpha_s$ one has instead a sum over
all quark and antiquark flavors, as in \eqref{small-b-exp}.  The
contribution from gluon distributions in \eqref{F-def} is absent for
chiral-odd distributions.  The kernels $K_q$ and $K_g$ are of course
not the same for the different functions in \eqref{tails-2} and
\eqref{tails-3}, and we use $\mathcal{F}$ in a generic sense.  We have
explicitly calculated the hard-scattering kernels $H_2$ and $H_3$
defined by \eqref{H23-def} and verified that they give nonzero
contributions for the functions given as arguments of $\mathcal{F}$ in
\eqref{tails-2} and \eqref{tails-3}.  By $\ldots$ we have denoted
contributions from three-parton correlation functions.  We have not
listed the twist-three distributions parameterizing $\Phi_3(x)$ as
arguments of $\mathcal{F}$, because they can (up to quark mass
suppressed terms) be related to the functions in
$\Phi_{\partial,3}(x)$ and to quark-gluon-quark correlation functions,
as we remarked after Eq.~\eqref{mom-notation}.  An exception is
$e(x)$, which has no counterpart in $\Phi_{\partial,3}(x)$.
Furthermore, there is no distribution in $\Phi_2$ or
$\Phi_{\partial,3}$ that is multiplied by $S_L$ and both $T$-odd and
chiral-odd.  At order $\alpha_s$ and $1/p^3$ the high-$p_T$ behavior
of $e_L^{}(x,p_T^2)$ can hence only be generated from the $\Phi_{A,3}$
term in $\Phi(x,p_T)$.

The relations in \eqref{tails-2} only involve collinear functions of
twist two and those in \eqref{tails-3} only collinear functions of
twist three, corresponding to the respective order in the $1/p$
expansion of the correlation function $\Phi(x,p_T)$.  On the other
hand, there are $p_T$-dependent functions of twist two and twist three
in both \eqref{tails-2} and \eqref{tails-3}.  In other words, the
twist of the collinear distributions and the $p_T$-dependent
distributions in the high-$p_T$ limit need not be the same.

By power counting, the $p_T$-dependent distributions $f_L^\perp$,
$g_{}^\perp$, $e_{T}^{}$, $e_{T}^\perp$ can receive a contribution
from the $1/p^2$ part of $\Phi(x,p_T)$, but explicit calculation at
order $\alpha_s$ gives a zero result.  This readily follows from our
discussion above \eqref{tails-2} because these distributions are
$T$-odd.  Their high-$p_T$ behavior therefore starts at order
$\alpha_s^2 /p^2$ and reads
\begin{align}
x f_L^\perp &\sim \dfrac{1}{\bm{p}_T^2}\, \alpha_s^2\, \ff{g_1^{}} \,,
&
x g_{}^\perp &\sim \dfrac{1}{\bm{p}_T^2}\, \alpha_s^2\, \ff{f_1^{}} \,,
\nonumber \\
x e_T^{} &\sim \dfrac{1}{\bm{p}_T^2}\, \alpha_s^2\, \ff{h_1^{}} \,,
&
x e_T^\perp &\sim \dfrac{1}{\bm{p}_T^2}\, \alpha_s^2\, \ff{h_1^{}} \,.
  \label{tails-4}
\end{align}
There will also be contributions to these functions from $\Phi(x,p_T)$
order $\alpha_s^{} /p^4$, where the necessary $T$-odd effects can come
from lowest-order graphs with three- or four-parton correlation
functions.
\new{By power counting in $1/p$, these contributions are subleading
  compared with ones in \eqref{tails-4}, although they appear at lower
  order in the $\alpha_s$ expansion.  We omit them in our subsequent
  discussion, but they can easily be restored.}

Explicit calculation also reveals that neither $h_{1T}^\perp(x,p_T)$
nor the combination $h_T^{}(x,p_T) + h_T^\perp(x,p_T)$ receives
contributions at order $\alpha_s$ and $1 /p^2$, although this would be
allowed by power counting.  We shall not investigate the reason of
this here, and simply write
\begin{align}
h_{1T}^\perp &\sim \dfrac{M^2}{\bm{p}_T^4}\, \alpha_s^2\, \ff{h_1} \,,
&
x h_T^{} + x h_T^\perp &\sim
     \dfrac{1}{\bm{p}_T^2}\, \alpha_s^2\, \ff{h_1^{}} \,.
  \label{tails-5}
\end{align}
\new{Again there will also be power suppressed contributions at lower
  order in $\alpha_s$, which go like $\alpha_s /p^6$ for $h_{1T}^\perp$
  and like $\alpha_s /p^4$ for $h_T^{} + h_T^\perp$.}
We caution that without a full calculation of the graphs with
multi-parton correlators we cannot exclude that the contributions to
the distributions given in \eqref{tails-3} vanish when all terms are
added up.  A corresponding caveat applies to the $\alpha_s^2$
contributions in \eqref{tails-4} and \eqref{tails-5}.  For
$\Phi(x,p_T)$ at order $\alpha_s$ and $1/p^2$ we give complete and
explicit results in section~\ref{sec:tails}.  As for the $1/p^3$ part
of $\Phi(x,p_T)$, the explicit calculation in
\cite{Ji:2006vf,Koike:2007dg} gives $f_{1T}^{\perp}(x,p_T^2) \sim (M^2
/\bm{p}_T^4)\, \alpha_s\ms \ff{G_F, \widetilde{G}_F}$, where
$\mathcal{F}$ now denotes two-variable convolutions of the form
\begin{align}
  \label{conv-two-var}
\int \frac{\de\xb_1}{\xb_1}\, \frac{\de\xb_2}{\xb_2}\;
  K(\xb_1, \xb_2)\,
  G_F\biggl( \frac{x}{\xb_1}, \frac{x}{\xb_2} \biggr) \,.
\end{align}
Given that $f_{1T}^{\perp (1) [\pm]}(x) = \mp \frac{\pi}{2}\,
G_F(x,x)$, the structure of our result for $f_{1T}^{\perp}(x,p_T^2)$
in \eqref{tails-3} is hence consistent with the full
calculation.\footnote{%
  The relation between $f_{1T}^{\perp (1)}$ and $G_F^{}$ can be
  obtained by combining \protect\eqref{pT-moment} in the present work
  with Eq.~(2) in \protect\cite{Eguchi:2006qz} and Eqs.~(29), (40) in
  \protect\cite{Boer:2003cm}.  Corresponding relations using different
  parameterizations have been given in \protect\cite{Boer:2003xz} and
  \protect\cite{Ji:2006ub}.}

At this point we briefly return to the question of ultraviolet
divergences in collinear correlation functions, which we mentioned
after \eqref{pt-int}.  With the high-$p_T$ behavior given in
\eqref{tails-2} and \eqref{tails-3} one explicitly sees that the
integral $\int \de^2 \bm{p}_T\, \Phi(x,p_T)$ diverges logarithmically
at high $p_T$, both for the twist-two and the twist-three part of
$\Phi(x,p_T)$.  The corresponding ultraviolet subtractions in the
collinear correlator $\Phi(x)$ result in a logarithmic dependence on
the subtraction scale $\mu$ for all distributions in \eqref{eq:phix}.
This dependence is described by DGLAP equations, whose evolution
kernels are closely related to the kernels appearing in the
convolutions of \eqref{tails-2} and \eqref{tails-3}.
With \eqref{tails-3} one also finds that the integral $\int \de^2
\bm{p}_T^{}\, p_T^\alpha\, \Phi_2^{}(x,p_T)$ diverges logarithmically.
This leads to DGLAP equations for the $\bm{p}_T^2$ moments of
twist-two distributions in the parameterization \eqref{pT-moment} of
$\Phi_\partial(x)$, which have been investigated in
\cite{Henneman:2005th}.  In contrast, the integral $\int \de^2
\bm{p}_T^{}\, p_T^\alpha\, \Phi_3^{}(x,p_T)$ diverges quadratically in
$p_T$ according to \eqref{tails-2} and \eqref{tails-4}.  In a proper
definition for $\bm{p}_T^2$-moments of twist-three distributions, such
as $f^{\perp (1)}(x)$ or $g^{\perp (1)}(x)$, one would hence have to
deal with power-like divergences.

In the dimensional analysis following \eqref{basis-vectors} we have
ignored that the hard-scattering kernels also depend on the
regularization parameter $\zeta$, which is Lorentz invariant and has
mass dimension two.  In applications of low-$q_T$ factorization one
needs $\zeta$ comparable to the large scale, as already mentioned in
section~\ref{sec:CS-fact}, so that we can restrict our attention to
$\zeta \gg p_T^2$.  Terms in the hard-scattering kernels going with a
positive power of $p_T^2 /\zeta$ are then negligible.  In contrast,
terms with a positive power of $\zeta /p_T^2$ would lead to a faster
$p_T^2$ falloff than derived in this section.  They would correspond
to power-like rapidity divergences in $\Phi(x,p_T)$.  In the explicit
calculations at order $1/p^2$ in section~\ref{sec:tails} we will not
encounter such terms, obtaining only a modification of the power-laws
in \eqref{tails-2} by logarithms $\ln(\zeta /p_T^2)$.  A corresponding
statement holds for the calculation of $f_{1T}^\perp(x,p_T^2)$ in
\cite{Ji:2006vf,Koike:2007dg}.

The high-$k_T$ behavior of the fragmentation correlator
$\Delta(z,k_T)$ can be obtained in full analogy to the case of
$\Phi(x,p_T)$.  One can readily obtain results by crossing the
hard-scattering graphs calculated for the distribution functions,
replacing $x \to 1/z$ and $p_T \to k_T$.  This gives
\begin{align}
D_1 &\sim \dfrac{1}{\bm{k}_T^2} \, \alpha_s\, \ff{D_1} \,,
&
\frac{D^\perp}{z} &\sim \dfrac{1}{\bm{k}_T^2} \, \alpha_s\,
                        \ff{D_1} \,,
\nonumber \\[0.1em]
H_1^\perp &\sim \dfrac{M^2}{\bm{k}_T^4} \, \alpha_s\,
                \ff{H_1^{\perp (1)}, \ldots} \,,
&
\frac{G^\perp}{z} &\sim \dfrac{1}{\bm{k}_T^2} \,\alpha_s^2\,
                        \ff{D_1} \,,
\nonumber \\[0.1em]
\frac{H}{z} &\sim \dfrac{1}{\bm{k}_T^2} \, \alpha_s\,
                  \ff{H_1^{\perp (1)}, \ldots} \,,
&
\frac{E}{z} &\sim \dfrac{1}{\bm{k}_T^2} \, \alpha_s\,
                  \lff{\frac{E}{z},\, \ldots} \,.
\label{tails-frag}
\end{align}
Compared with their analogs \eqref{F-def}, the convolutions
\begin{align}
\ff{D} &= \frac{1}{z^2}\,
       \Bigl[ D^a \otimes K_q + D^g \otimes K_g \Bigr]
\end{align}
have an additional factor $1/z^2$, which reflects the corresponding
factor in \eqref{eq:deltaz}.


\subsection{Results for structure functions}
\label{sec:lowtm}

Let us begin this section by recalling the expressions for SIDIS
structure functions at low $q_T$ in terms of
transverse-momentum-dependent distribution and fragmentation
functions.  Extending earlier work in
\cite{Mulders:1995dh,Boer:1997nt}, the study \cite{Bacchetta:2006tn}
has given a complete set of results at leading and first subleading
order in $1/Q$, i.e., at twist-two and twist-three accuracy.  A
detailed investigation of color gauge invariance and the appropriate
choice of gauge links has been given in \cite{Boer:2003cm}.  The
calculations just quoted take into account tree-level graphs, where
gluons are restricted to be collinear to the target or to the observed
hadron $h$ and only appear when they are attached to the distribution
or fragmentation correlators (see Fig.~2 in \cite{Bacchetta:2006tn}).

For a compact presentation of the results, we introduce the unit
vector $\hat{\bm{h}} = -\bm{q}_T /|\bm{q}_T|$ and the
transverse-momentum convolution
\begin{equation}
  \label{tree-con}
\mathcal{C}\bigl[ w f D \bigr]
= \sum_a x\ms e_a^2 \int \de^2 \bm{p}_T\,  \de^2 \bm{k}_T^{}
\, \delta^{(2)}\bigl(\bm{p}_T - \bm{k}_T^{} + \bm{q}_T \bigr)
\,w(\bm{p}_T,\bm{k}_T^{})\,
f^a(x,p_T^2)\,D^a(z,k_T^2) ,
\end{equation}
where $w(\bm{p}_T,\bm{k}_T^{})$ is an arbitrary function and the sum
runs over quarks and antiquarks.  The results for the structure
functions appearing in (\ref{e:crossmaster}) then read
\cite{Bacchetta:2006tn}
\begin{align} 
\label{F_UUT}
\hspace{-3mm}
F_{UU ,T}
& 
= \mathcal{C}\bigl[ f_1 D_1 \bigr],
\phantom{\biggl[ \biggr]}
\\[0.1em]
\label{F_UUL}
\hspace{-3mm}
F_{UU ,L}
& 
= \mathcal{O}\biggl(\frac{M^2}{Q^2}, \frac{q_T^2}{Q^2}\biggr),
\phantom{\biggl[ \biggr]}
\\[0.1em] 
\label{F_UUcosphi}
\hspace{-3mm}
F_{UU}^{\cos\phi_h}
& 
= \frac{2M}{Q}\,\mathcal{C}\biggl[
   - \frac{\hat{\bm{h}}\cdott \bm{k}_T^{}}{M_h}
   \biggl(x  h\, H_{1}^{\perp } 
   + \frac{M_h}{M}\,f_1 \frac{\tilde{D}^{\perp }}{z}\biggr)
   - \frac{\hat{\bm{h}}\cdott \bm{p}_T}{M}
     \biggl(x f^{\perp } D_1
   + \frac{M_h}{M}\,h_{1}^{\perp } \frac{\tilde{H}}{z}\biggr)\biggr],
\\[0.1em]
\label{F_UUcos2phi}
\hspace{-3mm}
F_{UU}^{\cos 2\phi_h}\!
& 
= \mathcal{C}\biggl[
   - \frac{2\, \bigl( \hat{\bm{h}}\cdott \bm{k}_T^{} \bigr)
   \,\bigl( \hat{\bm{h}}\cdott \bm{p}_T \bigr)
    -\bm{k}_T^{}\cdott \bm{p}_T}{M M_h}
    h_{1}^{\perp } H_{1}^{\perp }\biggr],
\\[0.1em]
\label{F_LUsinphi} 
\hspace{-3mm}
F_{LU}^{\sin\phi_h}
& 
=\frac{2M}{Q}\,\mathcal{C}\biggl[ 
- \frac{\hat{\bm{h}}\cdott \bm{k}_T^{}}{M_h}
   \biggl(x  e \, H_1^{\perp } 
   +\frac{M_h}{M}\,f_1\frac{\tilde{G}^{\perp }}{z}\biggr)
   \!+\! \frac{\hat{\bm{h}}\cdott \bm{p}_T}{M}
   \biggl(x  g^{\perp }  D_1 
   + \frac{M_h}{M}\, h_1^{\perp } \frac{\tilde{E}}{z} \biggr)\biggr],
\\[0.1em]
\label{F_ULsinphi} 
\hspace{-3mm}
F_{UL}^{\sin\phi_h}
&
 = \frac{2M}{Q}\,\mathcal{C}\biggl[
   - \frac{\hat{\bm{h}}\cdott \bm{k}_T^{}}{M_h}
    \biggl(x  h_L  H_1^{\perp } 
   \!+\! \frac{M_h}{M}\,g_{1L}\frac{\tilde{G}^{\perp } }{z}\biggr)
   \!+\! \frac{\hat{\bm{h}}\cdott \bm{p}_T}{M}
    \biggl(x f_{L}^{\perp }  D_1 
   - \frac{M_h}{M}\, h_{1L}^{\perp } \frac{\tilde{H}}{z}\biggr)\biggr],\,
\\[0.1em]
\hspace{-3mm}
F_{UL}^{\sin 2\phi_h}\!
&
 = \mathcal{C}\biggl[
   -\frac{2\,\bigl( \hat{\bm{h}}\cdott \bm{k}_T^{} \bigr)
   \,\bigl( \hat{\bm{h}}\cdott \bm{p}_T \bigr)
   -\bm{k}_T^{}\cdott \bm{p}_T}{M M_h}
    h_{1L}^{\perp } H_{1}^{\perp }\biggr],
\displaybreak
\\[0.1em]
\hspace{-3mm}
F_{LL}
&
 =\mathcal{C}\bigl[ g_{1L}  D_1\bigr], 
\phantom{\biggl[ \biggr]}
\\[0.1em]
\hspace{-3mm}
F_{LL}^{\cos \phi_h}
& =
 \frac{2M}{Q}\,\mathcal{C}\biggl[ \frac{\hat{\bm{h}}\cdott\bm{k}_T^{}}{M_h}
   \biggl(x e_L  H_1^{\perp }
   - \frac{M_h}{M}\,g_{1L}   \frac{\tilde{D}^{\perp }}{z}\biggr)
   - \frac{\hat{\bm{h}}\cdott \bm{p}_T}{M}
   \biggl(x  g_L^{\perp }   D_1
   +  \frac{M_h}{M}\,h_{1L}^{\perp } \frac{\tilde{E}}{z}\biggr)\biggr],
\end{align} 
\begin{align} 
F_{UT ,T}^{\sin\lf(\phi_h -\phi_S\rg)}
\!& =
\mathcal{C}\biggl[-\frac{\h\cdott\bm{p}_T}{M} f_{1T}^{\perp } D_1\biggr],
\label{e:sivers}
\\[0.1em]
F_{UT ,L}^{\sin\lf(\phi_h -\phi_S\rg)}
\!&= \mathcal{O}\biggl(\frac{M^2}{Q^2}, \frac{q_T^2}{Q^2}\biggr),
\phantom{\biggl[ \biggr]}
\\[0.1em]
F_{UT}^{\sin\lf(\phi_h +\phi_S\rg)}
\!& =
 \mathcal{C}\biggl[-\frac{\h\cdott\bm{k}_T^{}}{M_h}
                    h_{1} H_1^{\perp }\biggr],
\label{e:collins}
\\[0.1em]
F_{UT}^{\sin\lf(3\phi_h -\phi_S\rg)}
\!& = 
   \mathcal{C}\biggl[
   \frac{2\, \bigl(\h\cdott \bm{p}_T \bigr)\, 
        \bigl( \bm{p}_T\cdott\bm{k}_T^{} \bigr)
   +\bm{p}_T^2\, \bigl(\h\cdott \bm{k}_T^{} \bigr)
   -4\, (\h\cdott\bm{p}_T)^2 \, (\h\cdott\bm{k}_T^{})}{2 M^2 M_h}
    \,h_{1T}^{\perp }   H_1^{\perp }\biggr],\,\,\,
\label{e:F_UTsin3phi}
\\[0.3em]
F_{UT}^{\sin \phi_S }
& = \frac{2M}{Q}\,\mathcal{C}\biggl\{
   \biggl(x  f_T   D_1
   - \frac{M_h}{M} \, h_{1}  \frac{\tilde{H}}{z}\biggr) 
\nonumber \\
 & \qquad
\hspace{-9mm}
   - \frac{\bm{k}_T^{}\cdott \bm{p}_T}{2 M M_h}\,
     \biggl[\biggl(x  h_{T}  H_{1}^{\perp } 
   + \frac{M_h}{M} g_{1T} \,\frac{\tilde{G}^{\perp }}{z}\biggr)
   -  \biggl(x  h_{T}^{\perp }  H_{1}^{\perp } 
   - \frac{M_h}{M} f_{1T}^{\perp } \,\frac{\tilde{D}^{\perp }}{z}
   \biggr) \biggr]\biggr\},
\\[0.3em] 
F_{UT}^{\sin\lf(2\phi_h -\phi_S\rg)}
\!& 
= \frac{2M}{Q}\,\mathcal{C}\biggl\{
   \frac{2\, (\hat{\bm{h}}\cdott \bm{p}_T)^2 -\bm{p}_T^2}{2 M^2}\,
   \biggl(x  f_T^{\perp }   D_1
   - \frac{M_h}{M} \, h_{1T}^{\perp }  \frac{\tilde{H}}{z}\biggr)
\nonumber \\
 & \qquad
\hspace{4mm}  - \frac{2\, \bigl( \hat{\bm{h}}\cdott \bm{k}_T^{} \bigr)
   \, \bigl( \hat{\bm{h}}\cdott \bm{p}_T \bigr)
   -\bm{k}_T^{}\cdott \bm{p}_T}{2 M M_h}\,
   \biggl[\biggl(x  h_{T}  H_{1}^{\perp } 
   + \frac{M_h}{M} g_{1T} \,\frac{\tilde{G}^{\perp }}{z}\biggr)
\nonumber \\
 & \qquad
\hspace{53mm}   
        +  \biggl(x  h_{T}^{\perp }  H_{1}^{\perp } 
   - \frac{M_h}{M} f_{1T}^{\perp } \,\frac{\tilde{D}^{\perp }}{z}
   \biggr) \biggr]\biggr\},
\label{e:dUT}
\\[0.1em]
F_{LT}^{\cos(\phi_h -\phi_S)}
\!&
 =\mathcal{C}\biggl[ \frac{\h\cdott\bm{p}_T}{M} g_{1T}
D_1 \biggr] , 
\\[0.3em]
F_{LT}^{\cos \phi_S}
&
 = \frac{2M}{Q}\,\mathcal{C}\biggl\{
   -\biggl(x  g_T   D_1
   + \frac{M_h}{M} \, h_{1}  \frac{\tilde{E}}{z}\biggr) 
\nonumber \\
 & \qquad
\hspace{-9mm}   +\frac{\bm{k}_T^{}\cdott \bm{p}_T}{2 M M_h}\,
   \biggl[\biggl(x  e_{T}  H_{1}^{\perp } 
   - \frac{M_h}{M} g_{1T} \,\frac{\tilde{D}^{\perp }}{z}\biggr)
   +  \biggl(x  e_{T}^{\perp }  H_{1}^{\perp } 
   + \frac{M_h}{M} f_{1T}^{\perp } \,\frac{\tilde{G}^{\perp }}{z}
   \biggr) \biggr]\biggr\},
\\[0.3em] 
F_{LT}^{\cos(2\phi_h - \phi_S)}
\!&
 = \frac{2M}{Q}\,\mathcal{C}\biggl\{
   -\frac{2\,(\hat{\bm{h}}\cdott \bm{p}_T)^2 -\bm{p}_T^2}{2 M^2}\,
   \biggl(x  g_T^{\perp }   D_1
   + \frac{M_h}{M} \, h_{1T}^{\perp }  \frac{\tilde{E}}{z}\biggr)
\nonumber \\
 & \qquad
\hspace{4mm} + \frac{2\, \bigl( \hat{\bm{h}}\cdott \bm{k}_T^{} \bigr)
   \, \bigl( \hat{\bm{h}}\cdott \bm{p}_T \bigr)
   -\bm{k}_T^{}\cdott \bm{p}_T}{2 M M_h}\,
   \biggl[\biggl(x  e_{T}  H_{1}^{\perp } 
   - \frac{M_h}{M} g_{1T} \,\frac{\tilde{D}^{\perp }}{z}\biggr)
\nonumber \\
 & \qquad
\hspace{53mm}   -  \biggl(x  e_{T}^{\perp }  H_{1}^{\perp } 
   + \frac{M_h}{M} f_{1T}^{\perp } \,\frac{\tilde{G}^{\perp }}{z}
   \biggr) \biggr]\biggr\} .
\label{e:last-fun}
\end{align}
In the entries for $F_{UU,L}^{}$ and $F_{UT,L}^{\sin(\phi_h-\phi_S)}$
we have indicated that these structure functions come out to be zero
when the calculation includes only terms up to order $1/Q$.
The fragmentation functions with a tilde are given by
\begin{align}
\frac{\tilde{D}^{\perp}}{z}  &= \frac{D^{\perp}}{z} -  D_1,
\label{e:Dtilde}
\\
\frac{\tilde{G}^{\perp}}{z}  &= \frac{G^{\perp}}{z}
                                - \frac{m}{M_h}\,H_1^{\perp},
\\
\frac{\tilde{E}}{z}  &= \frac{E}{z} -  \frac{m}{M_h}\,D_1,
\\
\frac{\tilde{H}}{z}  &= \frac{H}{z}
                        + \frac{\bm{k}_T^2}{M_h^2}\, H_1^{\perp}.
\end{align}
Using \eqref{tails-frag} and neglecting the small contributions
proportional to the quark mass~$m$, we readily see that the behavior
for $k_T \gg M$ is the same for corresponding functions with and
without a tilde.

The tree-level calculations in
\cite{Mulders:1995dh,Boer:2003cm,Bacchetta:2006tn} do not take into
account soft gluon exchange or virtual corrections involving hard
loops, so that the soft and hard factors we encountered in
\eqref{FUUTconv} and \eqref{Wsidis-full} do not appear in the
convolution \eqref{tree-con}.  Detailed investigations of
factorization for SIDIS with measured $q_T$ have recently been given
in \cite{Ji:2004wu,Ji:2004xq} and
\cite{Collins:2004nx,Collins:2007ph}, extending the seminal work of
Collins and Soper \cite{Collins:1981uk}.  The factorization formulae
discussed in these papers have the form \eqref{Wsidis-full} and are
valid at all orders in $\alpha_s$ but restricted to the leading order
in $1/Q$.  Although a number of subtle issues remain to be fully
clarified \cite{Collins:2007ph}, we will use \eqref{Wsidis-full} in
the following.  Since we aim at deriving expressions at lowest
nonvanishing order in $\alpha_s$, we can neglect the hard factor
$|H|^2 = 1 + O(\alpha_s)$.  The convolution in \eqref{tree-con} should
then be extended to
\begin{equation} \begin{split} 
    \label{soft-con}
\mathcal{C}\bigl[ w f D \bigr] = 
\sum_a x\ms e_a^2 
& \int \de^2 \bm{p}_T\,  \de^2 \bm{k}_T^{}\, \de^2 \bm{l}_T^{}\,
\delta^{(2)}\bigl(\bm{p}_T - \bm{k}_T^{} + \bm{l}_T^{} + \bm{q}_T \bigr)
\\
& \times w(\bm{p}_T,\bm{k}_T^{})\,
f^a(x,p_T^2)\, D^a(z,k_T^2)\, U(l_T^2) \,.
\end{split} \end{equation} 
At high transverse momentum $l_T \gg M$ the soft factor behaves as
$U(l_T^2) \sim \alpha_s /\ms\bm{l}_T^2$, with a coefficient we shall
give in \eqref{e:soft-tail} below.  Our normalization convention is
\begin{equation}
  \label{U-norm}
\int \de^2 \bm{l}_T^{}\, U(l_T^2) = 1 + O(\alpha_s) \,,
\end{equation}
where it is understood that the integral must be suitably regularized
at large $l_T$.

\new{Whether Collins-Soper factorization can be extended to structure
  functions that are of order $1/Q$ is not known.  We note that the study
  of color gauge invariance in \cite{Boer:2003cm} was limited to
  $q_T$-integrated observables in this case, and that a problem with
  twist-three factorization has been found in a spectator model
  calculation \cite{Gamberg:2006ru}.}
In the following we adopt the working hypothesis that the twist-two
factorization formula can simply be taken over at twist-three accuracy,
using the convolution \eqref{soft-con} also for evaluating the high-$q_T$
behavior of the $1/Q$ suppressed structure functions in \eqref{F_UUT} to
\eqref{e:last-fun}.  We will return to this point at the end of
section~\ref{sec:tailstructure}.

We now show how to calculate the high-$q_T$ behavior of the
convolution \eqref{soft-con}.  At order $\alpha_s$, only one of the
factors $f(x,p_T^2)$, $D(z,k_T^2)$, $U(l_T^2)$ can be taken at high
transverse momentum.  Let us first consider the simple case where
$w(\bm{p}_T,\bm{k}_T^{}) = 1$.  In the region where $p_T$ is large, we
use the $\delta$ function in \eqref{soft-con} to perform the $p_T$
integral and approximate $\bm{p}_T = \bm{k}_T^{} - \bm{l}_T^{} -
\bm{q}_T \approx - \bm{q}_T$ in $f(x,p_T^2)$.  The remaining integrals
over $k_T$ and $l_T$ can then be carried out independently.  According
to \eqref{eq:deltaz} and our discussion after
\eqref{small-b-relation}, the integral over $k_T$ gives a collinear
fragmentation function, up to \mbox{$\alpha_s$-corrections} that can
be neglected to our accuracy.  Likewise, the integral over $l_T$ gives
unity up to \mbox{$\alpha_s$-corrections} according to \eqref{U-norm}.
Since we are considering the region where $k_T$ and $l_T$ are small
compared with $q_T$ the integrals over these momenta should be
suitably cut off, as is required for \eqref{eq:deltaz} and
\eqref{U-norm} to make sense.
Repeating these arguments for the cases where $k_T$ or $l_T$ are
large, we obtain
\begin{equation} \begin{split} 
& \int \de^2 \bm{p}_T\,  \de^2 \bm{k}_T^{}\, \de^2 \bm{l}_T^{}\,
\delta^{(2)}\bigl(\bm{p}_T - \bm{k}_T^{}
                + \bm{l}_T^{} + \bm{q}_T \bigr)\,
f(x,p_T^2)\, D(z,k_T^2)\, U(l_T^2)
\\
& \quad \approx
  f(x,q_T^2)\, \frac{D(z)}{z^2}
+ f(x)\, D(z,q_T^2)
+ f(x)\, \frac{D(z)}{z^2}\; U(q_T^2) \,.
\label{e:FUUTexp}
\end{split} \end{equation} For nontrivial functions
$w(\bm{p}_T,\bm{k}_T^{})$ the calculation is slightly more involved.
Instead of approximating e.g.\ $\bm{p}_T = \bm{k}_T^{} - \bm{l}_T^{} -
\bm{q}_T \approx - \bm{q}_T$, we need to Taylor expand the functions
of $\bm{p}_T$ around $-\bm{q}_T$.  We take as an example the
convolution $\mathcal{C}\bigl[ \bigl(\bm{k}_T^{}\cdott \bm{p}_T
\bigr)\, h_{1}^{\perp} H_{1}^{\perp} \bigr]$ appearing in
$F_{UU}^{\cos 2\phi_h}$ and consider the region where $p_T$ is large.
We perform the integral over $p_T$ using the $\delta$ function and
obtain
\begin{align}
  \label{taylor-expand}
 \int & \de^2 \bm{k}_T^{}\, \de^2 \bm{l}_T^{}\;
    H_{1}^{\perp}(z, k_T^2)\, U(l_T^2)\;
   \bigl( \bm{k}_T^2 - \bm{k}_T^{} \cdott \bm{l}_T^{}
                     - \bm{k}_T^{} \cdott \bm{q}_T^{} \bigr)\;
   h_{1}^{\perp}\bigl(x,
          (\bm{k}_T^{} - \bm{l}_T^{} - \bm{q}_T^{})^2 \ms\bigr)
\nonumber \\
& \approx
\int \de^2 \bm{k}_T^{}\, \de^2 \bm{l}_T^{}\;
   H_{1}^{\perp}(z, k_T^2)\, U(l_T^2)
\nonumber \\
& \qquad \times 
   \bigl( \bm{k}_T^2 - \bm{k}_T^{} \cdott \bm{l}_T^{}
                     - \bm{k}_T^{} \cdott \bm{q}_T^{} \bigr)\;
   \biggl[ h_{1}^{\perp}\bigl(x, q_T^2) 
       - 2 \bigl( \bm{k}_T^{}\cdott \bm{q}_T
                - \bm{l}_T^{} \cdott \bm{q}_T^{} \bigr)\,
           \frac{\partial}{\partial \bm{q}_T^2}\,
             h_{1}^{\perp}\bigl(x, q_T^2) \biggr] + \ldots
\nonumber \\[0.2em]
& \approx
  \int \de^2 \bm{k}_T^{}\, H_{1}^{\perp}(z, k_T^2)\,
  \biggl[\ms \bm{k}_T^2\, h_{1}^{\perp}\bigl(x, q_T^2)
        + 2 \bigl( \bm{k}_T^{}\cdott \bm{q}_T \bigr)^2
          \frac{\partial}{\partial \bm{q}_T^2}\,
             h_{1}^{\perp}\bigl(x, q_T^2) \biggr] + \ldots
\nonumber \\[0.2em]
& = 2M_h^2\, \frac{H_{1}^{\perp (1)}(z)}{z^2}\, 
   \biggl[ h_{1}^{\perp}\bigl(x, q_T^2)
         + \bm{q}_T^2\, \frac{\partial}{\partial \bm{q}_T^2}\,
             h_{1}^{\perp}\bigl(x, q_T^2) \biggr] + \ldots
\end{align}
where both terms in square brackets behave as $1/q_T^4$.  The $\ldots$
represent contributions from the regions where $k_T$ or $l_T$ is
large, which are of the same order in $1/q_T$.

As we see in \eqref{tails-4}, \eqref{tails-5}, and \eqref{tails-frag},
the leading power behavior of some distribution or fragmentation
functions comes with a factor $\alpha_s^2$.  At this order, one must
also take into account regions of integration in \eqref{soft-con}
where two out of the three momenta $p_T$, $k_T$, $l_T$ are large, but
it turns out that these do not contribute to the $\alpha_s^2$ terms
given in the following.
Using the high-transverse-momentum behavior in \eqref{tails-2} to
\eqref{tails-5} and \eqref{tails-frag}, we obtain
\begin{align}
F_{UU,T} &\sim \frac{1}{q_T^2}\, \alpha_s\, \ff{f_1 D_1} \,,
\label{e:tailFUUT}
 \\
F_{UU}^{\cos\phi_h} &\sim \frac{1}{Q\ms q_T}\, \alpha_s\,
                          \ff{f_1 D_1} \,,
 \\
F_{UU}^{\cos 2\phi_h} &\sim \frac{M^2}{q_T^4}\,
     \alpha_s\, \ff{h_1^{\perp (1)}\ms H_1^{\perp (1)}, \ldots} \,,
\label{e:tailFUUcos2phi}
 \\
F_{LU}^{\sin\phi_h} &\sim 
   \frac{1}{Q\ms q_T}\, \alpha_s^2\, \ff{f_1 D_1} \,,
\label{e:tailFLUsinphi}
 \\
F_{UL}^{\sin\phi_h} &\sim 
   \frac{1}{Q\ms q_T}\, \alpha_s^2\, \ff{g_1 D_1} \,,
\label{e:tailFULsinphi}
 \\  \displaybreak[1]
F_{UL}^{\sin 2\phi_h} &\sim \frac{M^2}{q_T^4}\,
     \alpha_s\, \ff{h_{1L}^{\perp (1)}\ms H_1^{\perp (1)}, \ldots} \,,
\label{e:tailFULsin2phi}
 \\
F_{LL} &\sim \frac{1}{q_T^2}\, \alpha_s\, \ff{g_1 D_1} \,,
 \\
F_{LL}^{\cos\phi_h} &\sim \frac{1}{Q\ms q_T}\, \alpha_s\,
                          \ff{g_1 D_1} \,,
 \\[0.3em]
F_{UT,T}^{\sin(\phi_h-\phi_S)} &\sim \frac{M}{q_T^3} \,
      \alpha_s\, \ff{f_{1T}^{\perp (1)}\ms D_1^{},\, \ldots} \,,
\label{e:tailFUTT}
 \\
F_{UT}^{\sin(\phi_h+\phi_S)} &\sim \frac{M}{q_T^3} \,
      \alpha_s\, \ff{h_1^{}\ms H_1^{\perp (1)},\, \ldots} \,,
 \\
F_{UT}^{\sin(3\phi_h-\phi_S)} &\sim 
    \frac{M}{q_T^3} \, \alpha_s^2\,
      \ff{h_1^{}\ms H_1^{\perp (1)}, \ldots} \,,
\label{e:tailFUTsin3phi}
 \\
F_{UT}^{\sin\phi_S} &\sim \frac{M}{Q\ms q_T^2} \,
      \alpha_s\, \ff{f_{1T}^{\perp (1)}\ms D_1^{},\,
                     h_1^{}\ms H_1^{\perp (1)}, \ldots} \,,
 \\
F_{UT}^{\sin(2\phi_h-\phi_S)} &\sim \frac{M}{Q\ms q_T^2} \,
      \alpha_s\, \ff{f_{1T}^{\perp (1)}\ms D_1^{}, \ldots} \,,
\label{e:tailFUTsin2phi}
 \\
F_{LT}^{\cos(\phi_h-\phi_S)} &\sim \frac{M}{q_T^3} \,
      \alpha_s\, \ff{g_{1T}^{(1)}\ms D_1^{},\, \ldots} \,,
\label{e:tailFLTcosphi}
 \\
F_{LT}^{\cos\phi_S} &\sim \frac{M}{Q\ms q_T^2} \,
      \alpha_s\, \lff{g_{1T}^{(1)}\ms D_1^{},\,
                      h_1^{}\ms \frac{E}{z},\, \ldots} \,,
 \\
F_{LT}^{\cos(2\phi_h-\phi_S)} &\sim \frac{M}{Q\ms q_T^2} \,
       \alpha_s\, \ff{g_{1T}^{(1)}\ms D_1^{}, \ldots} \,.
\label{e:tailFLTcos2phi}
\end{align}
Here either the parton distributions or the fragmentation functions
are convoluted with kernels $K_{i}$ or $L_{i}$\,:
\begin{align}
\ff{f D} &= \frac{1}{z^2}\, \sum_{a,i} e_a^2\;
   \Bigl[\ms \bigl( K_i \otimes f^i \bigr)(x) \, D^a(z)
           + f^a(x) \, \bigl( D^i \otimes L_i \bigr)(z) \ms\Bigr] \,,
\label{e:convolution3}
\end{align} 
where the sum runs over quarks and antiquarks for $a$ and over quarks,
antiquarks and gluons for $i$.  As we will see in
section~\ref{sec:tails}, these kernels contain logarithms of $Q/q_T$.
Their origin is the dependence of $f_1(x,p_T^2)$ or $D_1(z,k_T^2)$ on
$\zeta$ or $\zeta_h$, which we tacitly omitted in \eqref{soft-con}.
When resummed to all orders in $\alpha_s$ in the way we sketched in
section~\ref{sec:example}, these logarithms can lead to a substantial
modification of the power laws in \eqref{e:tailFUUT} to
\eqref{e:tailFLTcos2phi}.  A numerical study of these effects on
azimuthal asymmetries in Drell-Yan production has been performed in
\cite{Boer:2001he}.

We note that for the $1/Q$ suppressed structure functions in
\eqref{e:tailFUUT} to \eqref{e:tailFLTcos2phi}, contributions from
$U(l_T^2)$ taken at $\bm{l}{}_T \approx -\bm{q}{}_T$ are power
suppressed or have the same power behavior as contributions where
either $\bm{p}{}_T \approx -\bm{q}{}_T$ or $\bm{k}_T \approx
\bm{q}{}_T$.  For these structure functions, the power behavior at
high $q_T$ hence remains the same if we simply ignore the soft factor
and work with the tree-level convolution \eqref{tree-con} instead of
\eqref{soft-con}.


\section{Comparing results at intermediate $q_T$}
\label{sec:comparison}

We can now compare the results for the region $M \ll q_T \ll Q$
obtained in the low-$q_T$ calculation of the previous section with
those obtained in the high-$q_T$ calculation.  As we mentioned in
section \ref{sec:large}, not all structure functions have been
calculated in the high-$q_T$ picture.  For the cases where results are
available, we find
\begin{align}
\label{e:high_FUU_power}
F_{UU,T}^{} &\sim \frac{1}{q_T^2}\, \alpha_s\, \ff{f_1 D_1}\,,
\\
\label{e:high_FUUL_power}
F_{UU,L}^{} &\sim \frac{1}{Q^2}\, \alpha_s\, \ff{f_1 D_1}\,,
\phantom{\frac{1}{q_T^2}}
 \\
\label{e:high_FUUcosphi_power}
F_{UU}^{\cos\phi_h} &\sim \frac{1}{Q q_T}\, \alpha_s\, \ff{f_1 D_1}\,,
\\
\label{e:high_FUUcos2phi_power}
F_{UU}^{\cos 2\phi_h} & \sim \frac{1}{Q^2}\, \alpha_s\, \ff{f_1 D_1}\,,
\\
\label{e:high_FLUsinphi_power}
F_{LU}^{\sin\phi_S} &\sim \frac{1}{Q q_T}\, \alpha_s^2\, \ff{f_1 D_1}\,,
\\
\label{e:high_FLL_power}
F_{LL}^{} &\sim \frac{1}{q_T^2}\, \alpha_s\, \ff{g_1 D_1}\,,
\\
\label{e:high_FLLcosphi_power}
F_{LL}^{\cos\phi_h} &\sim \frac{1}{Q q_T}\, \alpha_s\, \ff{g_1 D_1}\,,
\\[0.2em]
F_{UT,T}^{\sin(\phi_h-\phi_S)} &\sim \frac{M}{q_T^3} \,
      \alpha_s\, \ff{G_F^{}\ms D_1^{},\widetilde{G}_F^{}\ms D_1^{}},
\label{e:high_FUTT}
\\
F_{UT,L}^{\sin(\phi_h-\phi_S)} &\sim \frac{M}{Q^2 q_T} \,
      \alpha_s\, \ff{G_F^{}\ms D_1^{}},
\label{e:high_FUTL}
\\
F_{UT}^{\sin(\phi_h+\phi_S)} &\sim \frac{M}{q_T^3} \,
      \alpha_s\, \ff{h_1\ms\widehat{E}_F} ,
\label{e:high_FUTcollins}
\\
F_{UT}^{\sin(3\phi_h-\phi_S)} &\sim \frac{M}{Q^2 q_T} \,
      \alpha_s\, \ff{G_F^{}\ms D_1^{},\widetilde{G}_F^{}\ms D_1^{}},
\label{e:high_FUTsin3phi}
\\
F_{UT}^{\sin\phi_S} &\sim \frac{M}{Q\ms q_T^2} \,
      \alpha_s\, \ff{G_F^{}\ms D_1^{},\widetilde{G}_F^{}\ms D_1^{},
                     h_1\ms \widehat{E}_F} ,
\label{e:high_FUTsinphiS}
\\
F_{UT}^{\sin(2\phi_h-\phi_S)} &\sim \frac{M}{Q\ms q_T^2} \,
      \alpha_s\, \ff{G_F^{}\ms D_1^{},\widetilde{G}_F^{}\ms D_1^{} }.
\label{e:high_FUTsin2phi}
\end{align} 
The symbol $\mathcal{F}$ in \eqref{e:high_FUU_power} to
\eqref{e:high_FLLcosphi_power} has the same meaning as in
\eqref{e:convolution3}, whereas for the terms involving the
three-parton correlation functions $G_{F}$ and $\widetilde{G}_F$ we
have
\begin{align}
\ff{G D} &= \frac{1}{z^2}\, \sum_{a,i} e_a^2\;
   \Bigl[\ms \bigl( K_i \otimes G^i \ms\bigr)(x) \, D^a(z)
        + G^a(x,x) \, \bigl( D^i \otimes L_i \bigr)(z) \ms\Bigr] \,,
\label{e:convolution4}
\end{align} 
where the two-variable convolution $\bigl( K_i \otimes G^i
\ms\bigr)(x)$ is of the form \eqref{conv-two-var}.  The terms
involving the three-parton fragmentation function $\widehat{E}_F$ in
\eqref{e:high_FUTcollins} and \eqref{e:high_FUTsinphiS} are defined in
analogy to \eqref{e:convolution4}.

The results in \eqref{e:high_FUU_power} to
\eqref{e:high_FUUcos2phi_power} and \eqref{e:high_FLL_power} to
\eqref{e:high_FLLcosphi_power} are directly taken from our expressions
\eqref{e:high_FUU} to \eqref{e:high_LLcosphi}, whereas the result for
$F_{LU}^{\sin\phi_h}$ in \eqref{e:high_FLUsinphi_power} has been
extracted from the calculation in \cite{Hagiwara:1983cq}.  The form of
the $F_{UT}$ structure functions in \eqref{e:high_FUTT} to
\eqref{e:high_FUTsin2phi} can be obtained by taking the limit $q_T \ll
Q$ of the results of Eguchi et al.~\cite{Eguchi:2006qz,Eguchi:2006mc},
with the caveats discussed in section~\ref{sec:large}.
We note that the results of \cite{Eguchi:2006qz,Eguchi:2006mc} also
contain terms involving the product $h_1 \widehat{E}_F$ in
$F_{UT,T}^{\sin(\phi_h-\phi_S)}$, as well as terms involving $G_F D_1$
or $\widetilde{G}_F D_1$ in $F_{UT}^{\sin(\phi_h+\phi_S)}$.  However,
these contributions behave like $M/ (Q^2\ms q_T)$ for $q_T \ll Q$ and
are thus power suppressed compared with the terms given in
\eqref{e:high_FUTT} and \eqref{e:high_FUTcollins}.

Let us first discuss the unpolarized structure functions.  Comparing
the high-$q_T$ results \eqref{e:high_FUU_power} to
\eqref{e:high_FUUcos2phi_power} with the low-$q_T$ results
\eqref{e:tailFUUT} to \eqref{e:tailFUUcos2phi}, we find that at
intermediate $q_T$ the power behavior of both $F_{UU,T}^{}$ and
$F_{UU}^{\cos\phi_h}$ agrees in the two calculations.  We shall see in
section \ref{sec:tailstructure} that in the case of $F_{UU,T}$ this
agreement extends to the explicit expression of the structure function
at order $\alpha_s$.
By contrast, the leading power behavior obtained for
$\smash{F_{UU}^{\cos 2\phi_h}}$ in the intermediate region is not the
same in the low- and the high-$q_T$ calculations.  In fact, the two
results \eqref{e:high_FUUcos2phi_power} and \eqref{e:tailFUUcos2phi}
describe two different physical mechanisms, since the low-$q_T$
calculation involves chiral-odd distribution and fragmentation
functions, whereas the high-$q_T$ calculation involves chiral-even
ones.  Finally, the longitudinal structure function $F_{UU,L}$ only
appears at order $1/Q^2$ in the low-$q_T$ calculation and is hence
beyond the accuracy of the results given in section~\ref{sec:lowtm}.
We remark that it is far from clear whether small-$q_T$ factorization
still holds at twist-four level, given that even the twist-three case
is not fully understood.

At this point we wish to discuss the calculation of the unpolarized
structure functions at low transverse momentum in the parton model
\cite{Anselmino:2005nn}, where intrinsic transverse momentum is
included in distribution and fragmentation functions and the
kinematics is taken such that the quarks in the parton-level
subprocess $\gamma^* q\to q$ are on shell.  Using Eqs.~(4) and (32) of
\cite{Anselmino:2005nn} and expanding in $1/Q$, we obtain $F_{UU,T} =
\mathcal{C}\bigl[ f_1 D_1 \bigr]$ as in \eqref{F_UUT}, whereas up to
relative corrections in $1/Q$ the other unpolarized structure
functions read
\begin{align}
  \label{Cahn}
F_{UU}^{\cos\phi_h} &= - \frac{2M}{Q}\, \mathcal{C}\biggl[
     \frac{\hat{\bm{h}}\cdott \bm{p}_T}{M}\, f_1 D_1 \biggr] \,,
\nonumber \\[0.2em]
F_{UU}^{\cos 2\phi_h} &= \frac{4M^2}{Q^2}\, \mathcal{C}\biggl[
     \frac{2\, (\hat{\bm{h}}\cdott \bm{p}_T)^2 -\bm{p}_T^2}{2 M^2}\,
     f_1 D_1 \biggr] \,,
&
F_{UU,L}^{} &= \frac{4M^2}{Q^2}\, \mathcal{C}\biggl[
     \frac{\bm{p}_T^2}{M^2}\, f_1 D_1 \biggr] \,, 
\end{align}
with the tree-level convolution $\mathcal{C}$ defined in
\eqref{tree-con}.  The modulations in $\cos \phi_h$ and $\cos 2\phi_h$
obtained in this calculation are often referred to as Cahn effect
\cite{Cahn:1978se,Cahn:1989yf}.
Taking the limit $q_T \gg M$ of the expressions in \eqref{Cahn} we
find the same power behavior as in the high-$q_T$ expressions
\eqref{e:high_FUUL_power} to \eqref{e:high_FUUcos2phi_power}.
However, the high-$q_T$ behavior of \eqref{Cahn} comes only from the
high-$p_T$ tail of $f_1$ but not from the high-$k_T$ tail of $D_1$.
It hence only involves terms of the form $(K_i \otimes f_1^i)\,
D_1^a$, with the same kernels $K_i$ for $F_{UU}^{\cos\phi_h}$,
$F_{UU}^{\cos 2\phi_h}$, and $F_{UU,L}^{}$.  This readily implies that
at intermediate $q_T$ the parton-model results \eqref{Cahn} do
\emph{not} match with the explicit results \eqref{e:high_FUUL} to
\eqref{e:high_FUUcos2phi} of the high-$q_T$ calculation.

We remark that the high-$q_T$ limit of the full twist-three result
\eqref{F_UUcosphi} for $F_{UU}^{\cos\phi_h}$ comes from the
chiral-even terms
\begin{equation}
  \label{F_UUcosphi_part}
- \frac{2M}{Q}\,\mathcal{C}\biggl[
    \frac{\hat{\bm{h}}\cdott \bm{k}_T^{}}{M}\,
      f_1 \frac{\tilde{D}^{\perp }}{z}
  + \frac{\hat{\bm{h}}\cdott \bm{p}_T}{M}
      x f^{\perp } D_1 \biggr] \,.
\end{equation}
As observed in \cite{Bacchetta:2006tn}, this coincides with the parton
model result \eqref{Cahn} if one makes the Wandzura-Wilczek
approximation, i.e., if one sets to zero the functions
$\tilde{D}^\perp = D^\perp - z D_1$ and $x \tilde{f}^\perp = x f^\perp
- f_1$, which are related to quark-gluon-quark correlation functions
by the equation of motion for the quark field.  We will see in section
\ref{sec:tails} that for $q_T \gg M$ these functions are in fact
\emph{not} negligible compared with $D_1$ and $f_1$, so that the
approximations leading to \eqref{Cahn} are not adequate in this limit.
In a similar way, one may understand the parton model results for
$F_{UU,L}^{}$ and $F_{UU}^{\cos 2\phi_h}$ as part of the (unknown)
complete twist-four expression in a low-$q_T$ calculation.  They have
the correct power behavior to match the results \eqref{e:high_FUUL}
and \eqref{e:high_FUUcos2phi} of the high-$q_T$ calculation, but do
not reproduce all terms in these results.
\new{We note that $F_{UU}^{\cos 2\phi_h}$ has the form \eqref{low-ex2}
  discussed in the introduction.  The term with coefficient $l_{2,4}$ is
  given by the low-$q_T$ result \eqref{e:tailFUUcos2phi}, the term with
  $h_{2,4}$ by the high-$q_T$ expression \eqref{e:high_FUUcos2phi},
  whereas the parton-model result \eqref{Cahn} contributes to the
  subleading term $l_{4,2}$ in the low-$q_T$ power counting scheme.} 

\new{Several phenomenological analyses, for instance those in
  \cite{Mendez:1978vr,Konig:1982uk,Chay:1991nh,%
    Oganessyan:1997jq,Anselmino:2006rv}, have used the parton model
  expressions for the unpolarized structure functions together with the
  high-$q_T$ results \eqref{e:high_FUU} to \eqref{e:high_FUUcos2phi}.  We
  point out that in these papers a Gaussian behavior $f_1(x,p_T^2) \propto
  \exp[- a\ms p_T^2]$ and $D_1(z,k_T^2) \propto \exp[-A\ms k_T^2]$ is
  assumed for the distribution and fragmentation functions appearing in
  the parton model calculation.  Such an approach differs from the one
  taken in the present work, where the power-law behavior of
  $f_1(x,p_T^2)$ and $D_1(z,k_T^2)$ at large transverse momentum is
  retained and explicitly calculated using perturbation theory.}

Turning now to polarized observables, we find that the structure
functions $F_{LL}^{}$ and $F_{LL}^{\cos \phi_h}$ have the same power
behavior in the high- and low-$q_T$ calculations, as do their
unpolarized counterparts $F_{UU,T}^{}$ and $F_{UU}^{\cos \phi_h}$.  As
in the unpolarized case, we will see in section
\ref{sec:tailstructure} that $F_{LL}^{}$ matches exactly at order
$\alpha_s$ in the two calculations.
Our low-$q_T$ result \eqref{e:tailFLUsinphi} for the $T$-odd structure
function $F_{LU}^{\sin\phi_S}$ has the correct structure to match the
limit \eqref{e:high_FLUsinphi_power} of the calculation at high $q_T$
and order $\alpha_s^2$ in \cite{Hagiwara:1983cq}.  One may expect that
our low-$q_T$ result \eqref{e:tailFULsinphi} for $F_{UL}^{\sin\phi_S}$
would also match with a high-$q_T$ calculation at the same order, but
were are not aware of such a calculation in the literature.

For transverse polarization observables we compare
Eqs.~\eqref{e:high_FUTT} to \eqref{e:high_FUTsin2phi} with
\eqref{e:tailFUTT} to \eqref{e:tailFUTsin2phi} and see that four out
of six structure functions have a matching power behavior, namely
$F_{UT,T}^{\sin(\phi_h-\phi_S)}$, $F_{UT}^{\sin(\phi_h+\phi_S)}$,
$F_{UT}^{\sin\phi_S}$, and $F_{UT}^{\sin(2\phi_h-\phi_S)}$.  The
distribution and fragmentation functions appearing in the respective
results are compatible as well, given that $\smash{f_{1T}^{\perp
    (1)}}$ is related with $G_F$ and $H_{\smash{1}}^{\perp (1)}$ with
$\widehat{E}_F$.  As already mentioned in section \ref{sec:polar}, the
explicit low- and high-$q_T$ calculations of
${F_{UT,T}^{\sin(\phi_h-\phi_S)}}$ in \cite{Ji:2006br,Koike:2007dg}
found exact matching at order $\alpha_s$ for this observable.
Looking at the remaining two $F_{UT}$ structure functions, we see that
$\smash{F_{UT,L}^{\sin(\phi_h-\phi_S)}}$ is beyond the accuracy of the
low-transverse-momentum results.  This is just as for $F_{UU,L}$,
which is the only other structure function in \eqref{e:crossmaster}
that involves purely longitudinal polarization of the virtual photon
\cite{Bacchetta:2006tn}.
The structure function $\smash{F_{UT}^{\sin(3\phi_h-\phi_S)}}$ does
not match in the low- and high-$q_T$ calculations.  As in the case of
$F_{UU}^{\cos 2\phi_h}$, the low-transverse-momentum result involves
chiral-odd functions, whereas the high-transverse-momentum expression
involves chiral-even ones.  The low-$q_T$ result
\eqref{e:tailFUTsin3phi} for $F_{UT}^{\sin(3\phi_h-\phi_S)}$ can
potentially match a high-transverse-momentum calculation at twist
three and order $\alpha_s^2$, and the high-$q_T$ result
\eqref{e:high_FUTsin3phi} could match with a low-$q_T$ calculation at
twist four.  Both types of calculation are beyond the current state of
the art.

To the best of our knowledge, $F_{LT}^{\cos(\phi_h-\phi_S)}$,
$F_{LT}^{\cos\phi_S}$ and $F_{LT}^{\cos(2\phi_h-\phi_S)}$ have not
been computed in the high-$q_T$ approach.  {}From the overall factor
$M$ in \eqref{e:tailFLTcosphi} to \eqref{e:tailFLTcos2phi} we can only
conclude that these low-$q_T$ results can potentially match with those
of a high-$q_T$ calculation at twist-three accuracy.

In table~\ref{tab:overview} we collect the results for the leading
power behavior of all structure functions we have discussed.  We
notice that for several observables the twist of the low-$q_T$ and the
high-$q_T$ calculation is not the same, which is reminiscent of a
similar observation we made for the high-$p_T$ behavior of
distribution functions in section~\ref{sec:power}.

\TABLE[t]{
\renewcommand{\arraystretch}{1.3}
\caption{\label{tab:overview} 
  \new{Leading power behavior of SIDIS structure functions in
  the intermediate region $M\ll q_T \ll Q$, corresponding to the
  expansions in \protect\eqref{low-inter} and \protect\eqref{hi-inter},
  respectively.} 
  Empty fields indicate
  that no calculation is available.  The specification of twist $4$
  for $F^{}_{UU,L}$ and $F^{\sin(\phi_h-\phi_S)}_{UT,L}$ reflects that
  these observables are zero when calculated at twist-two and
  twist-three accuracy.}
\begin{tabular}{|l|ccc|ccc|c|} \hline
 & \multicolumn{3}{|c|}{low-$q_T$ calculation}
 & \multicolumn{3}{|c|}{high-$q_T$ calculation}
 & \new{leading powers} \\
observable & twist & order & power & twist & order & power
 & match \\ \hline
$F^{}_{UU,T}$ & 2 & $\alpha_s$ & $1/q_T^{2}$
              & 2 & $\alpha_s$ & $1/q_T^{2}$ & yes \\
$F^{}_{UU,L}$ & 4 & &
              & 2 & $\alpha_s$ & $1/Q^{2}$ &  \\
$F^{\cos\phi_h}_{UU}$ & 3 & $\alpha_s$ & $1/(Q\ms q_T)$
                      & 2 & $\alpha_s$ & $1/(Q\ms q_T)$ & yes \\
$F^{\cos 2\phi_h}_{UU}$ & 2 & $\alpha_s$ & $1/q_T^{4}$
                        & 2 & $\alpha_s$ & $1/Q^{2}$ & no \\
$F^{\sin\phi_h}_{LU}$ & 3 & $\alpha_s^2$ & $1/(Q\ms q_T)$
                      & 2 & $\alpha_s^2$ & $1/(Q\ms q_T)$ & yes \\
$F^{\sin\phi_h}_{UL}$ & 3 & $\alpha_s^2$ & $1/(Q\ms q_T)$
                      & & & &  \\
$F^{\sin 2\phi_h}_{UL}$ & 2 & $\alpha_s$ & $1/q_T^{4}$ & & & &  \\
$F^{}_{LL}$ & 2 & $\alpha_s$ & $1/q_T^{2}$
            & 2 & $\alpha_s$ & $1/q_T^{2}$ & yes \\
$F^{\cos\phi_h}_{LL}$ & 3 & $\alpha_s$ & $1/(Q\ms q_T)$
                      & 2 & $\alpha_s$ & $1/(Q\ms q_T)$ & yes \\
$F^{\sin(\phi_h-\phi_S)}_{UT,T}$ & 2 & $\alpha_s$ & $1/q_T^{3}$
                                 & 3 & $\alpha_s$ & $1/q_T^{3}$ & yes \\
$F^{\sin(\phi_h-\phi_S)}_{UT,L}$ & 4 & &
                      & 3 & $\alpha_s$ & $1/(Q^2\ms q_T)$ &  \\
$F^{\sin(\phi_h+\phi_S)}_{UT}$ & 2 & $\alpha_s$ & $1/q_T^{3}$
                               & 3 & $\alpha_s$ & $1/q_T^{3}$ & yes \\
$F^{\sin(3\phi_h-\phi_S)}_{UT}$
                      & 2 & $\alpha_s^2$ & $1/q_T^{3}$
                      & 3 & $\alpha_s$   & $1/(Q^{2}\ms q_T)$ & no \\
$F^{\sin\phi_S}_{UT}$ & 3 & $\alpha_s$ & $1/(Q\ms q_T^{2})$
                      & 3 & $\alpha_s$ & $1/(Q\ms q_T^{2})$ & yes \\
$F^{\sin(2\phi_h-\phi_S)}_{UT}$
                      & 3 & $\alpha_s$ & $1/(Q\ms q_T^2)$
                      & 3 & $\alpha_s$ & $1/(Q\ms q_T^2)$ & yes \\
$F^{\cos(\phi_h-\phi_S)}_{LT}$ & 2 & $\alpha_s$ & $1/q_T^{3}$ & & & & \\
$F^{\cos\phi_S}_{LT}$ & 3 & $\alpha_s$ & $1/(Q\ms q_T^{2})$ & & & & \\
$F^{\cos(2\phi_h-\phi_S)}_{LT}$
                      & 3 & $\alpha_s$ & $1/(Q\ms q_T^{2})$ & & & & \\ 
\hline
\end{tabular}
}


\subsection{Interpolating from low to high $q_T$}
\label{sec:interpol}

Let us now see how one can practically proceed when the leading terms
in the low- and high-$q_T$ descriptions of an observable do not match
in the intermediate region.  As an example we take the unpolarized
structure function $F_{UU}^{\cos 2\phi_h}$.  We denote its low-$q_T$
approximation given in \eqref{F_UUcos2phi} by $L_{UU}^{\cos 2\phi_h}$,
and its high-$q_T$ approximation \eqref{e:high_FUUcos2phi} by
$H_{UU}^{\cos 2\phi_h}$.  Since in the intermediate region the two
expressions describe distinct contributions to the cross section, one
may consider to use
\begin{equation}
\label{F_UUcos-interp}
F_{UU}^{\cos 2\phi_h} \approx
  L_{UU}^{\cos 2\phi_h} + H_{UU}^{\cos 2\phi_h}
\end{equation}
as an approximation for this observable.  The quality of this
approximation can be assessed from the power behavior of its terms in
the different regions:
\begin{alignat}{3}
  \label{LUUcos-pow-lo}
L_{UU}^{\cos 2\phi_h} &\sim q_T^2/M^4 &\qquad & \text{for}~ q_T \lsim M \,,
\\ 
  \label{LUUcos-pow-hi}
L_{UU}^{\cos 2\phi_h} &\sim M^2/ q_T^4 && \text{for}~ q_T \gg M \,,
\\
  \label{HUUcos-pow}
H_{UU}^{\cos 2\phi_h} &\sim 1/Q^2 && \text{for all}~ q_T \,,
\end{alignat}
where the behavior in \eqref{LUUcos-pow-lo} reflects that
$L_{UU}^{\cos 2\phi_h}$ must vanish like $q_T^2$ for $q_T^{}\to 0$ due
to angular momentum conservation \cite{Diehl:2005pc}.  In the
intermediate region $M\ll q_T\ll Q$ both terms in
\eqref{F_UUcos-interp} are required: together they give an
approximation with relative corrections of order $M^2/q_T^2$ or
$q_T^2/Q^2$.  The relative weight of the two terms in this region is
$L_{UU}^{\cos 2\phi_h} /H_{UU}^{\cos 2\phi_h} \sim M^2/q_T^2 \times
Q^2/q_T^2$ and thus varies from values above to values below~1.
As an aside, let us comment on the use of a
transverse-momentum-dependence like $h_1^\perp(x,p_T^2) \propto \exp[-
c\ms p_T^2]$ and $H_1^\perp(z,k_T^2) \propto \exp[-C k_T^2]$, which
is often taken in phenomenological analyses.  Whereas at small
transverse momentum a Gaussian behavior of distribution and
fragmentation functions is found to give a good description of data in
many situations, it misses the perturbative tails of these functions.
As a result it does not give a good approximation of $F_{UU}^{\cos
  2\phi_h}$ at intermediate $q_T$.  For $M \ll q_T \,\lsim\,
\sqrt{MQ\rule{0pt}{1.9ex}}$ the contribution \eqref{LUUcos-pow-hi}
from the perturbative tails is actually dominant, and for
$\sqrt{MQ\rule{0pt}{1.9ex}} \,\lsim\, q_T \ll Q$ it is only suppressed
compared with \eqref{HUUcos-pow} by a factor much larger than
$M^2/q_T^2$.

For large $q_T \sim Q$ the ansatz \eqref{F_UUcos-interp} can be used
as well: the low-$q_T$ calculation is not valid in this region, but
the term $L_{UU}^{\cos 2\phi_h}$ is power suppressed by $M^2/Q^2$
compared with the leading term $H_{UU}^{\cos 2\phi_h}$, which itself
provides an approximation of $F_{UU}^{\cos 2\phi_h}$ up to $M^2/Q^2$
corrections.  Adding $L_{UU}^{\cos 2\phi_h}$ in this region hence does
not spoil the accuracy of the description.  Likewise, the high-$q_T$
calculation is not justified for $q_T \sim M$, but in this region the
term $H_{UU}^{\cos 2\phi_h}$ is suppressed by $M^2/Q^2$ compared with
the correct approximation $L_{UU}^{\cos 2\phi_h}$.  However, one
cannot use \eqref{F_UUcos-interp} for $q_T\to 0$ since $H_{UU}^{\cos
  2\phi_h}$ does not vanish like $q_T^2$.  To repair this, one may
instead take
\begin{equation}
  \label{F_UUcos-inter-rho}
F_{UU}^{\cos 2\phi_h} \approx L_{UU}^{\cos 2\phi_h}
  + \rho\biggl( \frac{q_T^2}{M^2} \biggr)\, H_{UU}^{\cos 2\phi_h}
\end{equation}
with an interpolating function $\rho(r)$ that satisfies $\rho(r)\sim
r$ for $r\to 0$ and $\rho(r) - 1 \sim r^{-1}$ for $r\gg 1$.  A simple
choice is $\rho(r) = r /(1+r)$, but obviously there are other
possibilities.

An often considered observable is the azimuthal asymmetry
\begin{equation}
A_{UU}^{\cos 2\phi_h} = \frac{\varepsilon F_{UU}^{\cos 2\phi_h}}{%
  F^{}_{UU,T} + \varepsilon F^{}_{UU,L}} \,.
\end{equation}
Depending on $q_T$ we can approximate its denominator using the
high-$q_T$ expression $H_{UU,T} + \varepsilon H_{UU,L}$ from
\eqref{e:high_FUU} and \eqref{e:high_FUUL} or the low-$q_T$ result
$L_{UU,T}$ given in \eqref{F_UUT}.  Since $F_{UU,L}$ is suppressed by
$1/Q^2$ for $q_T \ll Q$, we do not need the unknown low-$q_T$
expression for this structure function.  Using
\begin{alignat}{3}
L_{UU,T} &\sim 1/M^2 &\qquad & \text{for}~ q_T \lsim M \,,
\\
L_{UU,T} &\sim 1/q_T^2 && \text{for}~ q_T \gg M \,,
\\
H_{UU,T} + \varepsilon H_{UU,L} &\sim 1/q_T^2 && \text{for all}~ q_T
\end{alignat}
together with \eqref{LUUcos-pow-lo} to \eqref{HUUcos-pow}, we find
that
\begin{equation}
  \label{A_UUcos-interp}
A_{UU}^{\cos 2\phi_h} \approx
  \frac{\varepsilon L_{UU}^{\cos 2\phi_h}}{L^{}_{UU,T}}
+ \frac{\varepsilon H_{UU}^{\cos 2\phi_h}}{%
  H^{}_{UU,T} + \varepsilon H^{}_{UU,L}}
\end{equation}
gives a good approximation of the asymmetry in the full $q_T$ range.
In the intermediate region, the denominators of the two terms in
\eqref{A_UUcos-interp} coincide up to terms of order $M^2/q_T^2$ or
$q_T^2/Q^2$ and approximate $F_{UU,T} + \varepsilon F_{UU,L}$ with
that precision.  As discussed above, both the low-$q_T$ and the
high-$q_T$ contributions are important in the intermediate region
$M\ll q_T \ll Q \,$ (where again one finds that with a Gaussian ansatz
for the transverse-momentum-dependence of distribution and
fragmentation functions, the low-$q_T$ term would not be correctly
described).  For $q_T \sim Q$ the low-$q_T$ term is power suppressed
and may hence be kept in \eqref{A_UUcos-interp}.  For $q_T \lsim M$,
the high-$q_T$ term in the asymmetry is suppressed by a relative
factor of $M^2/Q^2$ compared with the low-$q_T$ term and does not
degrade the quality of the approximation \eqref{A_UUcos-interp} in the
limit $q_T\to 0$.  An additional suppression factor as in
\eqref{F_UUcos-inter-rho} is therefore not required.  Recalling the
discussion after Eq.~\eqref{last-splitting-fct}, we can understand why
$\varepsilon H_{UU}^{\cos 2\phi_h} /(H^{}_{UU,T} + \varepsilon
H^{}_{UU,L})$ has the correct $q_T\to 0$ limit required by angular
momentum conservation: the propagator factors $1/q_T^2$ that lead to
an unphysical behavior of the individual structure functions cancel in
this ratio.

Let us finally remark that the discussion in this subsection is at the
level of power counting arguments.  When using
\eqref{F_UUcos-inter-rho} or \eqref{A_UUcos-interp} in practice, one
can explicitly check whether the terms that are out of their region of
validity (the $L$ terms for $q_T \sim Q$ and the $H$ terms for $q_T
\lsim M$) are numerically small compared with the leading ones.


\section{Integrating over $q_T$}
\label{sec:int-obs}

\subsection{Behavior of integrated and weighted observables}
\label{sec:int-first}

Up to now we have focused on the $q_T$-dependence of the structure
functions $F(Q,q_T)$.  As we mentioned in the introduction,
observables that are integrated over $q_T$, with or without a
weighting factor $(q_T/M)^p$, can be preferable to observables
differential in $q_T$ for experimental reasons.  Without dwelling on
such practical issues, we now use our results of the previous sections
for discussing the theoretical interpretation of integrated
observables.  
As a shorthand notation we introduce
\begin{equation}
\Big\langle \!\!\Big\langle
    \Bigl(\frac{q_T}{M}\Bigr)^p\ms F(Q,q_T) 
\Big\rangle \!\!\Big\rangle
  = \pi z^2 \int_{0}^{q_{\text{max}}^2} \de q_T^2\;
    \Bigl(\frac{q_T}{M}\Bigr)^p\ms F(Q,q_T)  \,,
\end{equation}
where $q_{\text{max}}$ is the upper kinematic limit of $q_T$, to be
treated as a quantity of order $Q$ in the power counting.  The
prefactor has been chosen for later convenience---note that $\pi
z^2\ms \de q_T^2$ corresponds to $\de^2 P_{h\perp}$.

To make the notion of ``intermediate transverse momentum'' more
precise, we introduce two scales $\Gamma M^2$ and $\gamma Q^2$ such
that $\Gamma \gg 1$, $\gamma \ll 1$, and $\Gamma M^2 < \gamma Q^2$.
In the intermediate region $\Gamma M^2 < q_T^2 < \gamma Q^2$ the
results of both the low-$q_T$ and the high-$q_T$ calculations are then
valid, and one can use their respective limiting expressions given in
sections \ref{sec:lowtm} and~\ref{sec:comparison}.
It is easy to determine the power-law behavior of the contributions
from the regions $q_T^2 < \Gamma M^2$ and $q_T^2 > \gamma Q^2$ to an
integrated observable.  For a single term in the general low-$q_T$ and
high-$q_T$ expansions \eqref{low-expand} and \eqref{hi-expand}, we
obtain
\begin{align}
  \label{low-int}
\frac{1}{M^2} \int_{0}^{\Gamma M^2} \de q_T^2\, 
  \biggl[\frac{q_T}{M}\biggr]^{p}\,
  \biggl[\frac{q_T}{Q}\biggr]^{n-2}
  l_{n} \biggl(\frac{M}{q_T}\biggr)
&\;\sim\; \biggl[\frac{M}{Q}\biggr]^{n-2} \,,
\\[0.2em]
  \label{hi-int}
\frac{1}{M^2} \int_{\gamma Q^2}^{q_{\text{max}}^2} \de q_T^2\,
  \biggl[\frac{q_T}{M}\biggr]^{p}\,
  \biggl[\frac{M}{q_T}\biggr]^{n}
  h_{n} \biggl(\frac{q_T}{Q}\biggr)
&\;\sim\; \biggl[\frac{M}{Q}\biggr]^{n-2-p}
\end{align}
from straightforward dimensional analysis.  The $Q$-dependence of the
integrals can thus be established without knowledge of the functions
$l_n(q_T/M)$ and $h_n(q_T/Q) \ms$: it is directly determined by the
twist $n$ in the low-$q_T$ case \eqref{low-int}, and by the twist $n$
and the weighting power $p$ in the high-$q_T$ case \eqref{hi-int}.  We
observe in particular that for $p=0$, i.e.\ without weighting, the
twist-two terms in both the low- and high-$q_T$ calculations give
contributions to the integral that stay constant for $Q\to \infty$,
whereas higher-twist terms die out in that limit.  For $p>0$ the
contribution from the high-$q_T$ region is enhanced: a twist-two term
in the low-$q_T$ calculation will then only dominate the integral
over all $q_T$ if for the observable in question a sufficient number
of terms with low twist in the high-$q_T$ result are zero.

As a preparation for the discussion of azimuthal and polarization
asymmetries let us first take a closer look at the familiar structure
functions $F_{UU,T}$ and $F_{UU,L}$.  With the behavior $F_{UU,T} \sim
1/q_T^2$ in the intermediate region (obtained in both the low- and
high-$q_T$ calculations), we obtain
\begin{equation}
  \label{FUUT-medium}
\int_{\Gamma M^2}^{\gamma Q^2} \de q_T^2\, F_{UU,T}^{} \;\sim\;
  \ln\biggl[ \dfrac{\gamma}{\Gamma}\, \dfrac{Q^2}{M^2} \biggr] \,.
\end{equation}
For the integral in the low-$q_T$ domain $q_T^2 < \Gamma M^2$ we have
the generic power-law behavior given in \eqref{low-int} with $n=2$ and
$p=0$.  Using in addition that $F_{UU,T} \sim 1/q_T^2$ at the upper
end of the integration region, we have
\begin{equation}
  \label{FUUT-low}
\int_0^{\Gamma M^2} \de q_T^2\, F_{UU,T}^{}
   \;\sim\; \ln\frac{\Gamma}{\Gamma_0}
\end{equation}
with some number $\Gamma_0 \sim 1$.  Likewise, we can use that
$F_{UU,T} \sim 1/q_T^2$ at the lower end of the integration region in
the high-$q_T$ domain $q_T^2 > \gamma Q^2$, and get
\begin{equation}
  \label{FUUT-high}
\int_{\gamma Q^2}^{q_{\text{max}}^2} \de q_T^2\, F_{UU,T}^{} \;\sim\;
  \ln\frac{\gamma_0}{\gamma}
\end{equation}
with some $\gamma_0 \sim 1$.  In the complete integral
$\intobs{F_{UU,T}}$ all three regions in \eqref{FUUT-medium} to
\eqref{FUUT-high} thus contribute at leading power in $1/Q$, and the
dependence on the artificial separation parameters $\Gamma$ and
$\gamma$ cancels as it should.  We note that, since we are concerned
with power behavior in this section, we have not taken into account
logarithms of $Q/q_T$ in the high- or low-$q_T$ results for
$F_{UU,T}$, which would modify the logarithms on the r.h.s.\ of
\eqref{FUUT-medium} to \eqref{FUUT-high}.

To calculate the integrated structure function one must not double
count the contributions from the low-$q_T$ and high-$q_T$ calculations
in the intermediate region.  Since the results of the two calculations
coincide there, one may simply switch from one to the other
description at a suitable point, say at $q_T^2 = \gamma Q^2$.  We can
now make contact with the standard description of $q_T$-integrated
SIDIS in the collinear factorization framework, where
$\intobs{F_{UU,T}}$ is expressed in terms of the collinear functions
$f_1(x)$ and $D_1(z)$.  Let us in this framework take $\mu^2 = \gamma
Q^2$ for the factorization scale and consider the Born graph as well
as the real and virtual $\alpha_s$-corrections, i.e., the one-loop
graphs where a gluon either does or does not cross the final-state
cut.
Loosely speaking, the Born term then corresponds to the sum of
\eqref{FUUT-medium} and \eqref{FUUT-low}, and the real corrections to
\eqref{FUUT-high}.  The virtual corrections correspond to the hard
factor $|H|^2$ in the Collins-Soper factorization formula
\eqref{FUUTconv}, which we neglected in section~\ref{sec:power} when
extracting results at lowest order in $\alpha_s$.  The logarithm of
$\gamma Q^2$ in \eqref{FUUT-medium} corresponds to the scale
dependence of the collinear distribution and fragmentation functions,
whereas the $\gamma$-dependence in \eqref{FUUT-high} corresponds to an
explicit logarithm $\ln(Q^2 /\mu^2)$ in the real corrections.  At the
technical level, however, collinear factorization is typically
implemented by using dimensional regularization instead of a
transverse-momentum cutoff.  The real corrections are then integrated
down to $q_T=0$, whereas collinear distribution and fragmentation
functions are defined from integrals $\int \de^{2-\varepsilon}\ms
\bm{p}_T^{}\, f_1(x,p_T^2)$ and $\int \de^{2-\varepsilon}\ms
\bm{k}_T^{}\, D_1(x,k_T^2)$ over the full transverse-momentum region.
Subtractions defined for instance by the $\overline{\text{MS}}$
prescription are then performed, which on one hand ensure that there
is no double counting and on the other hand remove terms corresponding
to logarithmic divergences in the physical limit $\varepsilon\to 0$.
Since the incoming and outgoing parton momenta are approximated as
collinear to the associated hadron momenta, the Born term and the
virtual corrections appear with a factor $\delta^{(2)}(\bm{q}_T)$ in
the calculation.

For the longitudinal structure function $F_{UU,L}$ the situation is
quite different.  Using the same procedure as for $F_{UU,T}$ we obtain
\begin{alignat}{2}
  \label{FUUL-medium}
\int_{\Gamma M^2}^{\gamma Q^2} \de q_T^2\, F_{UU,L}^{}
  &\;\sim\; \gamma\,
   \biggl( 1 - \frac{\Gamma M^2}{\gamma Q^2} \biggr)
  &&\;\sim\;  \gamma \,,
\\
  \label{FUUL-high}
\int_{\gamma Q^2}^{q_{\text{max}}^2} \de q_T^2\, F_{UU,L}^{}
  &\;\sim\;  \gamma_0 - \gamma
  &&\;\sim\;  1
\end{alignat}
from the result \eqref{e:high_FUUL_power} of the high-$q_T$
calculation.  For the second step in \eqref{FUUL-medium} we have
assumed that $(\Gamma M^2) /(\gamma Q^2)$ is sufficiently small
compared to $1$---otherwise the corresponding integral becomes small
simply because its integration region shrinks to zero.  After the
second step in \eqref{FUUL-high}, the dependence on $\gamma$ no longer
explicitly cancels in the sum of the two integrals, but this leads to
no inconsistency because \eqref{FUUL-medium} is negligible compared
with \eqref{FUUL-high}.
One easily sees that the parton-model approximation \eqref{Cahn},
whose power behavior agrees with the high-$q_T$ result in the
intermediate region, gives a result suppressed as $M^2/Q^2$ when
integrated over the low-$q_T$ domain $q_T^2 < \Gamma M^2$.  We thus
find that the integrated structure function $\intobs{F_{UU,L}}$ is
dominated by large $q_T \sim Q$ and can be calculated from the
high-$q_T$ result alone.  Moreover, one can integrate this result down
to $q_T=0$, since the contribution from $q_T^2 < \Gamma M^2$ is power
suppressed by $M^2/Q^2$ and thus of the same order as the accuracy of
the result in the high-$q_T$ region.  Put differently, one can use the
high-$q_T$ result extrapolated to $q_T^2 < \Gamma M^2$ instead of the
(unknown) low-$q_T$ result when evaluating the integrated longitudinal
structure function.
This is just what is done in the standard calculation using collinear
factorization, where the first nonvanishing contribution to this
observable starts at order $\alpha_s$.  The integration over all $q_T$
of the high-$q_T$ expression for $F_{UU,L}$ is convergent and simply
removes the $\delta$ function in the analog of \eqref{e:FUUThigh}. No
subtraction is necessary, and correspondingly no dependence on the
factorization scale $\mu$ arises at order $\alpha_s$.

\TABLE{
  \caption{\label{tab:int} Behavior of selected observables integrated
    over different regions of $q_T^2$.  It is assumed that $\Gamma\gg
    1$, $\gamma \ll 1$ and that $(\Gamma M^2) /(\gamma Q^2)$ is
    sufficiently small compared to $1$.  In cases where the low-$q_T$
    and high-$q_T$ calculations do not match in the intermediate
    region, their respective contributions are given in separate rows.
    The low-$q_T$ entry for $F_{UU,L}$ corresponds to the parton-model
    approximation in \protect\eqref{Cahn}.}
\renewcommand{\arraystretch}{1.9}
\begin{tabular}{|l@{\hspace{4pt}}l|c@{\hspace{18pt}}c@{\hspace{18pt}}c|}
\hline
  & & low $q_T$ & intermediate $q_T$ & high $q_T$ \\
\multicolumn{2}{|l|}{$f(q_T)$}
  & ${\displaystyle\int\limits_0^{\Gamma M^2}
     \de q_T^2\, f(q_T)}$
  & ${\displaystyle\int\limits_{\Gamma M^2}^{\gamma Q^2}
     \de q_T^2\, f(q_T)}$
  & ${\displaystyle\int\limits_{\gamma Q^2}^{q_{\text{max}}^2}
     \de q_T^2\, f(q_T)}$ \\[1.5em] \hline
\multicolumn{2}{|l|}{$F_{UU,T}^{}$}
  & $\ln\Gamma$
  & $\ln\biggl[ \dfrac{\gamma}{\Gamma}\, \dfrac{Q^2}{M^2} \biggr]$
  & $\ln\dfrac{1}{\gamma}$ \\
\multicolumn{2}{|l|}{$F_{UU,L}^{}$}
  & $\dfrac{M^2}{Q^2}\, \Gamma$
  & $\gamma$ & $1$ \\
\multicolumn{2}{|l|}{$F_{UU}^{\cos\phi_h}$}
  & $\dfrac{M}{Q}\ms \sqrt{\smash[b]{\Gamma}}$
  & $\sqrt{\gamma \rule{0pt}{1.4ex}}$ & $1$ \\
\multicolumn{2}{|l|}{$\dfrac{q_T}{M}\ms F_{UU}^{\cos\phi_h}$}
  & $\dfrac{M}{Q}\, \Gamma$
  & $\gamma\, \dfrac{Q}{M}$ & $\dfrac{Q}{M}$ \\
$F_{UU}^{\cos 2\phi_h}$ & (low $q_T$)
  & $1$ & $\dfrac{1}{\Gamma}$ & \\
\phantom{$F_{UU}^{\cos 2\phi_h}$} & (high $q_T$)
  & & $\gamma$ & $1$ \\
$\dfrac{q_T^2}{M^2}\ms F_{UU}^{\cos 2\phi_h}$ & (low $q_T$)
  & $\ln\Gamma$
  & $\ln\biggl[ \dfrac{\gamma}{\Gamma}\, \dfrac{Q^2}{M^2} \biggr]$ & \\
\phantom{$\dfrac{q_T^2}{M^2}\, F_{UU}^{\cos 2\phi_h}$} & (high $q_T$)
  &  & $\gamma^2\, \dfrac{Q^2}{M^2}$ & $\dfrac{Q^2}{M^2}$ \\[0.4em]
\hline
%
%
$F_{UT,T}^{\sin(\phi_h-\phi_S)}$, & $F_{UT}^{\sin(\phi_h+\phi_S)}$
  & $1$ & $\dfrac{1}{\sqrt{\Gamma}}$
        & $\dfrac{1}{\sqrt{\gamma}} \dfrac{M}{Q}$ \\
$\dfrac{q_T}{M}\ms F_{UT,T}^{\sin(\phi_h-\phi_S)}$,
  & $\dfrac{q_T}{M}\ms F_{UT}^{\sin(\phi_h+\phi_S)}$ & $\ln\Gamma$
  & $\ln\biggl[ \dfrac{\gamma}{\Gamma}\, \dfrac{Q^2}{M^2} \biggr]$
  & $\ln\dfrac{1}{\gamma}$ \\
$F_{UT}^{\sin\phi_S}$, & $F_{UT}^{\sin(2\phi_h-\phi_S)}$
  & $\dfrac{M}{Q} \ln{\Gamma}$ & $\dfrac{M}{Q}\,
        \ln\biggl[ \dfrac{\gamma}{\Gamma}\, \dfrac{Q^2}{M^2} \biggr]$
  & $\dfrac{M}{Q} \ln\dfrac{1}{\gamma}$ \\
$\dfrac{q_T^2}{M^2}\ms F_{UT}^{\sin\phi_S}$,
  & $\dfrac{q_T^2}{M^2}\ms F_{UT}^{\sin(2\phi_h-\phi_S)}$
  & $\dfrac{M}{Q}\, \Gamma$ & $\gamma\, \dfrac{Q}{M}$
  & $\dfrac{Q}{M}$ \\[0.4em]
\hline
\end{tabular}
}

Let us now turn to the structure functions that describe the
$\phi_h$-dependence of the unpolarized cross section.  In
table~\ref{tab:int} we see that the integrated structure function
$\intobs{F_{UU}^{\cos\phi_h}}$ is dominated by large transverse
momenta $q_T$, whereas the region where the low-$q_T$ calculation is
valid contributes only as a power correction of order $M/Q$.  The
condition $(\Gamma M^2) /(\gamma Q^2) < 1$ implies
$\sqrt{\smash[b]{\Gamma}} M/Q < \sqrt{\gamma\rule{0pt}{1.4ex}} \ll 1$,
so that the factor $\sqrt{\smash[b]{\Gamma}}$ cannot compensate the
suppression by $M/Q$.  An important consequence is that
$\intobs{F_{UU}^{\cos\phi_h}}$ is \emph{not} a good observable to
study the transverse-momentum-dependent distribution and fragmentation
functions appearing in the low-$q_T$ result \eqref{F_UUcosphi}.  An
appropriate observable for this purpose is the structure function
differential in $q_T$.  If integration over $q_T$ is required by
statistics, one should impose a suitable upper cutoff on the integral.
According to table~\ref{tab:int}, the dependence of the integral on
this cutoff is not negligible and must hence explicitly be kept in the
theoretical calculation.  Note that in order not to introduce an
artificial $\phi_h$-dependence, the cutoff should be imposed on
$q_T^2$, or equivalently on $P_{h\perp}^2$, but not on a transverse
momentum w.r.t.\ the lepton beam axis.

Integrated observables which are weighted with a suitable power of
$q_T/M$ have the desirable property that the transverse-momentum
convolutions \eqref{tree-con} in the low-$q_T$ results factorize into
separate integrals over either distribution or fragmentation functions
\cite{Kotzinian:1996cz,Boer:1997nt}.  With $\h = - \bm{q}{}_T /q_T$
one readily finds from \eqref{F_UUcosphi} that $\int \de q_T^2\,
(q_T^{} /M)\ms F_{UU}^{\cos\phi_h}$ formally factorizes into terms
involving the $p_T^2$-moments $f^{\perp (1)}(x)$ and
$h_{\smash{1}}^{\perp (1)}(x)$ and corresponding $k_T^2$-moments of
fragmentation functions.  However, this deconvolution only takes place
if one integrates over all $q_T$ up to infinity.  This is clearly
inadequate because $(q_T^{} /M)\ms F_{UU}^{\cos\phi_h}$ becomes
constant for $q_T \gg M$.  A reflection of this is the fact that the
$p_T^2$-moment $f^{\perp (1)}(x)$ involves a quadratic divergence at
large $p_T$, as we already noted in section~\ref{sec:power}.
Moreover, we see in table~\ref{tab:int} that the contribution from the
low-$q_T$ region to $\intobs{(q_T^{} /M)\ms F_{UU}^{\cos\phi_h}}$ is
power suppressed by $M^2/Q^2$ compared with the contribution from $q_T
\sim Q$, so that this observable is even less well suited to study
small $q_T$ than the unweighted structure function.  Conversely, the
weighted structure function is a \emph{good} observable for studying
large $q_T$.  The high-$q_T$ expression for $F_{UU}^{\cos\phi_h}$
depends on the same collinear functions $f_1(x)$ and $D_1(z)$ as
$F_{UU,T}$ but involves different hard-scattering kernels, so that
$F_{UU}^{\cos\phi_h}$ provides an additional observable if one aims,
for instance, at separating the fragmentation functions for different
quark and antiquark flavors and the gluon, or at testing the adequacy
of the theoretical description.  Up to corrections of order $M^2/Q^2$
one can evaluate $\intobs{(q_T^{} /M)\ms F_{UU}^{\cos\phi_h}}$ from
the high-$q_T$ result alone, which in addition may be integrated down
to $q_T=0$.  One then obtains a simple expression, just as in the
analogous case of $\intobs{F_{UU,L}^{}}$.
The unweighted integral $\intobs{F_{UU}^{\cos\phi_h}}$ is less
attractive for studying the high-$q_T$ result since the contribution
from the low-$q_T$ region is only suppressed by $M/Q$.  To evaluate
that contribution is difficult in practice as it contains
transverse-momentum-dependent distribution and fragmentation functions
that are poorly known.  If the weighted integral and the differential
structure function are affected with large experimental uncertainties,
one may instead have to consider the integral of $F_{UU}^{\cos\phi_h}$
with a lower cutoff on $q_T$.  This was for instance done in
\cite{Breitweg:2000qh} and \cite{Chay:1991nh,Chay:1997qy}.

As discussed in the previous section, the structure function
$F_{UU}^{\cos 2\phi_h}$ receives contributions from the low-$q_T$ and
high-$q_T$ calculations which do not match in the intermediate region
and have distinct dynamical origins, given that they respectively
involve chiral-odd and chiral-even distribution and fragmentation
functions.  As we see in table~\ref{tab:int}, both mechanisms
contribute to the integrated structure function at leading power, with
only moderate contributions from intermediate $q_T$.  For calculating
the integrated structure function it is appropriate to add the
contributions from the two mechanisms.  Furthermore, it is consistent
to perform the $q_T$-integral over the entire kinematical region for
both mechanisms, i.e.\ without introducing cutoff parameters, given
that contributions from regions where the approaches are not valid
(low $q_T$ for the high-$q_T$ calculation and vice versa) are
power-suppressed by $M^2/Q^2$.  This is similar to the case of the
interpolation formula \eqref{F_UUcos-interp} discussed in the previous
subsection, but for the integrated structure function the unphysical
behavior of the high-$q_T$ result in the limit $q_T\to 0$ does not
matter, at least at the level of power counting.
The weighted structure function $\intobs{(q_T^{}/M)^2 F_{UU}^{\cos
    2\phi_h}}$ has been proposed for obtaining a low-$q_T$ result in
terms of the moments $h_{\smash{1}}^{\perp (1)}(x)$ and
$H_{\smash{1}}^{\perp (1)}(z)$ of the Boer-Mulders and the Collins
functions, without any convolution of transverse-momentum-dependent
factors \cite{Boer:1997nt}.  According to table~\ref{tab:int} this
observable is, however, dominated by $q_T \sim Q$ and only sensitive
to $h_{\smash{1}}^{\perp (1)}(x)$ and $H_{\smash{1}}^{\perp (1)}(z)$
at the level of $M^2/Q^2$ corrections.  Such contributions are not
under control in the integrated observable, because uncalculated
corrections of the same size appear in the high-$q_T$ region as well.
Neglecting $M^2/Q^2$ corrections, one can evaluate
$\intobs{(q_T^{}/M)^2 F_{UU}^{\cos 2\phi_h}}$ as an integral of the
high-$q_T$ expression over the full $q_T$ domain.  In the same way as
$\intobs{(q_T^{}/M)\ms F_{UU}^{\cos\phi_h}}$, this provides an
independent observable sensitive to the twist-two functions $f_1(x)$
and $D_1(z)$.


\subsection{Polarization dependence}

Among the many observables for polarized SIDIS, the structure
functions $\smash{F_{UT,T}^{\sin(\phi_h-\phi_S)}}$ and
$\smash{F_{UT}^{\sin(\phi_h+\phi_S)}}$ have received particular
attention in the recent literature.  According to the low-$q_T$
results \eqref{e:sivers} and \eqref{e:collins}, they provide access to
the Sivers function $f_{1T}^\perp$ in the first and to the
transversity distribution $h_1^{}$ and the Collins function
$H_1^\perp$ in the second case \cite{Collins:1993kk}.  Both structure
functions have been found to be of significant size in HERMES
measurements on a proton target \cite{Airapetian:2004tw}.

As we see in table~\ref{tab:int}, the integrated structure function
$\intobs{F_{UT,T}^{\sin(\phi_h-\phi_S)}}$ is dominated by the
low-$q_T$ region and can hence be used for extracting information
about $f_{1T}^\perp(x,p_T^2)$.  The high-$q_T$ region is however only
suppressed by $M/Q$, so that it may be of advantage to impose an upper
cutoff on the $q_T$ integral in such analysis.
The weighted structure function $\intobs{(q_T^{}/M)\ms
  F_{UT,T}^{\sin(\phi_h-\phi_S)}}$ receives contributions from both
high and low $q_T$ at leading order in $M/Q$.  One can thus compute
the weighted integral by switching from one to the other formulation
at some $q_T$.  To achieve a factorization of the transverse-momentum
convolution in the low-$q_T$ expression, one should however integrate
it over all $q_T$ up to infinity.  Since $(q_T^{}/M)\ms
F_{UT,T}^{\sin(\phi_h-\phi_S)}$ behaves as $1/q_T^2$ for $q_T^{} \gg
M$, a suitable regularization is required.  This suggests a procedure
akin to the description of $\intobs{F_{UU,T}}$ in collinear
factorization, which we reviewed in the previous subsection.  As
dimensional regularization preserves rotation invariance in the
transverse plane, the integral over all $q_T$ of the weighted
low-$q_T$ result \eqref{e:sivers} turns into the product
\begin{equation}
  \label{sivers-mom}
\intobs{(q_T^{}/M)\, F_{UT,T}^{\sin(\phi_h-\phi_S)}}
 = - 2\ms \sum_a x\ms e_a^2\;
   f_{1T}^{a \perp \smash{(1)}}(x;Q)\, D_1^a(z;Q)
\end{equation}
of collinear functions defined in the $\overline{\text{MS}}$ scheme.
One can trade $f_{1T}^{\perp (1)}(x)$ for the twist-three function
$G_F(x,x)$, which appears in the high-$q_T$ calculation
\cite{Eguchi:2006mc,Ji:2006br,Koike:2007dg}.  In order to integrate
the high-$q_T$ result down to $q_T=0$, one must extend it to
$4-\varepsilon$ dimensions and perform the necessary
$\overline{\text{MS}}$ subtractions.  Adding graphs with virtual
corrections to the hard-scattering subprocess (which give the hard
factor $|H|^2$ in the Collins-Soper formalism) one will obtain a
complete NLO result in $\alpha_s$.  Such a procedure would be the
analog of a standard NLO computation for integrated observables within
collinear factorization at twist-two level.
Note that \eqref{sivers-mom} gives a consistent approximation of the
weighted structure function at LO in $\alpha_s$, in analogy to the
well-known tree-level expression $\intobs{F_{UU,T}} = \sum_a x\ms
e_a^2\; f_{1}^{a}(x)\, D^a_1(z)$. The factorization scale $\mu$ of the
functions in \eqref{sivers-mom} has been set to $Q$ in order to avoid
large logarithms of $Q/\mu$ appearing in the $\alpha_s$-corrections.
To leading order, the logarithmic $Q$ dependence of the weighted
structure function then follows from the evolution equations for
$D_1^{}(z)$ and $f_{1T}^{\perp (1)}(x)$.  The latter have been
investigated in \cite{Henneman:2005th}.

The situation for the structure function
$F_{UT}^{\sin(\phi_h+\phi_S)}$ is the same as for
$F_{UT,T}^{\sin(\phi_h-\phi_S)}$ since the power behavior of these
observables coincides in both the low- and high-$q_T$ calculations.
The evaluation of the weighted structure function in collinear
factorization gives
\begin{equation}
  \label{collins-mom}
\intobs{(q_T^{}/M_h)\, F_{UT}^{\sin(\phi_h+\phi_S)}}
 = 2\ms \sum_a x\ms e_a^2\;
   h_1^a(x;Q)\, H_{1}^{a \perp \smash{(1)}}(z;Q)
\end{equation}
for the Born term.  The $k_T^2$-moment $H_{\smash{1}}^{\perp (1)}(z)$
is related to the twist-three fragmentation function $\widehat{E}_F$
appearing at order $\alpha_s$.  We note that according to the
high-$q_T$ results in \cite{Eguchi:2006qz,Eguchi:2006mc} the
$\alpha_s$-corrections to both \eqref{sivers-mom} and
\eqref{collins-mom} involve each of the twist-three functions $G_F$,
$\widetilde{G}_F$, and $\widehat{E}_F$.

According to table~\ref{tab:int}, the integrated structure functions
$\intobs{F_{UT}^{\sin\phi_S}}$ and
$\intobs{F_{UT}^{\sin(2\phi_h-\phi_S)}}$ receive comparable
contributions from all regions of $q_T$.  Weighting the structure
functions with $(q_T/M)^2$ one obtains integrals that can be evaluated
in the high-$q_T$ formalism up to corrections of order $M^2/Q^2$,
similarly to the case of $\intobs{(q_T^{}/M)\ms F_{UU}^{\cos\phi_h}}$
we discussed in the previous subsection.  The high-$q_T$ expressions
computed in \cite{Eguchi:2006qz,Eguchi:2006mc} imply that
$\intobs{(q_T^{}/M)^2\ms F_{UT}^{\sin(2\phi_h-\phi_S)}}$ is sensitive
to $G_F$ and $\widetilde{G}_F$, whereas $\intobs{(q_T^{}/M)^2\ms
  F_{UT}^{\sin\phi_S}}$ also depends on $\widehat{E}_F$.  Whether
these observables are large enough to be measured in practice is, of
course, a different question.

{}From \eqref{e:tailFUTsin3phi} we can infer that the integral of
$F_{UT}^{\sin(3\phi_h-\phi_S)}$ receives a contribution of order $1$
from low $q_T$, whereas the high-$q_T$ result
\eqref{e:high_FUTsin3phi} is suppressed by $M/Q$.  According to
\eqref{e:F_UTsin3phi} the integrated structure function
$\intobs{F_{UT}^{\sin(3\phi_h-\phi_S)}}$ may hence be used to
extract information on $h_{1T}^\perp(x,p_T)$ and on the Collins
fragmentation function, with the same caveat we discussed for
$\intobs{F_{UT,T}^{\sin(\phi_h-\phi_S)}}$.
To obtain an integral that is dominated by the high-$q_T$ result up to
$M^2/Q^2$ corrections, one must weight
$\smash{F_{UT}^{\sin(3\phi_h-\phi_S)}}$ with $(q_T/M)^3$.

Let us now turn to observables that involve longitudinal polarization.
Similarly to $F_{UU}^{\cos\phi_h}$, the lepton-helicity-dependent
structure function $F_{LU}^{\sin\phi_h}$ for an unpolarized target
receives a contribution of order $M/Q$ from low $q_T$ and of order
unity from high $q_T$.  It is therefore in principle suitable for
investigating the high-$q_T$ result of Hagiwara et
al.~\cite{Hagiwara:1983cq}.  However, the contribution from large
$q_T$ comes with a factor $\alpha_s^2$ in this case, which may not be
sufficient for neglecting power-suppressed contributions from low
$q_T$ in practice.  From this point of view, it would be advantageous
to weight the structure function with $q_T/M$, or to integrate over
$q_T$ starting from a lower cutoff.

Finally, the structure functions $F_{LL}^{}$ and $F_{LL}^{\cos\phi_h}$
have the same power behavior as their unpolarized counterparts
$\smash{F_{UU,T}^{}}$ and $\smash{F_{UU}^{\cos\phi_h}}$, and their
discussion is analogous to the one in the previous subsection.  In
particular, the weighted integral $\intobs{(q_T^{}/M)\ms
  F_{LL}^{\cos\phi_h}}$ depends on the polarized parton densities
$g_1$ and, if measurable with sufficient accuracy, could be used in
addition to the well-known observable $\intobs{F_{LL}^{}}$ for
disentangling the contributions from different quark and antiquark
flavors and from the gluon.


\section{From low to intermediate $q_T\ms$: explicit calculation} 
\label{sec:tails}

In this section we compute the high-transverse-momentum tails of the
quark distributions in \eqref{tails-2} and of the analogous
fragmentation functions.  These are the functions which appear at
lowest order in the $1/p_T$ expansion of section~\ref{sec:power} and
are hence expressed in terms of collinear functions of twist two.
While in section~\ref{sec:comparison} we identified observables whose
\emph{power behavior} agrees in the low- and high-$q_T$ calculations,
we will then be able to check for selected structure functions whether
agreement is also found for their explicit expressions.

\subsection{High-$p_T$ tails of distribution functions}
\label{sec:taildistribution}

Let us begin with the quark distribution functions.  We work in the
original scheme of Collins and Soper \cite{Collins:1981uk}, using a
spacelike axial gauge with the singularities of the gluon propagator
regulated by the principal value prescription.  The only Feynman
diagrams to be evaluated are then those depicted in
Fig.~\ref{f:rungdistAG}a and b.  
\new{For further discussion and a comparison with the calculation in
  Feynman gauge, we refer to appendix~\ref{app:FG}.}

\FIGURE{
\includegraphics[width=11cm]{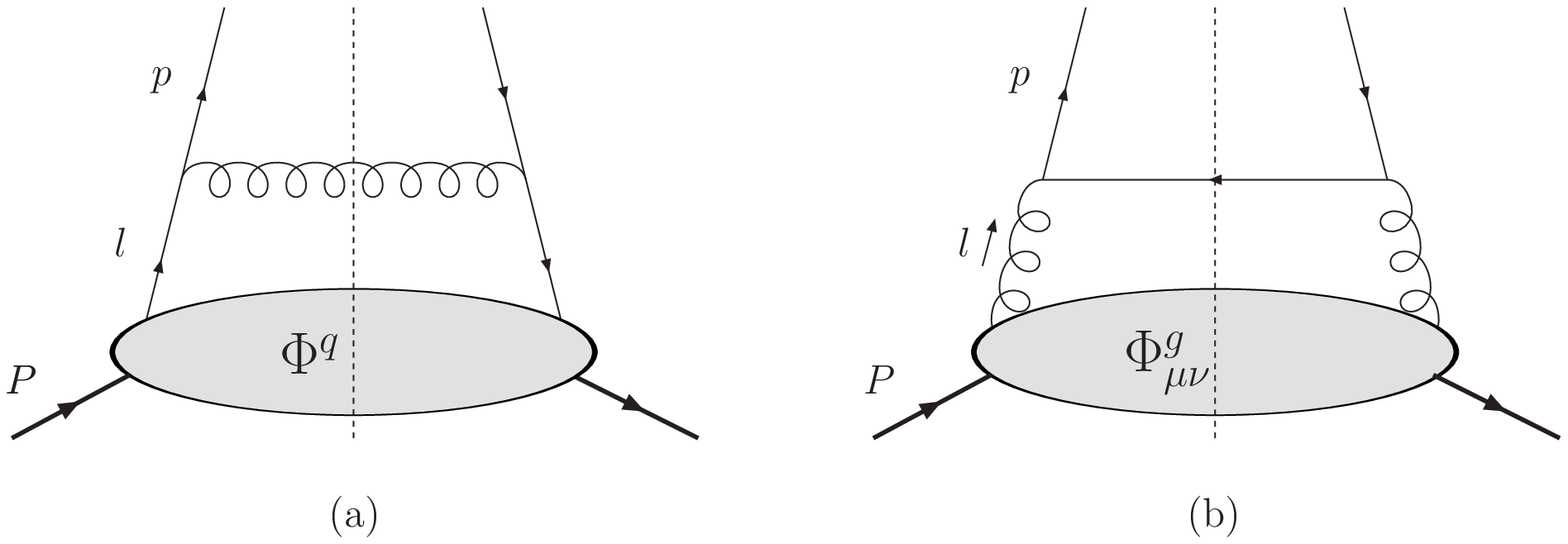}
\caption{\label{f:rungdistAG} Diagrams for the calculation of the
  leading high-$p_T$ behavior of the quark-quark correlator
  $\Phi(x,p_T)$ in axial gauge $A\cdot v =0$.}
}

The contribution of the quark-to-quark term shown in
Fig.~\ref{f:rungdistAG}a reads
\begin{equation} \begin{split} 
\Phi^{q}(x,p_T) \,\Bigr|_{\text{(\protect\ref{f:rungdistAG}a)}}
& =
\frac{4\pi\alpha_s}{(2 \pi)^3}\,C_F 
\int \de p^- \int \de l^+\,
\delta\bigl((l-p)^2\bigr)\,\theta(l^+-p^+)
\\ & \quad \times
d^{\mu \nu}(l-p;v)\;
  \frac{\pslash\,}{p^2}\, \gamma_{\nu}\ms
  \Phi_2^q \Bigl(\frac{x}{\xb}\Bigr)
  \gamma_{\mu}\ms \frac{\pslash\,}{p^2}
  \;\Biggr|_{\,l^- =0, \,\bm{l}_T =\bm{0}_T} \,,
\label{e:startdist0}
\end{split} \end{equation} where it is understood that $p^+ = x P^+$
and $l^+ = p^+ /\hat{x}$.  As explained in section~\ref{sec:power},
the restriction to leading order in $1/p_T$ allows us to set $l^-$ and
$\bm{l}_T$ to zero when calculating the hard-scattering subprocess,
and to retain only the twist-two part $\Phi^q_2(x/\hat{x})$ of the
collinear quark-quark correlator at the bottom of the graph.  The
gluon polarization sum in $A\cdot v =0$ gauge is given by
\begin{equation}
d^{\mu \nu}(q;v)= - g^{\mu \nu}
  + \frac{q^{\mu} v^{\nu} + q^{\nu} v^{\mu}}{q \cdott v}
  - \frac{q^{\mu} q^{\nu}}{(q \cdott v)^2}\, v^2 \,,
\label{e:gluonpolAG}
\end{equation} 
where the singularities at $q\cdott v =0$ are to be regulated by the
principal value prescription.
Using the $\delta$-function to perform the $p^-$ integration,
\begin{equation}
  \label{delta-p-minus}
\delta\bigl((l-p)^2\bigr) \,\Bigr|_{\,l^- =0, \,\bm{l}_T =\bm{0}_T}
= \frac{\xb}{2p^+ (1-\xb)} \,\delta\biggl( p^-
    + \frac{\bm{p}_T^2}{2p^+}\, \frac{\xb}{1-\xb} \biggr)
\end{equation} 
we obtain
\begin{equation}
\Phi^q(x,p_T) \,\Bigr|_{\text{(\protect\ref{f:rungdistAG}a)}}
= \frac{\alpha_s}{(2 \pi)^2}\, C_F\, \frac{1}{\bm{p}_T^4}\,
  \int_x^1 \frac{\de\xb}{\xb}\,  (1-\xb)\;
  d^{\mu \nu}(\bar{l}-\bar{p};v)\,
  \pbarslash\, \gamma_{\nu}\ms
  \Phi_2^q \Bigl(\frac{x}{\xb}\Bigr)
  \gamma_{\mu} \pbarslash \,,
\label{e:startdist}
\end{equation}
where we have introduced the notation
\begin{align} 
  \label{p-approx}
\bar{p} &=  p^+ n_{+}
   - \frac{\bm{p}_T^2}{2p^+}\, \frac{\xb}{1-\xb}\, n_{-} + p_T \,,
\\
  \label{l-approx}
\bar{l} &= \frac{\ms p^+}{\xb}\, n_{+}
\end{align} 
for the approximated momenta in the hard-scattering kernel.  We note
that the virtuality of the upper quark legs
\begin{equation}
  \label{p-bar-squared}
\bar{p}^{\ms 2} = - \frac{\bm{p}_T^2}{1-\xb}
\end{equation}
is always spacelike.
The gauge fixing vector can be written as
\begin{equation}
  \label{gauge-vector}
v = v^-\ms n_- - \frac{2 (P^+)^2\, v^-}{\zeta}\, n_+
\end{equation}
with
\begin{equation}
  \label{zeta-def}
  \zeta = - \frac{(2 P\cdott v)^2}{v^2} 
        = - \frac{2 (P^+)^2\, v^-}{v^+} \,,
\end{equation}
where in the second step we have neglected $M^2$ compared with
$\zeta$.  We therefore have
\begin{equation} 
  \label{axial-denominator}
(\bar{l}-\bar{p})\cdott v = (\bar{l}-\bar{p})^+ v^-
   - \frac{2 (p^+)^2}{x^2 \zeta}\, (\bar{l}-\bar{p})^- v^-
= \frac{1}{\xb}\ms
  \biggl[ 1-\xb - \eta\, \frac{\xb^2}{1-\xb} \biggr]\, p^+ v^- \,,
\end{equation} 
where we have introduced the parameter
\begin{equation}
  \label{eta-def}
\eta = \frac{\bm{p}_T^2}{x^2\ms \zeta}
     = - \frac{\bm{p}_T^2}{2 (p^+)^2}\, \frac{v^+}{v^-} \,.
\end{equation}
We now decompose the gluon polarization sum as
\begin{equation}
  \label{gluon-prop-dec}
d^{\mu \nu}_{\phantom{()}}(\bar{l}-\bar{p};v) =
  \sum_{i=1}^4 d^{\mu \nu}_{(i)}(\bar{l}-\bar{p};v) \,,
\end{equation}
with
\begin{align}
  \label{d-1}
d^{\mu \nu}_{(1)}(\bar{l}-\bar{p};v) &= 
   \frac{1-\xb}{1-\xb - \eta\ms \frac{\;\xb^2}{1-\xb}}\;
   d^{\mu \nu}(\bar{l}-\bar{p};n_-) \,,
\\[0.2em]
d^{\mu \nu}_{(2)}(\bar{l}-\bar{p};v) &= \eta\,
   \frac{\xb^2}{1-\xb}\, \frac{g^{\mu \nu}}{%
     1-\xb - \eta\ms \frac{\;\xb^2}{1-\xb}} \,,
\\[0.2em]
d^{\mu \nu}_{(3)}(\bar{l}-\bar{p};v) &= -\eta\,
   \frac{2\xb p^+}{\bm{p}_T^2}\, 
   \frac{(\bar{l}-\bar{p})^{\mu}\ms n_+^{\nu}+
         (\bar{l}-\bar{p})^{\nu}\ms n_+^{\smash{\mu}}}{%
     1-\xb - \eta\ms \frac{\;\xb^2}{1-\xb}} \,,
\\[0.2em]
d^{\mu \nu}_{(4)}(\bar{l}-\bar{p};v) &= \eta\,
   \frac{4\xb^2}{\bm{p}_T^2}\,
   \frac{(\bar{l}-\bar{p})^{\mu}\ms (\bar{l}-\bar{p})^{\nu}}{%
     \bigl[\ms 1-\xb - \eta\ms \frac{\;\xb^2}{1-\xb} \,\bigr]^2} \,.
\end{align} 
Notice that the first term \eqref{d-1} is proportional to the
polarization sum $d^{\mu \nu}(\bar{l}-\bar{p};n_-)$ one would use when
calculating in light-cone gauge $A\cdot n_- =0$.  We will see shortly
that the prefactor in \eqref{d-1} regulates the divergence at $\xb =1$
which would arise in that gauge.
{}From the parameterization \eqref{eq:phix} we readily see that the
twist-two part of the quark-quark correlator satisfies $\nslash_+\ms
\Phi_2^q = \Phi_2^q \nslash_+ =0$, so that terms with
$n_+^{\smash{\mu}}$ or $n_+^{\nu}$ in $d^{\mu\nu}$ vanish when
inserted into \eqref{e:startdist}.  With $\bar{l}$ being proportional
to $n_+$, we hence need only the first two terms and the
$\bar{p}^{\mu} \bar{p}^{\nu}$ part of the last term in the
decomposition \eqref{gluon-prop-dec}.  This gives
\begin{equation} \begin{split}  
\Phi^q(x,p_T) \,\Bigr|_{\text{(\protect\ref{f:rungdistAG}a)}}
&= \frac{\alpha_s}{(2 \pi)^2}\, C_F\, \frac{1}{\bm{p}_T^2}\,
  \int_x^1 \frac{\de\xb}{\xb}\,
\\[0.2em] & \quad
\times\Biggl[
\frac{1-\xb}{(1-\xb)^2 - \eta\ms\xb^2}\;
  \frac{1}{\bm{p}_T^2}\, (1-\xb)^2\,
  d^{\mu \nu}(\bar{l}-\bar{p};n_-)\,
  \pbarslash\,\gamma_{\nu}\,
  \Phi_2^q  \Bigl(\frac{x}{\xb}\Bigr)\,
  \gamma_{\mu} \pbarslash
\\ & \qquad
+ \frac{\eta}{(1-\xb)^2 - \eta\ms\xb^2}\;
  \frac{\xb^2}{\bm{p}_T^2}\, (1-\xb)
  \pbarslash\,\gamma^{\mu}\,
  \Phi_2^q  \Bigl(\frac{x}{\xb}\Bigr)\,
  \gamma_{\mu} \pbarslash
\\ & \qquad
+ \frac{\eta\ms (1-\xb)}{[\ms (1-\xb)^2 - \eta\ms\xb^2 \ms]^2}\;
  4\xb^2\ms \Phi_2^q  \Bigl(\frac{x}{\xb}\Bigr)\,
\Biggr] \,.
\label{e:3terms}
\end{split} \end{equation}
To proceed we must determine the behavior of the different terms in
the limit $\xb\to 1$, where $\bar{p}^{\ms-} \sim (1-\xb)^{-1}$ becomes
singular.  Using the form \eqref{e:gluonpolAG} with $v$ replaced by
$n_-$, we obtain
\begin{align}
  \label{x-to-1-first}
(1-\xb)^2 \, &
d^{\mu \nu}(\bar{l}-\bar{p};n_-)\,
  \pbarslash\,\gamma_{\nu}\,
  \Phi_2^q\, \gamma_{\mu}\,\pbarslash
\nonumber \\
& = {}
 - (1-\xb)^2\, \pbarslash\,\gamma^{\mu}\,
    \Phi_2^q\, \gamma_{\mu}\,\pbarslash
 + \frac{\xb\ms \bm{p}_T^2}{p^+}\,
   \bigl( \pbarslash\ms \nslash_- \Phi_2^q
        + \Phi_2^q\ms \nslash_- \pbarslash \bigr) \,.
\end{align}
Since the minus-component of $\bar{p}$ drops out in $\pbarslash\ms
\nslash_-$ and $\nslash_- \pbarslash$, the expression in
\eqref{x-to-1-first} is finite for $\xb\to 1$.  For the second term in
\eqref{e:3terms} we have
\begin{equation}
(1-\xb) \pbarslash\,\gamma^{\mu}\,
  \Phi_2^q\, \gamma_{\mu}\,\pbarslash
= - (1-\xb) \pbarslash\,
  \bigl( f_1^q \nslash_+ - g_1^q\ms S_L^{} \gamma_5 \nslash_+ \bigr)
  \pbarslash
\end{equation}
after plugging in the parameterization of $\Phi_2^q$ from
\eqref{eq:phix}.  This contains a piece with two factors of
$\bar{p}^{\ms-}$, which is proportional to the Dirac matrices
$\nslash_-$ or $\gamma_5 \nslash_-$.  According to the decomposition
\eqref{phi-dec} it therefore does not contribute to the twist-two or
twist-three parts of the correlator $\Phi(x,p_T)$, on which we
concentrate here.  In the twist-four part of \eqref{e:startdist0} this
piece leads to a singularity at $\xb=1$, or in other words at
$\bar{p}^{\ms -} \to -\infty$, showing that at twist-four level the
$A\cdot v =0$ gauge is insufficient to render the integral over $p^-$
in $\Phi(x,p_T) = \int \de p^-\, \Phi(p)$ well defined.

In the following we take the limit $\bm{p}_T^2 \ll \zeta$,
corresponding to $\eta \ll 1$.  The motivation for this is that in
physical processes we need the correlator $\Phi(x,p_T;\zeta)$ for
$\bm{p}_T^2 \ll Q^2$ and $x^2 \zeta \sim Q^2$, as discussed in
section~\ref{sec:CS-fact}.  We note that in a frame where $xP^+ \sim
Q$ this corresponds to $v^+ \sim v^-$.  According to \eqref{eta-def}
the parameter $\sqrt{\eta\rule{0pt}{1.6ex}}$ is then proportional to
the small angle between the quark momentum $p$ and the hadron momentum
$P$, with a factor of proportionality of order $1$.  Notice that at
this point we introduce a hierarchy in size between $p_T$ and $P^+$,
which were not distinguished in the power counting of
section~\ref{sec:power}.  This is similar to what we have done with
the high-$q_T$ calculation of structure functions in
section~\ref{sec:large}\,: we started with the result
\eqref{e:FUUThigh}, which is derived without making a distinction
between the size of $q_T$ and $Q$, and in a second step we took its
limit for $q_T \ll Q$.

For $\eta\ll 1$ the first term in the square brackets of
\eqref{e:3terms} can be rewritten by using that for any function
$G(\xb)$ which is regular at $\xb=1$
\begin{equation}
    \label{first-distrib}
\lim_{\eta\to 0}\,
\text{PV}\!\int_x^1 \de\xb\; \frac{1-\xb}{(1-\xb)^2 - \eta\ms \xb^2}
  \; G(\xb)
= \int_x^1 \de\xb\; \frac{G(\xb)}{(1-\xb)_+}
  + \frac{1}{2}\ms G(1)\, \ln\frac{1}{|\ms \eta \ms|} \,,
\end{equation}  
where the plus-distribution is defined as in \eqref{plus-def}.  {}From
\begin{equation}
\lim_{\eta\to 0}\,
\text{PV}\!\int_x^1 \de\xb\; \frac{\eta}{(1-\xb)^2 - \eta\ms \xb^2}
  \; G(\xb)
= 0
\end{equation}
we see that the second term in \eqref{e:3terms} does not contribute in
the small-$\eta$ limit when restricted to the twist-two and
twist-three parts of $\Phi(x,p_T)$.  In contrast, the third term in
\eqref{e:3terms} does contribute, since
\begin{equation} 
  \label{third-distrib}
\lim_{\eta\to 0}\,
\text{PV}\!\int_x^1 \de\xb\;
\frac{\eta\ms (1-\xb)}{[\ms (1-\xb)^2 - \eta\ms\xb^2 \ms]^2}
  \; G(\xb)
= - \frac{1}{2}\ms G(1) \,.
\end{equation}
Therefore, our final result reads
\begin{equation} \begin{split}  
\Phi^q(x,p_T) \,\Bigr|_{\text{(\protect\ref{f:rungdistAG}a)}}
&= 
\frac{\alpha_s}{2 \pi^2}\,C_F\, \frac{1}{\bm{p}_T^2}
\int_x^1 \frac{\de\xb}{\xb}\; \biggl\{
\biggl[ \frac{1}{(1-\xb)_+}
      + \frac{1}{2}\ms \delta(1-\xb) \ln\frac{1}{\eta} \,\biggr]
\\ & \quad\times
\frac{(1-\xb)^2}{2 \bm{p}_T^2}\,
  d^{\mu \nu}(\bar{l}-\bar{p};n_-)\,
  \pbarslash\,\gamma_{\nu}\ms
  \Phi_2^q \Bigl(\frac{x}{\xb}\Bigr)
  \gamma_{\mu} \pbarslash
- \delta(1-\xb)\, \Phi_2^q  \Bigl(\frac{x}{\xb}\Bigr)
\biggr\}
\label{e:final}
\end{split} \end{equation}
to leading order in $1/p_T$,
where it is understood that we have restricted ourselves to the
twist-two and twist-three parts of the correlator on the l.h.s.  We
note at this point that if we work with a timelike axial gauge,
i.e.\ with negative $\zeta$ and $\eta$ in \eqref{gauge-vector} to
\eqref{eta-def}, we obtain the same result as in \eqref{e:final} with
$\ln(-\eta^{-1})$ instead of $\ln(\eta^{-1})$.  The polarization sum
$d^{\mu\nu}(\bar{l}-\bar{p};v)$ is then nonsingular in the whole
region $x\le \xb \le 1$, and the principal value prescription in
\eqref{first-distrib} to \eqref{third-distrib} is not required.  A
timelike vector $v$ was indeed used for the construction of
factorization by Ji et al.~\cite{Ji:2004wu}, whereas arguments in
favor of taking $v$ spacelike were given by Collins and Metz in
\cite{Collins:2004nx}.

The gluon-to-quark contribution to the correlation function comes from
the diagram in Fig.~\ref{f:rungdistAG}b.  Its calculation is simpler
than the previous one, due to the absence of a gluon polarization sum
in axial gauge.  Correspondingly, the result is independent of $\eta$.
The counterpart of the expression in \eqref{e:startdist} now reads
\begin{equation}
\Phi^q(x,p_T) \,\Bigr|_{\text{(\protect\ref{f:rungdistAG}b)}}
= \frac{\alpha_s}{(2 \pi)^2}\; T_R\, \frac{1}{\bm{p}_T^4}\,
\int_x^1 \frac{\de\xb}{\xb}\,  (1-\xb)\;
\Phi_2^{g,\mu \nu}  \Bigl(\frac{x}{\xb}\Bigr)\,
\pbarslash\,\gamma_{\nu}\ms \bigl(\lbarslash-\pbarslash\bigr)\ms
\gamma_{\mu} \pbarslash \,,
  \label{gluon-highpt}
\end{equation} 
where the twist-two part of the collinear gluon correlation function
is given by
\begin{equation}
 \Phi_2^{g, \mu \nu}(x)
= \frac{1}{2x P^+}\, \Big\{ -g_T^{\mu \nu} f_1^g(x)
   + \ii \epsilon_T^{\mu \nu}\ms S_L^{\phantom{\mu}}\ms 
     g_1^g(x) \ms\Bigr\} \,, 
\end{equation}
see e.g.~\cite{Meissner:2007rx}.  Inserting \eqref{p-approx} and
\eqref{l-approx} and using some Dirac algebra, one finds that the
integrand of \eqref{gluon-highpt} is finite at $\xb=1$.

{}From \eqref{e:final} and \eqref{gluon-highpt} we can easily project
out the contributions to the individual terms in the decomposition
\eqref{phi-dec} of $\Phi^q(x,p_T)$.  For the high-$p_T$ behavior of
the unpolarized distributions we obtain
\begin{align}
f_1^{q}(x,p_T^2)
&= \frac{\alpha_s}{2 \pi^2}\,
   \frac{1}{\bm{p}_T^2}\,
   \biggl[\ms \frac{L(\eta^{-1})}{2}\, f_1^q(x)
     - C_F\ms f_1^q(x) 
     + \bigl(P_{qq} \otimes f_1^q + P_{qg} \otimes f_1^g\bigr)(x)
   \biggr] \,,
\label{e:f1highkt}
\\
x f^{\perp q}(x,p_T^2)
&= \frac{\alpha_s}{2 \pi^2}\,
   \frac{1}{2 \bm{p}_T^2}\,
   \biggl[\ms \frac{L(\eta^{-1})}{2}\, f_1^q(x)
     + \bigl(P'_{qq} \otimes f_1^q + P'_{qg} \otimes f_1^g \bigr)(x)
   \biggr] \,,
\label{e:fperphighkt}
\end{align}
whereas for the polarized distributions we find
\begin{align}
g_{1L}^{q}(x,p_T^2)
&= \frac{\alpha_s}{2 \pi^2}\,
   \frac{1}{\bm{p}_T^2}\,
   \biggl[\ms \frac{L(\eta^{-1})}{2}\, g_1^q(x)
     - C_F\ms g_1^q(x)
     + \bigl(\Delta P_{qq} \otimes g_1^q
           + \Delta P_{qg} \otimes g_1^g\bigr)(x)
   \biggr] \,,
\\
x g_L^{\perp q}(x,p_T^2)
&= \frac{\alpha_s}{2 \pi^2}\,
   \frac{1}{2 \bm{p}_T^2}\,
   \biggl[\ms \frac{L(\eta^{-1})}{2}\, g_1^q(x)
     + \bigl(\Delta P'_{qq} \otimes g_1^q
           + \Delta P'_{qg} \otimes g_1^g\bigr)(x)
   \biggr] \,,
\end{align}
and
\begin{align}
h_1^{q}(x,p_T^2)
&= \frac{\alpha_s}{2 \pi^2}\,
   \frac{1}{\bm{p}_T^2}\,
   \biggl[\ms \frac{L(\eta^{-1})}{2}\, h_1^q(x)
     - C_F\ms h_1^q(x)
     + \bigl(\delta P_{qq} \otimes h_1^q\bigr)(x)
   \biggr] \,,
\label{e:h1highkt}
\\
x h_T^{\perp q}(x,p_T^2)
&= \frac{\alpha_s}{2 \pi^2}\,
   \frac{1}{2 \bm{p}_T^2}\,
   \biggl[\ms \frac{L(\eta^{-1})}{2}\, h_1^q(x)
     + \bigl(\delta P_{qq} \otimes h_1^q\bigr)(x)
   \biggr] \,,
\\
x h_T^{q}(x,p_T^2) &= - x h_T^{\perp q}(x,p_T^2) \,,
\phantom{\biggl[ \biggr]}
\label{e:hThighkt}
\end{align} 
where $L(\eta^{-1})$ is defined as in \eqref{e:sudakovleading}.  Here
\begin{equation} 
\delta P_{qq}(\xb) = C_F\,\biggl[\frac{2\xb}{(1-\xb)_{+}}
 + \frac{3}{2}\,\delta(1-\xb) \biggr]
\end{equation} 
is the leading-order DGLAP splitting function for the transversity
distribution~\cite{Artru:1990zv}, and the remaining splitting
functions are given in \eqref{splitting-fcts} to
\eqref{last-splitting-fct}.  The chiral-odd quark distributions in
\eqref{e:h1highkt} to \eqref{e:hThighkt} receive no contribution from
\eqref{gluon-highpt} because chirality is conserved for the quark line
in the graph of Fig.~\ref{f:rungdistAG}b.

The diagrams for the high-$p_T$ behavior of the antiquark correlation
function $\Phi^{\bar{q}}(x,p_T)$ are obtained from those in
Fig.~\ref{f:rungdistAG} by reversing the direction of the fermion
lines.  The results have the form of \eqref{e:f1highkt} to
\eqref{e:hThighkt}, with identical splitting functions and with all
quark distributions replaced by antiquark distributions.


\subsection{High-$k_T$ tails of fragmentation functions}
\label{sec:tailfragmentation}

The calculation of the high-transverse-momentum tails of quark
fragmentation functions proceeds in close analogy to the case of
distribution functions.  We nevertheless present the essential steps
in this subsection, so as to show that no problems occur when going
from a spacelike to a timelike situation.

\FIGURE[ht]{
\includegraphics[width=10cm]{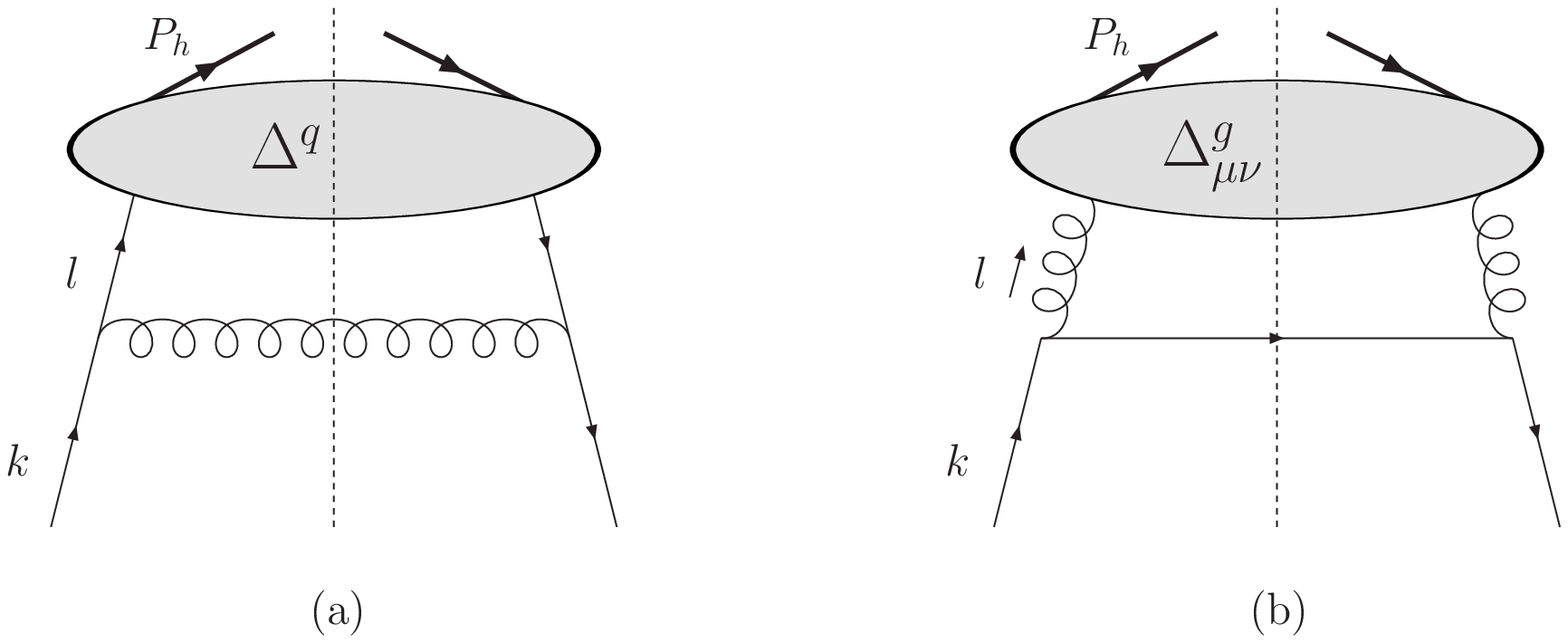}
\caption{\label{f:rungfrag} Diagrams for the calculation of the
  leading high-$k_T$ behavior of the quark-quark fragmentation
  correlator $\Delta(z,k_T)$ in axial gauge $A\cdot v =0$.}
}

The Feynman diagrams to be evaluated in $A\cdot v =0$ gauge are drawn
in Fig.~\ref{f:rungfrag}.  We first consider the quark-to-quark
contribution of Fig.~\ref{f:rungfrag}a.  The analog of the starting
expression \eqref{e:startdist0} now reads
\begin{equation} \begin{split}
    \label{e:startfrag0}
\Delta^q(z,k_T) \,\Bigr|_{\text{(\ref{f:rungfrag}a)}} 
&=
\frac{4\pi\alpha_s}{(2 \pi)^3}\,C_F \,\frac{1}{z}
\int \de k^+ \int \de l^-\,
\delta\bigl((k-l)^2\bigr)\,\theta(k^- -l^-)
\\ & \quad \times
d^{\mu \nu}(k-l;v)\,
  \frac{\kslash\,}{k^2}\, \gamma_{\mu}\,
  \frac{\zd}{z}\ms \Delta_2^q \Bigl(\frac{z}{\zd}\Bigr)
  \gamma_{\nu}\ms \frac{\kslash\,}{k^2}
  \;\Biggr|_{\,l^+ =0, \,\bm{l}_T =\bm{0}_T} \,,
\end{split} \end{equation} 
where $k^- = P_h^- /z$ and $l^- = \zd k^-$.  The factors $1/z$ and
$\zd /z$ in \eqref{e:startfrag0} arise from the definitions
\eqref{e:delta} and \eqref{eq:deltaz} of the fragmentation
correlators.  We perform the $k^+$-integration using
\begin{equation}
\delta\bigl((k-l)^2\bigr) \,\Bigr|_{\,l^+ =0, \,\bm{l}_T =\bm{0}_T}
= \frac{1}{2k^- (1-\zd)} \,\delta\biggl( k^+
   - \frac{\bm{k}_T^2}{2k^- (1-\zd)} \biggr)
\end{equation}
and obtain
\begin{equation} \begin{split} 
\Delta^q(z,k_T) \,\Bigr|_{\text{(\ref{f:rungfrag}a)}} 
&=
\frac{\alpha_s}{(2 \pi)^2}\, C_F\, \frac{1}{z^2\ms \bm{k}_T^4}\,
\int_z^1 \frac{\de\zd}{\zd}\,  (1-\zd)\;
  d^{\mu \nu}(\bar{k}-\bar{l};v)\,
\kbarslash\, \gamma_{\mu}\ms
\Delta_2^q \Bigl(\frac{z}{\zd}\Bigr)
\gamma_{\nu}  \kbarslash
\end{split} \end{equation}
with
\begin{align} 
\bar{k} &= \frac{\bm{k}_T^2}{2k^- (1-\zd)}\, n_{+} 
+ k^- n_{-} + k_T \,,
\\
\bar{l} &= \zd\ms k^- n_{-} \,.
\end{align}
The virtuality of the fragmenting quark
\begin{equation}
\bar{k}^2 = \frac{\zd\ms \bm{k}_T^2}{1-\zd}
\end{equation}
is always timelike, in contrast to its counterpart $\bar{p}^{\ms 2}$
in the distribution correlator.
For the calculation of the fragmentation correlator, it is useful to
write the gauge vector as
\begin{equation}
v = v^+ n_+ - \frac{2 (P_{\smash{h}}^-)^2\, v^+}{\zeta_h}\, n_-
\end{equation} 
with
\begin{equation}
  \label{zeta-h-def}
\zeta_h = - \frac{(2 P_{\smash{h}}^{\phantom{-}} \cdott v)^2}{v^2} 
        = - \frac{2 (P_{\smash{h}}^-)^2\, v^+}{v^-} \,.
\end{equation}
In analogy to \eqref{axial-denominator} and \eqref{eta-def} we can
then write
\begin{equation} 
(\bar{k}-\bar{l}) \cdot v =(\bar{k}-\bar{l})^- v^+
    - \frac{2(k^-)^2}{z^{-2}\ms \zeta_h}\, (\bar{k}-\bar{l})^+ v^+
= \biggl[1-\zd - \eta_h\, \frac{1}{1-\zd} \biggr]\, k^- v^+ \,,
\end{equation} 
where we have introduced
\begin{equation}
  \label{eta-h-def}
\eta_h = \frac{\bm{k}_T^2}{z^{-2}\ms \zeta_h}
       = - \frac{\bm{k}_T^2}{2 (k^-)^2}\, \frac{v^-}{v^+} \,.
\end{equation}
Taking the limit $\bm{k}_T^2 \ll \zeta_h$ and following similar steps
as in the previous subsection, we obtain
\begin{equation} \begin{split}
    \label{frag-a}
\Delta^q(z,k_T) \,\Bigr|_{\text{(\ref{f:rungfrag}a)}} 
&=
\frac{\alpha_s}{2 \pi^2}\,C_F\, \frac{1}{z^2\ms \bm{k}_T^2}\,
\int_z^1 \frac{\de\zd}{\zd}\; \biggl\{
\biggl[ \frac{1}{(1-\zd)_+}
      + \frac{1}{2}\ms \delta(1-\zd) \ln\frac{1}{\eta_h} \,\biggr]
\\ & \quad\times
\frac{(1-\zd)^2}{2 \bm{k}_T^2}\,
  d^{\mu \nu}(\bar{k}-\bar{l};n_+)\,
  \kbarslash\,\gamma_{\mu}\ms
  \Delta_2^q \Bigl(\frac{z}{\zd}\Bigr)
  \gamma_{\nu} \kbarslash\,
- \delta(1-\zd)\, \Delta_2^q \Bigl(\frac{z}{\zd}\Bigr)\,
\biggr\}
\end{split} \end{equation}  
to leading order in $1/k_T$,
where as in the case of distribution functions the result is
restricted to the twist-two and twist-three parts of the correlation
function on the l.h.s.
For the quark-to-gluon contribution, the diagram of
Fig.~\ref{f:rungfrag}b gives
\begin{equation}
  \label{frag-b}
\Delta^q(z,k_T) \,\Bigr|_{\text{(\ref{f:rungfrag}b)}} 
= \frac{\alpha_s}{(2 \pi)^2}\, C_F\, \frac{1}{z^2\ms \bm{k}_T^4}
  \int_z^1 \frac{\de\zd}{\zd}\, (1-\zd)\;
  \Delta_2^{g, \mu \nu} \Bigl(\frac{z}{\zd}\Bigr)\,
  \kbarslash\,\gamma_{\mu}\ms \bigl(\kbarslash-\lbarslash\bigr)\ms
  \gamma_{\nu} \kbarslash \,,
\end{equation}
where the twist-two part of the gluon fragmentation correlator is
parameterized by just one function,
\begin{equation}
 \Delta_2^{g, \mu \nu}(z)
= - \frac{z}{2 P_h^-}\; g_T^{\mu \nu} D_1^g(z) \,,
\end{equation}
because we consider an unpolarized hadron.  With the parameterization
\eqref{e:delta} of $\Delta^q(z,k_T)$ we obtain the high-$k_T$ behavior
\begin{align}
D_1^{q}(z,k_T^2) &=
\frac{\alpha_s}{2 \pi^2}\,
\frac{1}{z^2\ms {\bm k}{}_T^2}\,
  \biggl[\ms \frac{L( \eta_{\smash{h}}^{-1} )}{2}\, D_1^q(z)
   - C_F D_1^q(z)
   + \bigl(D_1^q \otimes P_{qq} + D_1^g \otimes P_{gq}\bigr)(z)
  \biggr] \,,
\raisetag{-0.2em}\label{e:D1highkt}
\\[0.4em]
\frac{D^{\perp q}(z,k_T^2)}{z} &=
\frac{\alpha_s}{2 \pi^2}\,
  \frac{1}{z^2\ms {\bm k}{}_T^2}\,
\biggl\{ \frac{L( \eta_{\smash{h}}^{-1} )}{4}\, D_1^q(z)
+ \int_z^1 \frac{\de\zd}{\zd}\, D_1^q\Bigl(\frac{z}{\zd}\Bigr)\;
   C_F \ms\biggl[\frac{1}{(1-\zd)_{+}}
             + \frac{3}{4}\,\delta(1-\zd) \biggr]
\nonumber \\[0.2em]
& \hspace{12.6em}
+ \int_z^1 \frac{\de\zd}{\zd}\, D_1^g\Bigl(\frac{z}{\zd}\Bigr)\;
   C_F\, \frac{2-\zd}{\zd} \biggr\}
\end{align}
from \eqref{frag-a} and \eqref{frag-b}.  For $\tilde{D}^{\perp q}$
this implies
\begin{align}
\frac{\tilde{D}^{\perp q}(z,k_T^2)}{z} &=
- \frac{\alpha_s}{2 \pi^2}\,
  \frac{1}{2 z^2\ms {\bm k}{}_T^2}\,
  \biggl[\ms \frac{L( \eta_{\smash{h}}^{-1} )}{2}\, D_1^q(z)
   - 2 C_F D_1^q(z)
   + \bigl(D_1^q \otimes P'_{qq} + D_1^g \otimes P'_{gq}\bigr)(z)
  \biggr] \,.
\raisetag{-0.3em}\label{e:Dperptildehighkt}
\end{align}
according to its definition \eqref{e:Dtilde}.  Analogous results with
the same kernels are obtained for the antiquark fragmentation
functions $D_1^{\bar{q}}$, $D^{\perp \bar{q}}_{\phantom{1}}$, and
$\tilde{D}^{\perp \bar{q}}_{\phantom{1}}$.


\subsection{Results for structure functions and their consequences}
\label{sec:tailstructure}

We are now ready to compute the behavior of the structure functions
$F_{UU,T}^{}$, $F_{LL}^{}$, $F_{UU}^{\cos\phi_h}$, and
$\smash{F_{LL}^{\cos\phi_h}}$ at intermediate transverse momentum.
For $F_{UU,T}$ we start from the low-$q_T$ result \eqref{F_UUT}, with
the convolution defined in~\eqref{soft-con}.  Using the expansion
\eqref{e:FUUTexp} we have
\begin{equation}
  \label{FUUT:high_generic}
F_{UU ,T} =
\sum_a x\ms e_a^2\,
\biggl[f_1^a(x,q_T^2)\, \frac{D_1^a(z)}{z^2}
     + f_1^a(x)\, D_1^a(z,q_T^2)
     + f_1^a(x)\, \frac{D_1^a(z)}{z^2}\; U(q_T^2)
\biggr]
\end{equation} 
for $M \ll q_T \ll Q$.
The high-transverse-momentum limits of $f_1^a(x,q_T^2)$ and
$D_1^a(z,q_T^2)$ are respectively given in \eqref{e:f1highkt} and
\eqref{e:D1highkt}.  For the corresponding limit of the soft factor
one obtains
\begin{equation} 
  U(q_T^2) = \frac{\alpha_s\ms C_F}{\pi^2}\, \frac{1}{{q}_T^2}
\label{e:soft-tail}
\end{equation} 
from \cite{Collins:1981uk}, as we show in appendix~\ref{app:sf}.
Given that $2P^+ P_{\smash{h}}^- = z Q^2 /x$ up to mass corrections,
the relations \eqref{zeta-def}, \eqref{eta-def} and
\eqref{zeta-h-def}, \eqref{eta-h-def} imply
\begin{align}
\sqrt{\zeta \zeta_h} &= \frac{z Q^2}{x} \,,
&
\sqrt{\eta\ms \eta_h\rule{0pt}{1.6ex}} &= \frac{q_T^2}{Q^2} \,,
  \label{eta-rel}
\end{align}
where in the second equation we have set $\bm{k}_T^2$ and $\bm{p}_T^2$
equal to $\bm{q}_T^2$, as appropriate for evaluating
\eqref{FUUT:high_generic}.
Putting the above results together, we obtain
%
%
\begin{align}
F_{UU ,T} &=
\frac{\alpha_s}{2 \pi^2}\,
   \frac{1}{z^2 q_T^2}\, \sum_a x\ms e_a^2
\nonumber \\
 & \quad \times \biggl\{
   \biggl[\ms \frac{L(\eta^{-1})}{2}\, f_1^a(x)
     - C_F\ms f_1^a(x) 
     + \bigl(P_{qq} \otimes f_1^a + P_{qg} \otimes f_1^g\bigr)(x)
   \biggr]\, D_1^a(z)
\nonumber \\
 & \qquad
+ f_1^a(x)\, \biggl[\ms \frac{L(\eta_{\smash{h}}^{-1})}{2}\, D_1^a(z)
   - C_F D_1^a(z)
   + \bigl(D_1^a \otimes P_{qq} + D_1^g \otimes P_{gq}\bigr)(z)
  \biggr]
\nonumber \\ &
 \qquad
+ 2 C_F\,f_1^a(x)\,D_1^a(z)
\biggr\}
\nonumber \\
&= \frac{1}{q_T^2}\, \frac{\alpha_s}{2\pi^2 z^2}\,
\sum_a x\ms e_a^2\,
\biggl[f_1^a(x)\,D_1^a(z)\,L\biggl( \frac{Q^2}{q_T^2} \biggr)
+ f_1^a(x)\, \bigl( D_1^a \otimes P_{qq}
                  + D_1^g \otimes P_{gq} \bigr)(z)
\nonumber \\
& \qquad
+ \bigl( P_{qq} \otimes f_1^a 
       + P_{qg} \otimes f_1^g \bigr)(x)\, D_1^a(z)
\biggr] \,,
  \label{FUUT-final}
\end{align} 
which is \emph{identical} with the result \eqref{e:high_FUU} of the
high-$q_T$ calculation.  The same agreement has been found by Ji et
al.~\cite{Ji:2006br}, who used the low-$q_T$ factorization scheme
specified in \cite{Ji:2004wu} instead of the original Collins-Soper
scheme \cite{Collins:1981uk}.
Note that the terms with $C_F f_1^a(x) \,D_1^a(z)$ cancel among the
different contributions in \eqref{FUUT-final}.  By virtue of
\eqref{eta-rel} the dependence on the gauge parameters $\eta$ and
$\eta_{\ms h}$ also cancels, as it should.  We remark that we obtain
the same final result if we take a timelike vector $v$ instead of a
spacelike one.  Both $\eta$ and $\eta_h$ are then negative, and
$L(\eta^{-1})$ and $L(\eta_{\smash{h}}^{-1})$ are replaced by
$L(-\eta^{-1})$ and $L(-\eta_{\smash{h}}^{-1})$, so that they still
add up to $2 L(Q^2 /q_T^2)$.
For $F_{LL}^{}$ we obtain a result analogous to \eqref{FUUT-final},
with the parton distributions $f_1$ replaced by $g_1$ and the
convolutions $P \otimes f_1$ by $\Delta P \otimes g_1$.  This result
exactly matches the expression \eqref{e:high_FLL} obtained in the
high-$q_T$ calculation.

We now turn to the structure function $F_{UU}^{\cos\phi_h}$.
According to \eqref{tails-3} and \eqref{tails-frag} the terms with
$h_{1}^{\perp}$ and $H_{1}^{\perp}$ in the low-$q_T$ expression
\eqref{F_UUcosphi} are power suppressed compared to the terms with
$f^\perp$ and $\tilde{D}^\perp$ when $q_T \gg M$.  For intermediate
$q_T$ we therefore have
\begin{equation}
  \label{FUUcos-expand}
F_{UU}^{\cos\phi_h} =
-\frac{2 q_T}{Q} \sum_a x\ms e_a^2\,
\biggl[
  x f^{\perp a} (x,q_T^2)\, \frac{D_1^a(z)}{z^2}
- f_1^a(x)\, \frac{\tilde{D}^{\perp a}(z,q_T^2)}{z}
\biggr]
\end{equation}
at leading power and leading order in $\alpha_s$.  In this case there
is no leading contribution from the soft factor taken at large
transverse momentum.  Proceeding as we did in \eqref{taylor-expand},
one finds that the leading term in the expansion of $U(l_T^2)$ around
$\bm{l}{}_T = -\bm{q}{}_T$ gives zero in the convolution
\eqref{F_UUcosphi} because it does not depend on a direction in the
transverse plane, whereas the next terms in the expansion only give
contributions that are power suppressed compared to those in
\eqref{FUUcos-expand}.  We therefore obtain the same result
\eqref{FUUcos-expand} if we omit the soft factor in the
transverse-momentum convolution \eqref{soft-con}.
Using the high-transverse-momentum limits \eqref{e:fperphighkt} and
\eqref{e:Dperptildehighkt} of $f^{\perp a}(x,q_T^2)$ and
$\tilde{D}^{\perp a}(z,q_T^2)$, we get
%
%
\pagebreak[4]
\begin{align}
F_{UU}^{\cos\phi_h}
&= - \frac{2 q_T}{Q}\,
  \frac{\alpha_s}{2\pi^2}\, \frac{1}{2 z^2 q_T^2}\,
  \sum_a x\ms e_a^2\,
\nonumber \\
 & \quad\times
\biggl\{
  \biggl[\ms
      \frac{L(\eta^{-1})}{2}\, f_1^a(x)
    + \bigl(P'_{qq} \otimes f_1^a + P'_{qg} \otimes f_1^g \bigr)(x)
  \biggr]\, D_1^a(z)
\nonumber \\
 & \qquad
+ f_1^a(x)\,
  \biggl[\ms
      \frac{L(\eta_{\smash{h}}^{-1})}{2}\, D_1^a(z)
   - 2 C_F D_1^a(z)
   + \bigl(D_1^a \otimes P'_{qq} + D_1^g \otimes P'_{gq}\bigr)(z)
  \biggr]
\biggr\}
\nonumber\\[0.2em]
&= - \frac{1}{Q\ms q_T}\, \frac{\alpha_s}{2\pi^2 z^2}
  \sum_a x\ms e_a^2\,
  \biggl[f_1^a(x)\,D_1^a(z)\, L\biggl( \frac{Q^2}{q_T^2} \biggr)
+ f_1^a(x)\, \bigl( D_1^a \otimes P_{qq}'
                    +  D_1^a \otimes P_{gq}'\bigr)(z)
\nonumber \\
 & \qquad
+ \bigl( P_{qq}' \otimes f_1^a
       + P_{qg}' \otimes f_1^g\bigr)(x)\, D_1^a(z)
- 2 C_F\ms f_1^a(x) \,D_1^a(z)
\biggr] \,,
  \label{e:FUUcosphimismatch}
\end{align}
which is \emph{not identical} to the high-$q_T$ result
\eqref{e:high_FUUcosphi} because of the extra term $2 C_F f_1^a(x)
\,D_1^a(z)$ in the brackets.  The same situation is found for
$F_{LL}^{\cos\phi_h}$, with $f_1$ replaced by $g_1$ and $P' \otimes
f_1$ by $\Delta P' \otimes g_1$

This disagreement has important consequences.  Since the leading terms
in the high-$q_T$ and the low-$q_T$ calculation of
$F_{UU}^{\cos\phi_h}$ have the same power behavior for $M \ll q_T \ll
Q$, their explicit expressions in that region \emph{must} agree if
both of them are calculated correctly.  This is clear, since both
calculations give the same term of a double expansion in $M /q_T$ and
$q_T /Q$, as given in \eqref{low-inter} and \eqref{hi-inter}.  We have
no reason to doubt the validity of the high-$q_T$ result
\eqref{e:high_FUUcosphi}, which comes from a twist-two calculation in
collinear factorization.  The same holds for the
high-transverse-momentum behavior of the functions $f^{\perp
  a}(x,q_T^2)$ and $\tilde{D}^{\perp a}(z,q_T^2)$ in
\eqref{e:fperphighkt} and \eqref{e:Dperptildehighkt}.  In contrast,
the low-$q_T$ expression we used for $F_{UU}^{\cos\phi_h}$ is a
twist-three result, for which no proof of factorization is available.
To obtain the expression in \eqref{FUUcos-expand} we have
\emph{assumed} that the tree-level result \eqref{F_UUcosphi} can be
generalized by taking over the convolution \eqref{soft-con}
established for the twist-two sector.  The comparison of
\eqref{e:FUUcosphimismatch} with \eqref{e:high_FUUcosphi} implies that
this assumption is incorrect.

Based on our finding, one may speculate how a correct twist-three
factorization formula will look like if factorization can be
established at that level.  Simple modification of the soft factor
$U(l_T^2)$ can obviously not yield agreement with the high-$q_T$
result since this factor does not appear in the limiting expression
\eqref{FUUcos-expand} for the reasons we explained above.  The
situation would be different if the soft factor were dependent on the
direction of $\bm{l}_T$, which would require it to have a nontrivial
structure in either Lorentz or Dirac space (through factors $l_T^\mu$
or $\lslash^{}_T$).  Such a dependence would go beyond the eikonal
approximation for the coupling of soft gluons to fast partons, which
may be necessary at subleading order in $1/Q$.  We shall not pursue
such speculations here.  Clearly, the requirement to match the
high-$q_T$ result \eqref{e:high_FUUcosphi} for $F_{UU}^{\cos\phi_h}$
at intermediate $q_T$ can be used as a consistency check for any
framework that extends Collins-Soper factorization to the twist-three
sector.

It is instructive to note that the low- and high-$q_T$ results
disagree by a term proportional to $f_1^a(x) \,D_1^a(z)$, where
neither the distribution nor the fragmentation function appears in a
convolution over longitudinal momentum fractions.  In the calculations
of the previous subsections, such terms arise from configurations
where a gluon has zero plus- or minus-momentum.  The correct treatment
of this phase space region is nontrivial already in proofs of
factorization at the twist-two level \cite{Ji:2004wu,Collins:2007ph},
so that it is not too surprising that this is where problems occur in
the naive extension to twist three which we have explored.

At this point we return to the issue of
transverse-momentum-resummation for $F_{UU}^{\cos\phi_h}$, which we
have briefly discussed in section~\ref{sec:polar}.  We can now
understand why the splitting functions $P'_{qq}$, $P'_{qg}$, and
$P'_{gq}$ in the high-$q_T$ result \eqref{e:high_FUUcosphi} are
different from the usual DGLAP kernels.  Up to $\delta$-function terms
they describe the high-transverse-momentum behavior of $f^\perp$ and
$\tilde{D}^\perp$, rather than the one of the more familiar functions
$f_1$ and $D_1$.  A corresponding remark applies to the $\cos\phi$
asymmetry in Drell-Yan production investigated in \cite{Boer:2006eq}.
If a low-$q_T$ factorization formula for these observables can be
established, it should also allow one to adapt the original CSS
procedure \cite{Collins:1984kg} for the resummation of large
logarithms $\ln (Q^2/q_T^2)$ at next-to-leading logarithmic accuracy
and beyond.  {}From this point of view, resummation for $F_{UU}^{\cos
  2\phi_h}$ and its analogs in Drell-Yan production or $e^+e^-$
annihilation appears rather daunting since it would require a
formulation of low-$q_T$ factorization at twist-four level,
extending the simple parton-model result in \eqref{Cahn} and putting
it on a rigorous footing.


\section{Summary}
\label{sec:sum}

The description of semi-inclusive deep inelastic scattering with
measured transverse momentum $q_T$ involves two theoretical
frameworks: at low $q_T$ one has a factorized representation in terms
of transverse-momentum-dependent distribution and fragmentation
functions, whereas at high $q_T$ standard collinear factorization can
be used.  We have systematically analyzed the relation between the two
descriptions at intermediate transverse momentum $M \ll q_T \ll Q$,
where both are applicable.  Depending on the specific observable, the
leading terms in the two descriptions may or may not coincide.

Using dimensional analysis and Lorentz invariance, we have derived the
general behavior at high $p_T$ for all transverse momentum-dependent
parton distributions of twist two or three.  The results, listed in
Eqs.~\eqref{tails-2} to \eqref{tails-5}, involve the convolution of
collinear parton distributions with hard-scattering kernels, which in
the simplest cases are closely related to the well-known DGLAP
splitting functions.  We have computed these kernels at leading order
in $\alpha_s$ for those cases where the collinear distributions are of
leading twist, obtaining the expressions \eqref{e:f1highkt} to
\eqref{e:hThighkt}.  With these results and their analogs for
transverse-momentum-dependent fragmentation functions we could
establish in Eqs.~\eqref{e:tailFUUT} to \eqref{e:tailFLTcos2phi} the
power behavior for $M \ll q_T \ll Q$ of all SIDIS structure functions
that appear in the low-$q_T$ description at twist-two or twist-three
accuracy, allowing for arbitrary polarization of target and beam.

In the high-$q_T$ description at order $\alpha_s$ one finds a
considerable simplification when taking the limit $q_T \ll Q\ms$: the
expressions of the structure functions then involve a convolution of
either the distribution or the fragmentation functions with
hard-scattering kernels, whereas the other function is evaluated at
the momentum fraction $x$ or $z$ fixed by the kinematics of the final
state.  For observables where the high-$q_T$ and low-$q_T$
calculations match, these kernels can be identified with the ones
describing the high-transverse-momentum behavior of the functions
appearing in the low-$q_T$ description.  In such a situation one can
use the procedure of Collins, Soper, and Sterman to resum large
logarithms of $Q^2/q_T^2$ to all orders in perturbation theory.  A
prerequisite for this is that the power behavior of the observable in
the low- and high-$q_T$ calculations must match.  We have compared the
corresponding powers for a wide range of observables, using our
low-$q_T$ results \eqref{e:tailFUUT} to \eqref{e:tailFLTcos2phi} and
their counterparts \eqref{e:high_FUU_power} to
\eqref{e:high_FUTsin2phi} for those structure functions that have been
evaluated in the high-$q_T$ formulation.  This comparison, compiled in
table~\ref{tab:overview}, is one of the main results of our work.

When the two formulations give the same power law at intermediate
$q_T$ for a given observable, their explicit results must agree
exactly because they describe the same term of a double expansion in
$M/q_T$ and $q_T/Q$.  This constitutes a nontrivial consistency check
for both the low- and high-$q_T$ calculations.  Confirming earlier
results in the literature, we have verified that there is such an
agreement for the unpolarized structure function $F_{UU,T}$, as well
as for its analog $F_{LL}$ for longitudinal beam and target
polarization.  By contrast, the structure function $F_{UU,L}$ for
longitudinal photon polarization only appears at twist four in the
low-$q_T$ framework, where a complete result is not available.  A
simple calculation in the parton model gives a power behavior which in
the intermediate region matches the one of the well-established
high-$q_T$ result but fails to reproduce its exact form.

A more involved picture arises for azimuthal asymmetries, even in
unpolarized scattering.  At low $q_T$ the structure function
$F_{UU}^{\cos 2\phi_h}$ is expressed in terms of the Boer-Mulders
function $h_1^\perp$ and the Collins fragmentation function
$H_1^\perp$, both of which are chiral-odd, whereas the high-$q_T$
expression involves the usual unpolarized distribution and
fragmentation functions $f_1$ and $D_1$, which are chiral-even.  The
two results thus describe different physical mechanisms, which is
consistent with our finding that at intermediate $q_T$ they have a
different power behavior.  In this region, the two results may hence
be added.  In practice, some arbitrariness is involved in deciding
what ``intermediate'' $q_T$ values are.  We have shown that the sum of
the high-$q_T$ and the low-$q_T$ expressions gives a valid
approximation for $F_{UU}^{\cos 2\phi_h}$ also at large $q_T$, where
the low-$q_T$ result cannot be trusted but is power suppressed
compared with the high-$q_T$ expression.  The latter, however, fails
to vanish in the limit $q_T\to 0$, as required by angular momentum
conservation, and should hence not be used at low $q_T$.  A more
favorable observable in this respect is the $\cos 2\phi_h$ asymmetry,
i.e., the ratio of $F_{UU}^{\cos 2\phi_h}$ and the $\phi_h$
independent part $F_{UU,T} + \varepsilon F_{UU,L}$ of the cross
section.  In this case, the sum \eqref{A_UUcos-interp} of the
expressions calculated for low and high $q_T$ gives a consistent
approximation for all transverse momenta, up to corrections of order
$M^2/q_T^2$ and $q_T^2/Q^2$.
The result of a parton-model calculation at low $q_T$, often referred
to as Cahn effect, has the same property for $F_{UU}^{\cos 2\phi_h}$
as it has for $F_{UU,L}^{}$: its power behavior agrees with the
high-$q_T$ result in the intermediate region, but its explicit
expression does not.  The parton-model result may hence only be
regarded as a partial estimate for the full but unknown twist-four
correction to $F_{UU}^{\cos 2\phi_h}$ at low $q_T$.

The description of the structure function $F_{UU}^{\cos\phi_h}$ is
more problematic: at high $q_T$ it can be evaluated in collinear
factorization at twist-two level, but at low $q_T$ it requires a
twist-three calculation, for which transverse-momentum-dependent
factorization at all orders in $\alpha_s$ has not been established.
As a working hypothesis we have taken the well-established result of a
tree-level calculation at low $q_T$ and assumed that the soft factor
which explicitly appears in the factorization theorem for twist-two
observables is also applicable at twist three.  This leads to an
expression that for intermediate $q_T$ agrees with the high-$q_T$
result in its power behavior and in the form of the hard-scattering
kernels, \emph{except} for a term proportional to $f_1(x)\ms D_1(z)$.
\new{We find this partial agreement encouraging, but it does show that our
  candidate factorization formula at twist three is incorrect as it
  stands, and that a proper analysis will have to devote special attention
  to gluons with vanishing plus- or minus-momentum and to the precise form
  of soft factors.}
We emphasize that the correct description of $F_{UU}^{\cos\phi_h}$ at
low $q_T$ is a prerequisite for applying the method of Collins, Soper,
and Sterman to resum large logarithms of $Q^2/q_T^2$.

The structure function $F_{UT,T}^{\sin(\phi_h-\phi_S)}$ for a
transversely polarized target presents a case where the low-$q_T$
calculation is of twist two, whereas the high-$q_T$ description is at
the twist-three level.  The explicit computations in
\cite{Ji:2006ub,Ji:2006vf,Ji:2006br,Koike:2007dg} find exact agreement
of the two descriptions at intermediate $q_T$ and thus validate both
frameworks.  One may expect that the same is true for
$F_{UT}^{\sin(\phi_h+\phi_S)}$, which at low $q_T$ is described in
terms of the Collins effect.

Observables that are integrated over $q_T$ are at times preferable to
differential ones from an experimental point of view.  We have shown
that some of them have the added virtue of admitting a relatively
simple description at the theory level, both for the complexity of the
expressions and for the number of distribution and fragmentation
functions on which they depend.  With the power-counting behavior
listed in table~\ref{tab:overview} one can readily determine to
which region of $q_T$ a given integrated observable is primarily
sensitive.  The results for selected observables are given in
table~\ref{tab:int}.
We find for instance that $\intobs{F_{UU}^{\cos 2\phi_h}}$ and
$\intobs{F_{UT}^{\sin\phi_S}}$ receive leading contributions from both
low and high $q_T$.  The integrated structure function
$\intobs{F_{UU}^{\cos\phi_h}}$ is dominated by large $q_T$, with
contributions from the low-$q_T$ region being suppressed by $M/Q$.
Conversely, both $\smash{\intobs{F_{UT,T}^{\sin(\phi_h-\phi_S)}}}$ and
$\smash{\intobs{F_{UT}^{\sin(\phi_h+\phi_S)}}}$ receive their dominant
contributions from low $q_T$, whereas the high-$q_T$ domain is
suppressed by $M/Q$.  They are hence sensitive to the Sivers function
in the first case, and to the transversity distribution and the
Collins fragmentation function in the second.  A suppression by $M/Q$
may, however, not be sufficient to simply neglect the corresponding
contributions in an analysis at experimentally achievable values of
$Q$.

A theoretically cleaner access to the high-$q_T$ region is through
observables that are weighted with an appropriate power of $q_T/M$.
We find in particular that $\intobs{(q_T/M)\ms F_{UU}^{\cos\phi_h}}$,
$\smash{\intobs{(q_T/M)^2\ms F_{UU}^{\cos 2\phi_h}}}$, and
$\smash{\intobs{(q_T/M)^2\ms F_{UT}^{\sin\phi_S}}}$ can be evaluated
from the high-$q_T$ results alone, up to corrections of order
$M^2/Q^2$, and that at the same accuracy one can extend the
integration down to $q_T=0$.  This leads to simple expressions,
similar to the one for the integrated longitudinal structure function
$\intobs{F_{UU,L}^{}}$.  The observables $\intobs{(q_T/M)\ms
  F_{UU}^{\cos\phi_h}}$ and $\smash{\intobs{(q_T/M)^2\ms F_{UU}^{\cos
      2\phi_h}}}$ are sensitive to the twist-two functions $f_1$ and
$D_1$ and may for instance be useful for separating the contributions
from different quark flavors, serving as complements to
$\intobs{F_{UU,T}^{}}$.  In contrast, $\smash{\intobs{(q_T/M)^2\ms
    F_{UT}^{\sin\phi_S}}}$ is sensitive to distribution and
fragmentation functions of twist three.

The weighted structure functions $\smash{\intobs{(q_T/M)\ms
    F_{UT,T}^{\sin(\phi_h-\phi_S)}}}$ and $\smash{\intobs{(q_T/M)\ms
    F_{UT}^{\sin(\phi_h+\phi_S)}}}$ play a special role in this
context.  They receive leading-power contributions from both low and
high $q_T$ and, as already pointed out in \cite{Boer:1997nt}, lead to
a deconvolution of the transverse-momentum integrals in the low-$q_T$
result.  We argued that they should permit a description in terms of
collinear functions of twist two and three, defined in the standard
$\overline{\text{MS}}$ scheme.  In this description, the low-$q_T$
expression gives the Born-level result, whereas the high-$q_T$
calculation of \cite{Eguchi:2006qz,Eguchi:2006mc,Koike:2007dg} gives
part of the $\alpha_s$ corrections.  If completed, such a description
would provide a full NLO result in $\alpha_s$ and be an extension to
twist-three level of the standard NLO calculation for
$\intobs{F_{UU,T}}$ within collinear factorization at twist-two
accuracy.  The leading-order expressions \eqref{sivers-mom} and
\eqref{collins-mom} for the weighted structure functions are analogs
of the familiar tree-level formula $\intobs{F_{UU,T}} = \sum_a x\ms
e_a^2\; f_{1}^{a}(x)\, D^a_1(z)$.  These expressions receive
corrections from the high-$q_T$ region which are of leading power but
suppressed by~$\alpha_s$.

Let us finally remark that the results we have discussed here carry
over to the analogous observables in the Drell-Yan process and in
$e^+e^-$ annihilation.  The SIDIS structure functions
$F_{UU}^{\cos\phi_h}$ and $F_{UU}^{\cos 2\phi_h}$ correspond for
instance to the $\cos\phi$ and $\cos 2\phi$ asymmetries in the angular
distribution of the lepton pair in unpolarized Drell-Yan production,
which have been measured \cite{Falciano:1986wk} and given rise to
several theoretical investigations, see e.g.\ the references in
\cite{Boer:2006eq}.  Furthermore, $F_{UU}^{\cos 2\phi_h}$ corresponds
to a $\cos 2\phi$ asymmetry for two-pion production in $e^+e^-$
annihilation, which has been measured by BELLE \cite{Abe:2005zx} and
provides the possibility for an independent determination of the
Collins fragmentation function \cite{Boer:1997mf}.  For a reliable
extraction, our discussion of matching low- and
high-transverse-momentum contributions should be of relevance.


\section*{Acknowledgments}

It is our pleasure to acknowledge valuable discussions with John
Collins and Werner Vogelsang.  We thank Jochen Bartels for helpful
remarks on the manuscript.
This research is part of the Integrated Infrastructure
Initiative ``Hadron Physics'' of the European Union under contract
number RII3-CT-2004-506078.  The work of M.D. is partially supported
by the Helmholtz Association, contract number VH-NG-004, and the work
of A.B. is partially supported by the SFB ``Particles, Strings and the
Early Universe''.  The Feynman diagrams in this paper were drawn using
JaxoDraw~\cite{Binosi:2003yf}.


\appendix


\section{Distribution functions at high $p_T\ms$: Feynman versus
  axial gauge}
\label{app:FG}

Our calculation in section~\ref{sec:taildistribution} is done in axial
gauge $A\cdot v =0$.  Let us see how the same calculation proceeds in
Feynman gauge.  In this case one must explicitly take into account the
gauge link $\mathcal{U}$ in the definition \eqref{e:phi} of the
correlation function $\Phi(x,p_T)$, which consists of sections
pointing along $v$ and a transverse section at
\new{infinity.  The detailed path of the gauge link reflects important
  physics, as shown for instance in \cite{Belitsky:2002sm,Boer:2003cm,%
    Bomhof:2004aw,Bomhof:2006dp,Collins:2004nx,Collins:2002kn}.}

\new{Let us}
consider the correlation function $\Phi^{[+]}(x,p_T)$ relevant for
SIDIS, whose gauge link $\mathcal{U}^+$ is closed at $a^- = +\infty$.
To evaluate the quark-to-quark contribution to the high-$p_T$ behavior
of $\Phi(x,p_T)$ at leading order in $1/p_T$, one has to take into
account the four diagrams shown in Fig.~\ref{f:rungdistFG}.  The
graphs with eikonal lines are due to gluons coupling to the gauge link
in the operator $\bar{\psi}_j(0)\, \mathcal{U}_{(0,\xi)}\,
\psi_i(\xi)$.  The corresponding Feynman rules read
\cite{Collins:1981uk,Collins:1982uw}
\begin{align}
\raisebox{-6mm}{\includegraphics[height=1.2cm]{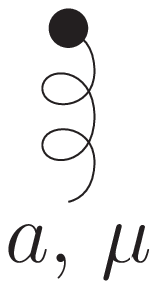}}
  &= \ii g \, t^{a}\ms v^{\mu} \,,
&
\includegraphics[width=1.6cm]{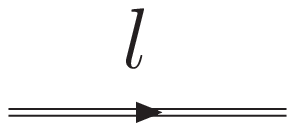}
  &= \frac{\ii}{l\cdot v +\ii \epsilon} \,,
\end{align} 
where the sign of $\ii \epsilon$ for the eikonal line corresponds to a
gauge link pointing to $a^- = +\infty$ if one takes $v^- >0$.  In cut
diagrams one must take the conjugate of these expressions for vertices
and propagators on the right of the final-state cut (indicated by the
dashed lines in Fig.~\ref{f:rungdistFG}).

\FIGURE[t]{
\includegraphics[width=11.7cm]{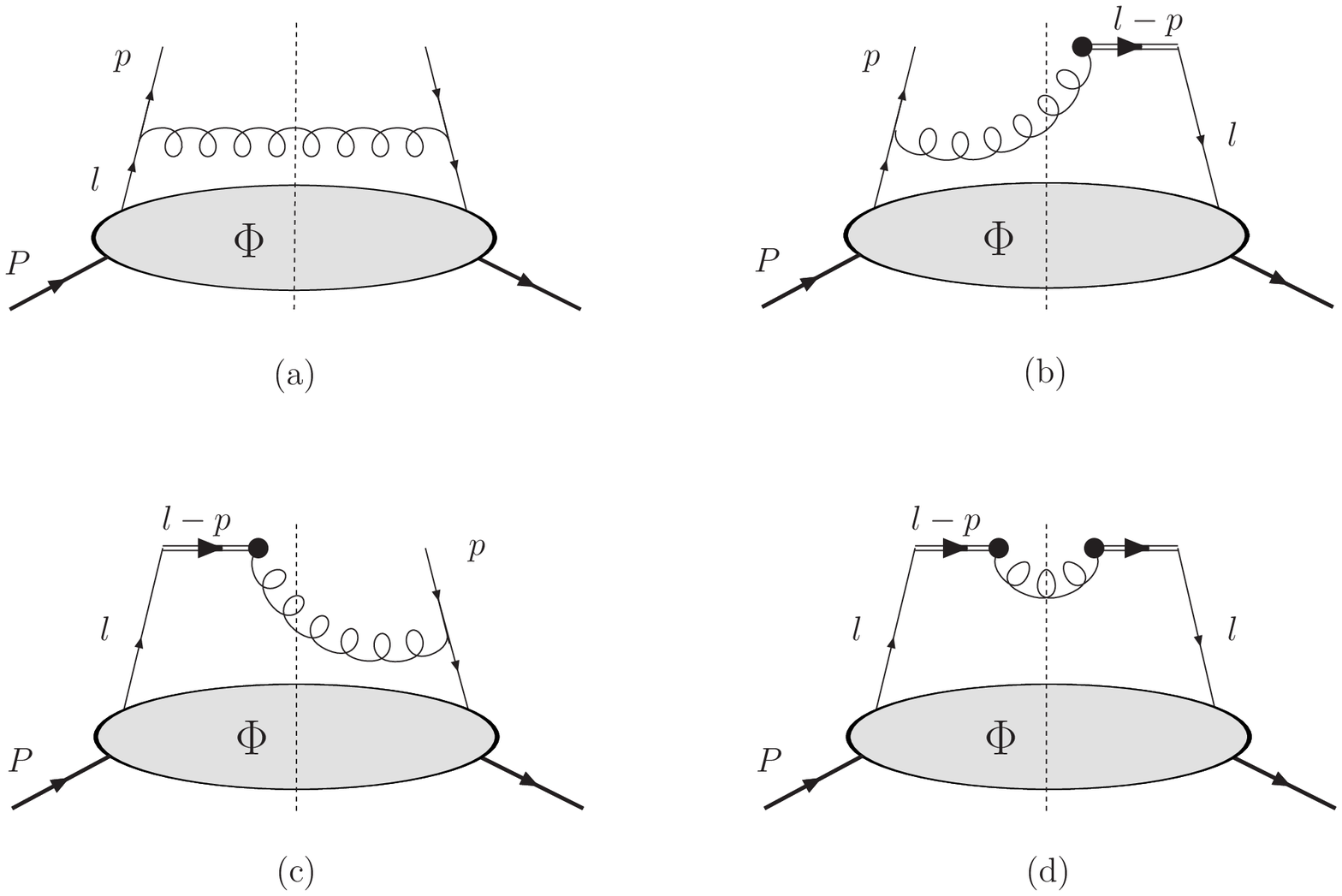}
\caption{\label{f:rungdistFG} Diagrams for the calculation of the
  leading high-$p_T$ behavior of the quark-quark correlator
  $\Phi(x,p_T)$ in Feynman gauge.}
}

After performing the integration over $p^-$ using the $\delta$
function in \eqref{delta-p-minus}, we have for the respective diagrams
\begin{align} 
\Phi^{[+] q}(x,p_T) \,\Bigr|_{\text{(\ref{f:rungdistFG}a)}}
 &=
- \frac{\alpha_s}{(2 \pi)^2}\,C_F 
  \int \frac{\de\xb}{\xb (1-\xb)}\;
  g^{\mu\nu}\,
  \frac{\pbarslash\,}{\bar{p}^{\ms 2}}\, \gamma_{\nu}\,
  \Phi_2^q \Bigl(\frac{x}{\xb}\Bigr)
  \gamma_{\mu}\, \frac{\pbarslash\,}{\bar{p}^{\ms 2}} \,,
\label{e:FGa}
\\[0.2em]
\label{e:FGb}
\Phi^{[+] q}(x,p_T) \,\Bigr|_{\text{(\ref{f:rungdistFG}b)}}
 &=
- \frac{\alpha_s}{(2 \pi)^2}\,C_F 
  \int \frac{\de\xb}{\xb (1-\xb)}\;
  \frac{\pbarslash\,}{\bar{p}^{\ms 2}}\ms \vslash\,
  \Phi_2^q \Bigl(\frac{x}{\xb}\Bigr)\,
  \frac{1}{(\bar{l}-\bar{p})\cdott v - \ii\epsilon\rule{0pt}{2.1ex}} \,,
\\[0.1em]
\Phi^{[+] q}(x,p_T) \,\Bigr|_{\text{(\ref{f:rungdistFG}c)}}
 &=
- \frac{\alpha_s}{(2 \pi)^2}\,C_F 
  \int \frac{\de\xb}{\xb (1-\xb)}\;
  \frac{1}{(\bar{l}-\bar{p})\cdott v + \ii\epsilon\rule{0pt}{2.1ex}}\;
  \Phi_2^q \Bigl(\frac{x}{\xb}\Bigr) \vslash\,
  \frac{\pbarslash\,}{\bar{p}^{\ms 2}} \,,
\\[0.1em]
\label{e:FGd}
\Phi^{[+] q}(x,p_T) \,\Bigr|_{\text{(\ref{f:rungdistFG}d)}}
 &=
- \frac{\alpha_s}{(2 \pi)^2}\,C_F 
  \int \frac{\de\xb}{\xb (1-\xb)}\;
  \Phi_2^q \Bigl(\frac{x}{\xb}\Bigr)\,
  \frac{v^2}{\bigl[ (\bar{l}-\bar{p})\cdott v + \ii\epsilon \bigr]
             \bigl[ (\bar{l}-\bar{p})\cdott v - \ii\epsilon \bigr]
             \rule{0pt}{2.1ex}}
\end{align}  
with $\bar{p}$ and $\bar{l}$ given in \eqref{p-approx} and
\eqref{l-approx}.  Since $\bar{p}^{\ms 2}$ is always spacelike
according to \eqref{p-bar-squared}, we have omitted the $\ii\epsilon$
in the quark propagators.

In the calculation using axial gauge only the first of the four
diagrams contributes, but instead of $g^{\mu\nu}$ in \eqref{e:FGa} we
then have to take
\begin{equation}
- d^{\mu \nu}(\bar{l}-\bar{p};v)= g^{\mu \nu}
 - \frac{(\bar{l}-\bar{p})^{\mu}\ms v^{\nu}
        +(\bar{l}-\bar{p})^{\nu}\ms v^{\mu}}{%
   (\bar{l}-\bar{p}) \cdott v \rule{0pt}{2.1ex}}
 + \frac{(\bar{l}-\bar{p})^{\mu} \ms (\bar{l}-\bar{p})^{\nu}}{%
   [ (\bar{l}-\bar{p})\cdott v \ms]^2 \rule{0pt}{2.1ex}}\, v^2 \,.
\label{e:gluonpolAG2}
\end{equation}
Each term in \eqref{e:gluonpolAG2} corresponds to one of the four
diagrams in the Feynman gauge calculation.  The correspondence between
the first term and diagram \ref{f:rungdistFG}a in Feynman gauge is
trivial.  To establish the correspondence between the second term and
diagram \ref{f:rungdistFG}b we use that
\begin{equation} 
\vslash\, \Phi_2^q \,
  (\lbarslash-\pbarslash)\, \frac{\pbarslash\,}{\bar{p}^{\ms 2}}
= \vslash\, \Phi_2^q \, \lbarslash \,
  \frac{\pbarslash\,}{\bar{p}^{\ms 2}} - \vslash\, \Phi_2^q
= - \vslash\, \Phi_2^q \,,
\end{equation} 
where in the second step we have used that the twist-two part of the
collinear quark correlator satisfies $\Phi_2^q\, \lbarslash =
\Phi_2^q\, \nslash_+ \, l^+ =0$.  In an analogous way one establishes
the correspondence between the last two terms in \eqref{e:gluonpolAG2}
with the respective contributions of diagrams \ref{f:rungdistFG}c and
\ref{f:rungdistFG}d in Feynman gauge.

A mismatch between the expressions in \eqref{e:FGb} to \eqref{e:FGd}
and the calculation in axial gauge is, however, the different
treatment of the singularities at $(\bar{l}-\bar{p})\cdott v =0$. 
\new{The principal value prescription we employed when using the spacelike
axial gauge of the original Collins-Soper paper \cite{Collins:1981uk}
differs from the $\ii\epsilon$ prescription for the different terms in the
Feynman gauge calculation, which arises from the structure of the Wilson
line $\mathcal{U}^+$ in the correlation function.  We note that the
integral in \eqref{e:FGd} is actually not well defined as it stands, since
the double pole at $(\bar{l}-\bar{p})\cdott v =0$ is pinched.  The
contribution from such unphysical poles must be absent in the physical
cross section and should hence cancel between the distribution function,
the fragmentation function, and the soft factor in the factorization
formula \eqref{Wsidis-full}.  How to implement this by regulating the
individual factors has so far not been addressed in the literature.
It is also currently unknown if and how the principal-value prescription
in axial gauge can be implemented in terms of Wilson lines for Feynman
gauge.  The discussion in \cite{Belitsky:2002sm} is for a light-cone
rather than an axial gauge and hence does not contain the problematic term
\eqref{e:FGd}.}

\new{We remark that corresponding problems did not appear in the Feynman
  gauge calculation of Ji et al.~\cite{Ji:2004wu}, where the vector $v$
  was chosen to be timelike.  If we do the same in our context, then the
  calculations in Feynman and axial gauge exactly coincide.  This is
  because $(\bar{l}-\bar{p})\cdott v$ remains positive in \eqref{e:FGb} to
  \eqref{e:FGd} according to \eqref{axial-denominator}, so that the
  singularity at $(\bar{l}-\bar{p})\cdott v =0$ is not reached in the loop
  integral.  As a consequence, the particular regularization of the
  axial-gauge propagator does not influence our results of
  section~\ref{sec:tails} if we take $v$ timelike.  Likewise, there is
  then no contribution from transverse segments of the gauge link at
  infinity, which involve a $\delta$ function in $(\bar{l}-\bar{p})\cdott
  v$.  This is not implausible, since the distribution functions
  considered in section~\ref{sec:tails} are $T$-even and must in
  particular be the same for the gauge links $\mathcal{U}^{+}$ and
  $\mathcal{U}^{-}$.}


\section{Integrated distribution functions and transverse-momentum
  cutoff}
\label{app:cutoff}

In this appendix we derive Eq.~\eqref{cutoff}, which relates two
different ways of regularizing the integral over the
transverse-momentum-dependent distribution $f_1(x,p_T^2)$.  More
precisely we show that with $\mu = b_0/b$ one has
\begin{align}
  \label{cute}
\int \de^2 \bm{p}\; e^{\ii \bm{b}\cdot \bm{p}}\, f(\bm{p}^2)
= \pi \int_{0}^{\infty} \de p^2 J_0(b\ms p)\, f(p^2)
= \pi \int_{0}^{\mu^2} \de {p}^2\, f(p^2)
  + \mathcal{O}\bigl( b^2 \lambda^2 \bigr)
\end{align}
for any function that can be expanded as
\begin{equation}
  \label{high-p-approx}
f(p^2) = \frac{c_2}{p^2} + \frac{c_4}{p^4} + \frac{c_6}{p^6}
         + \ldots
\end{equation}
for $p> \lambda$, where $c_2$, $c_4$, $c_6$, etc.\ are constants.  The
intermediate scale $\lambda$ can be taken just large enough for
\eqref{high-p-approx} to be valid, since corrections going for
instance like $M/\lambda$ do not appear.  The power corrections in
\eqref{cute} are understood as up to logarithms in $b^2 \lambda^2$.
For ease of notation we have written $p = |\bm{p}|$ and omitted the
subscript $T$.

To establish \eqref{cute} we split the integrals into the regions
$p<\lambda$ and $p>\lambda$.  In the first region we can write
\begin{equation}
\int_{0}^{\lambda^2} \de p^2 J_0(b\ms p)\, f(p^2)
  = \int_{0}^{\lambda^2} \de p^2 f(p^2)
    + \mathcal{O}\bigl( b^2 \lambda^2 \bigr) \,,
\end{equation}
using that the Bessel function admits a Taylor expansion $J_0(x) = 1 -
\frac{1}{4} x^2 + \ldots$ in even powers of $x$.  In the region
$p>\lambda$ we make use of the expansion \eqref{high-p-approx}.
Focusing first on the $1/p^2$ term, we write
\begin{equation}
\int_{\lambda^2}^{\infty} \frac{\de p^2}{p^2}\, J_0(b\ms p)
= 2 \int_{b \lambda}^\infty \frac{\de x}{x}\, J_0(x)
= - 2 \ln(b \lambda)\, J_0(b \lambda)
  + 2 \int_{b \lambda}^\infty \de x\ms \ln x\, J_1(x) \,,
\end{equation}
where in the second step we have integrated by parts.  We now use that
$\int_{0}^\infty \de x\ms \ln x\, J_1(x) = \ln 2 - \gamma_E$ and
$\int_{0}^y \de x\ms \ln x\, J_1(x) \sim y^2$ for $y\to 0$, where the
latter relation holds because $J_1(x) \sim x$ for $x\to 0$.  Recalling
that $b_0 = 2 e^{-\gamma_E}$ we obtain
\begin{equation}
\int_{\lambda^2}^{\infty} \frac{\de p^2}{p^2}\, J_0(b\ms p)
= 2\ms \bigl[ \ln(2 e^{-\gamma_E}) - \ln(b \lambda) \bigr]
  + \mathcal{O}\bigl( b^2 \lambda^2 \bigr)
= \int_{\lambda^2}^{\mu^2} \frac{\de p^2}{p^2}
  + \mathcal{O}\bigl( b^2 \lambda^2 \bigr) \,.
\end{equation}
For the $1/p^4$ term in \eqref{high-p-approx} we use again integration
by parts to write
\begin{equation}
\int_{\lambda^2}^{\infty} \frac{\de p^2}{p^4}\, J_0(b\ms p)
= 2 b^2 \int_{b \lambda}^\infty \frac{\de x}{x^3}\, J_0(x)
= \frac{J_0(b \lambda)}{\lambda^2}
  - b^2 \int_{b \lambda}^\infty \frac{\de x}{x^2}\, J_1(x) \,.
\end{equation}
Since the integrand in the last term behaves like $1/x$ for $x\to 0$,
we have
\begin{equation}
\int_{\lambda^2}^{\infty} \frac{\de p^2}{p^4}\, J_0(b\ms p)
= \frac{1}{\lambda^2} +  \mathcal{O}\bigl( b^2 \lambda^2 \bigr)
= \int_{\lambda^2}^{\mu^2} \frac{\de p^2}{p^4}
  + \mathcal{O}\bigl( b^2 \lambda^2 \bigr) \,.
\end{equation}
A similar argument can be given for terms going like $1/p^{2n}$ with
$n>2$, which completes the proof of \eqref{cute}.

As an illustration of our result let us consider the simple form
$f(p^2) = 1/(p^2 + \lambda^2)$.  The relevant integrals then are
\begin{align}
\int_0^{\infty} \frac{\de p^2}{p^2 + \lambda^2}\, J_0(b \lambda)
 &= 2 K_0(b \lambda) \,,
&
\int_0^{\mu^2}  \frac{\de p^2}{p^2 + \lambda^2}
 &= \ln\Bigl( 1 + \frac{\mu^2}{\lambda^2} \Bigr) \,.
\end{align}
With the behavior $K_0(x) = - \ln x + \ln b_0 + \mathcal{O}(x^2)$ of
the modified Bessel function at small~$x$ one readily finds that the
relation \eqref{cute} is satisfied.

As we have seen in section~\ref{sec:taildistribution}, a logarithmic
factor $\ln(\zeta /p^2)$ appears in the explicit calculation for the
high-$p_T$ behavior of distribution functions at order $\alpha_s$.
One can easily repeat the above arguments for the case where the
$1/p^2$ term in \eqref{high-p-approx} is multiplied by $\ln p^2$ and
the subleading terms by some power of $\ln p^2$.  Using that
$\int_{0}^\infty \de x\ms (\ln x)^2 J_1(x) = (\ln 2 - \gamma_E)^2$ one
finds that \eqref{cute} holds without modification also in this case.


\section{One-loop expression of the soft factor}
\label{app:sf}

In this appendix we show how to obtain the momentum-space expression
\eqref{e:soft-tail} of the soft factor in the Collins-Soper
factorization formula \cite{Collins:1981uk}.  The corresponding
expression in $b$-space is given in Eq.~(7.22) of
\cite{Collins:1981uk}.  With our definition \eqref{Ub} we obtain
$U(l_T^2)$ from this by setting $\varepsilon=0$ and omitting $\int
\de^2\bm{l}_T\; e^{\ii \bm{b}\cdot \bm{l}_T}$.  The result is
\begin{alignat}{3}
U(l_T^2) &=
-4 C_F\ms \alpha_s \int \frac{\de l^+ \de l^-}{(2 \pi)^2}\,
  \frac{\delta(l^2)\; \theta(l^+)}{l^+ l^-}\, d^{+-}(l;v)
\notag \\
&= 4 C_F\ms \alpha_s \int \frac{\de l^+ \de l^-}{(2 \pi)^2}\,
  \delta\bigl( 2 l^+ l^- - {\bm l}_T^2 \bigr)\; \theta(l^+)
  \,\frac{v^2}{(l \cdott v)^2}
\notag \\
&= \frac{C_F\ms \alpha_s}{\pi^2} \;\,
   \text{PV}\! \int_0^{\infty} \de l^+
   \,\frac{4\ms l^+ v^- /v^+}{%
     \bigl[\ms 2 (l^+)^2\, v^- /v^+ + {\bm l}_T^2 \bigr]^2}
&= \frac{C_F\ms \alpha_s}{\pi^2}\, \frac{1}{{\bm l}_T^2} \,,
\label{e:soft-tail2}
\end{alignat} 
where on the last line we have indicated that for a spacelike gauge
vector we need the principal value prescription to regulate the
integral, given that $v^- /v^+ < 0$.  In accordance with our footnote
on page \pageref{U-footnote}, the result of the integration is
independent of $v$.
Fourier transforming the result \eqref{e:soft-tail2} to $b$-space in
$2-\varepsilon$ transverse dimensions, we obtain
\begin{equation}
\frac{C_F\ms \alpha_s}{\pi^2}\, \mu^{\varepsilon}
\int \frac{\de^{2-\varepsilon}\, \bm{l}_T}{(2 \pi)^{-\varepsilon}}\;
   e^{\ii \bm{l}_T \cdot \bm{b}}\; \frac{1}{\bm{l}_T^2}
= {}- \frac{C_F\ms \alpha_s}{\pi}\,
  \biggl[ \ln\bigl(\mu^2 b^2 \pi\, e^\gamma \bigr)
        + \frac{2}{\varepsilon} \biggr]
\end{equation} 
in agreement with Eq.~(7.23) in \cite{Collins:1981uk}.


\providecommand{\href}[2]{#2}


\begin{thebibliography}{10}

\bibitem{Collins:1984kg}
J.~C. Collins, D.~E. Soper, and G.~Sterman, {\it {Transverse momentum
  distribution in Drell-Yan pair and $W$ and $Z$ boson production}},  {\em
  Nucl. Phys.} {\bf B250} (1985) 199.

\bibitem{Cahn:1978se}
R.~N. Cahn, {\it Azimuthal dependence in leptoproduction: a simple parton model
  calculation},  {\em Phys. Lett.} {\bf B78} (1978) 269.

\bibitem{Cahn:1989yf}
R.~N. Cahn, {\it Critique of parton model calculations of azimuthal dependence
  in leptoproduction},  {\em Phys. Rev.} {\bf D40} (1989) 3107.

\bibitem{Georgi:1977tv}
H.~Georgi and H.~D. Politzer, {\it Clean tests of QCD in $\mu p$ scattering},
  {\em Phys. Rev. Lett.} {\bf 40} (1978) 3.

\bibitem{Ji:2006ub}
X.~Ji, J.-W. Qiu, W.~Vogelsang, and F.~Yuan, {\it A unified picture for single
  transverse-spin asymmetries in hard processes},  {\em Phys. Rev. Lett.} {\bf
  97} (2006) 082002 [\href{http://xxx.lanl.gov/abs/hep-ph/0602239}{{\tt
  hep-ph/0602239}}].

\bibitem{Ji:2006vf}
X.~Ji, J.-W. Qiu, W.~Vogelsang, and F.~Yuan, {\it Single transverse-spin
  asymmetry in Drell-Yan production at large and moderate transverse momentum},
   {\em Phys. Rev.} {\bf D73} (2006) 094017
  [\href{http://xxx.lanl.gov/abs/hep-ph/0604023}{{\tt hep-ph/0604023}}].

\bibitem{Ji:2006br}
X.~Ji, J.-W. Qiu, W.~Vogelsang, and F.~Yuan, {\it Single-transverse spin
  asymmetry in semi-inclusive deep inelastic scattering},  {\em Phys. Lett.}
  {\bf B638} (2006) 178
  [\href{http://xxx.lanl.gov/abs/hep-ph/0604128}{{\tt hep-ph/0604128}}].

\bibitem{Koike:2007dg}
Y.~Koike, W.~Vogelsang, and F.~Yuan, {\it On the relation between mechanisms
  for single-transverse-spin asymmetries},
  {\em Phys.\ Lett.} {\bf B659} (2008) 878
  [\href{http://xxx.lanl.gov/abs/arXiv:0711.0636}{{\tt arXiv:0711.0636
  [hep-ph]}}].

\bibitem{Boer:2006eq}
D.~Boer and W.~Vogelsang, {\it Drell-Yan lepton angular distribution at small
  transverse momentum},  {\em Phys. Rev.} {\bf D74} (2006) 014004
  [\href{http://xxx.lanl.gov/abs/hep-ph/0604177}{{\tt hep-ph/0604177}}].

\bibitem{Berger:2007jw}
E.~L. Berger, J.-W. Qiu, and R.~A. Rodriguez-Pedraza, {\it Transverse momentum
  dependence of the angular distribution of the Drell-Yan process},  {\em Phys.
  Rev.} {\bf D76} (2007) 074006
  [\href{http://xxx.lanl.gov/abs/arXiv:0708.0578}{{\tt arXiv:0708.0578
  [hep-ph]}}].

\bibitem{Mulders:1995dh}
P.~J. Mulders and R.~D. Tangerman, {\it The complete tree-level result up to
  order $1/Q$ for polarized deep-inelastic leptoproduction},  {\em Nucl. Phys.}
  {\bf B461} (1996) 197
  [\href{http://xxx.lanl.gov/abs/hep-ph/9510301}{{\tt hep-ph/9510301}}];
  Erratum-ibid.\ {\bf B484} (1997) 538.

\bibitem{Boer:1997nt}
D.~Boer and P.~J. Mulders, {\it Time-reversal odd distribution functions in
  leptoproduction},  {\em Phys. Rev.} {\bf D57} (1998) 5780
  [\href{http://xxx.lanl.gov/abs/hep-ph/9711485}{{\tt hep-ph/9711485}}].

\bibitem{Boer:2003cm}
D.~Boer, P.~J. Mulders, and F.~Pijlman, {\it Universality of T-odd effects in
  single spin and azimuthal asymmetries},  {\em Nucl. Phys.} {\bf B667} (2003)
  201 [\href{http://xxx.lanl.gov/abs/hep-ph/0303034}{{\tt
  hep-ph/0303034}}].

\bibitem{Bacchetta:2006tn}
A.~Bacchetta, M.~Diehl, K.~Goeke, A.~Metz, P.~J. Mulders, and M.~Schlegel, {\it
  Semi-inclusive deep inelastic scattering at small transverse momentum},  {\em
  JHEP} {\bf 02} (2007) 093
  [\href{http://xxx.lanl.gov/abs/hep-ph/0611265}{{\tt hep-ph/0611265}}].

\bibitem{Gamberg:2006ru}
L.~P.~Gamberg, D.~S.~Hwang, A.~Metz, and M.~Schlegel,
  {\it Light-cone divergence in twist-3 correlation functions},
  {\em Phys. Lett.} {\bf B639} (2006) 508
  [\href{http://xxx.lanl.gov/abs/hep-ph/0604022}{{\tt hep-ph/0604022}}].

\bibitem{Arneodo:1986cf}
{EMC} Collaboration, M.~Arneodo {\em et~al.}, {\it Measurement of hadron
  azimuthal distributions in deep inelastic muon proton scattering},  {\em Z.
  Phys.} {\bf C34} (1987) 277;\\
%
{E665} Collaboration, M.~R. Adams {\em et~al.}, {\it Perturbative {QCD}
  effects observed in 490 {GeV} deep inelastic muon scattering},  {\em Phys.
  Rev.} {\bf D48} (1993) 5057;\\
%
{ZEUS} Collaboration, M.~Derrick {\em et~al.}, {\it Inclusive charged
  particle distributions in deep inelastic scattering events at {HERA}},  {\em
  Z. Phys.} {\bf C70} (1996) 1
  [\href{http://xxx.lanl.gov/abs/hep-ex/9511010}{{\tt hep-ex/9511010}}];\\
%
{H1} Collaboration, C.~Adloff {\em et~al.}, {\it Measurement of charged
  particle transverse momentum spectra in deep inelastic scattering},  {\em
  Nucl. Phys.} {\bf B485} (1997) 3
  [\href{http://xxx.lanl.gov/abs/hep-ex/9610006}{{\tt hep-ex/9610006}}];\\
%
{E665} Collaboration, M.~R. Adams {\em et~al.}, {\it Inclusive
  single-particle distributions and transverse momenta of forward produced
  charged hadrons in $\mu p$ scattering at 470 {GeV}},  {\em Z. Phys.} {\bf
  C76} (1997) 441;\\
%
{ZEUS} Collaboration, J.~Breitweg {\em et~al.}, {\it Measurement of
  multiplicity and momentum spectra in the current and target regions of the
  Breit frame in deep inelastic scattering at HERA},  {\em Eur. Phys. J.} {\bf
  C11} (1999) 251 [\href{http://xxx.lanl.gov/abs/hep-ex/9903056}{{\tt
  hep-ex/9903056}}];\\
%
H.~Mkrtchyan {\em et~al.}, {\it Transverse momentum dependence of
  semi-inclusive pion production},
  {\em Phys.\ Lett.}  {\bf B665} (2008) 20
  [\href{http://xxx.lanl.gov/abs/arXiv:0709.3020}{{\tt arXiv:0709.3020
  [hep-ph]}}].

\bibitem{Breitweg:2000qh}
{ZEUS} Collaboration, J.~Breitweg {\em et~al.}, {\it Measurement of
  azimuthal asymmetries in deep inelastic scattering},  {\em Phys. Lett.} {\bf
  B481} (2000) 199 [\href{http://xxx.lanl.gov/abs/hep-ex/0003017}{{\tt
  hep-ex/0003017}}];\\
%
{ZEUS} Collaboration, S.~Chekanov {\em et~al.}, {\it Measurement of
  azimuthal asymmetries in neutral current deep inelastic scattering
  at HERA},  {\em Eur. Phys. J.} {\bf C51} (2007) 289
  [\href{http://xxx.lanl.gov/abs/hep-ex/0608053}{{\tt hep-ex/0608053}}].

\bibitem{Avakian:2003pk}
{CLAS} Collaboration, H.~Avakian {\em et~al.}, {\it Measurement of
  beam-spin asymmetries for deep inelastic $\pi^+$ electroproduction},  {\em
  Phys. Rev.} {\bf D69} (2004) 112004
  [\href{http://xxx.lanl.gov/abs/hep-ex/0301005}{{\tt hep-ex/0301005}}];\\
%
{HERMES} Collaboration, A.~Airapetian {\em et~al.}, {\it Beam-spin
  asymmetries in the azimuthal distribution of pion electroproduction},  {\em
  Phys. Lett.} {\bf B648} (2007) 164
  [\href{http://xxx.lanl.gov/abs/hep-ex/0612059}{{\tt hep-ex/0612059}}].

\bibitem{Airapetian:1999tv}
{HERMES} Collaboration, A.~Airapetian {\em et~al.}, {\it Observation of a
  single-spin azimuthal asymmetry in semi-inclusive pion electro-production},
  {\em Phys. Rev. Lett.} {\bf 84} (2000) 4047
  [\href{http://xxx.lanl.gov/abs/hep-ex/9910062}{{\tt hep-ex/9910062}}];\\
%
{HERMES} Collaboration, A.~Airapetian {\em et~al.}, {\it Single-spin
  azimuthal asymmetries in electroproduction of neutral pions in semi-inclusive
  deep-inelastic scattering},  {\em Phys. Rev.} {\bf D64} (2001) 097101
  [\href{http://xxx.lanl.gov/abs/hep-ex/0104005}{{\tt hep-ex/0104005}}];\\
%
{HERMES} Collaboration, A.~Airapetian {\em et~al.}, {\it Measurement of
  single-spin azimuthal asymmetries in semi-inclusive electroproduction of
  pions and kaons on a longitudinally polarised deuterium target},  {\em Phys.
  Lett.} {\bf B562} (2003) 182
  [\href{http://xxx.lanl.gov/abs/hep-ex/0212039}{{\tt hep-ex/0212039}}];\\
%
{HERMES} Collaboration, A.~Airapetian {\em et~al.}, {\it Subleading-twist
  effects in single-spin asymmetries in semi-inclusive deep-inelastic
  scattering on a longitudinally polarized hydrogen target},  {\em Phys. Lett.}
  {\bf B622} (2005) 14 [\href{http://xxx.lanl.gov/abs/hep-ex/0505042}{{\tt
  hep-ex/0505042}}].

\bibitem{Airapetian:2004tw}
{HERMES} Collaboration, A.~Airapetian {\em et~al.}, {\it Single-spin
  asymmetries in semi-inclusive deep-inelastic scattering on a transversely
  polarized hydrogen target},  {\em Phys. Rev. Lett.} {\bf 94} (2005) 012002
  [\href{http://xxx.lanl.gov/abs/hep-ex/0408013}{{\tt hep-ex/0408013}}];\\
%
{HERMES} Collaboration, M.~Diefenthaler, {\it HERMES measurements of
  Collins and Sivers asymmetries from a transversely polarised hydrogen
  target}  \href{http://xxx.lanl.gov/abs/arXiv:0706.2242}{{\tt
  arXiv:0706.2242 [hep-ex]}}.

\bibitem{Alexakhin:2005iw}
{COMPASS} Collaboration, V.~Y. Alexakhin {\em et~al.}, {\it First
  measurement of the transverse spin asymmetries of the deuteron in
  semi-inclusive deep inelastic scattering},  {\em Phys. Rev. Lett.} {\bf 94}
  (2005) 202002 [\href{http://xxx.lanl.gov/abs/hep-ex/0503002}{{\tt
  hep-ex/0503002}}];\\
%
{COMPASS} Collaboration, E.~S. Ageev {\em et~al.}, {\it A new measurement
  of the Collins and Sivers asymmetries on a transversely polarised deuteron
  target},  {\em Nucl. Phys.} {\bf B765} (2007) 31
  [\href{http://xxx.lanl.gov/abs/hep-ex/0610068}{{\tt hep-ex/0610068}}];\\
%
{COMPASS} Collaboration, A.~Kotzinian, {\it Beyond Collins and Sivers:
  further measurements of the target transverse spin-dependent azimuthal
  asymmetries in semi-inclusive DIS from COMPASS},
  \href{http://xxx.lanl.gov/abs/arXiv:0705.2402}{{\tt arXiv:0705.2402
  [hep-ex]}}.

\bibitem{D'Alesio:2007jt}
U.~D'Alesio and F.~Murgia, {\it Azimuthal and single spin asymmetries in hard
  scattering processes},
  \href{http://xxx.lanl.gov/abs/arXiv:0712.4328}{{\tt arXiv:0712.4328
  [hep-ph]}}. 

\bibitem{Bacchetta:2004jz}
A.~Bacchetta, U.~D'Alesio, M.~Diehl, and C.~A. Miller, {\it Single-spin
  asymmetries: The Trento conventions},  {\em Phys. Rev.} {\bf D70} (2004)
  117504 [\href{http://xxx.lanl.gov/abs/hep-ph/0410050}{{\tt
  hep-ph/0410050}}].

\bibitem{Collins:1981uk}
J.~C. Collins and D.~E. Soper, {\it Back-to-back jets in {QCD}},  {\em Nucl.
  Phys.} {\bf B193} (1981) 381.

\bibitem{Collins:2003fm}
J.~C. Collins, {\it What exactly is a parton density?},  {\em Acta Phys.
  Polon.} {\bf B34} (2003) 3103
  [\href{http://xxx.lanl.gov/abs/hep-ph/0304122}{{\tt hep-ph/0304122}}].

\bibitem{Ji:2004wu}
X.~Ji, J.-P. Ma, and F.~Yuan, {\it QCD factorization for semi-inclusive
  deep-inelastic scattering at low transverse momentum},  {\em Phys. Rev.} {\bf
  D71} (2005) 034005 [\href{http://xxx.lanl.gov/abs/hep-ph/0404183}{{\tt
  hep-ph/0404183}}].

\bibitem{Collins:2007ph}
J.~C. Collins, T.~C. Rogers, and A.~M. Stasto, {\it Fully unintegrated parton
  correlation functions and factorization in lowest order hard scattering},
  {\em Phys.\ Rev.} {\bf D77} (2008) 085009
  [\href{http://xxx.lanl.gov/abs/arXiv:0708.2833}{{\tt arXiv:0708.2833
  [hep-ph]}}].

\bibitem{Nadolsky:1999kb}
P.~M. Nadolsky, D.~R. Stump, and C.~P. Yuan, {\it Semi-inclusive
  hadron production 
  at {HERA}: The effect of {QCD} gluon resummation},  {\em Phys. Rev.} {\bf
  D61} (2000) 014003 [\href{http://xxx.lanl.gov/abs/hep-ph/9906280}{{\tt
  hep-ph/9906280}}];\\
%
P.~M. Nadolsky, D.~R. Stump, and C.~P. Yuan, {\it Phenomenology of multiple
  parton radiation in semi-inclusive deep-inelastic scattering},  {\em Phys.
  Rev.} {\bf D64} (2001) 114011
  [\href{http://xxx.lanl.gov/abs/hep-ph/0012261}{{\tt hep-ph/0012261}}].

\bibitem{Kulesza:2002rh}
A.~Kulesza, G.~Sterman, and W.~Vogelsang, {\it Joint resummation in electroweak
  boson production},  {\em Phys. Rev.} {\bf D66} (2002) 014011
  [\href{http://xxx.lanl.gov/abs/hep-ph/0202251}{{\tt hep-ph/0202251}}].

\bibitem{Nadolsky:2001sf}
P.~M. Nadolsky, {\it Multiple parton radiation in hadroproduction at lepton
  hadron colliders},  \href{http://xxx.lanl.gov/abs/hep-ph/0108099}{{\tt
  hep-ph/0108099}}.

\bibitem{Catani:2000vq}
S.~Catani, D.~de~Florian, and M.~Grazzini, {\it Universality of non-leading
  logarithmic contributions in transverse momentum distributions},  {\em Nucl.
  Phys.} {\bf B596} (2001) 299
  [\href{http://xxx.lanl.gov/abs/hep-ph/0008184}{{\tt hep-ph/0008184}}].

\bibitem{Koike:2006fn}
Y.~Koike, J.~Nagashima, and W.~Vogelsang, {\it Resummation for polarized
  semi-inclusive deep-inelastic scattering at small transverse momentum},  {\em
  Nucl. Phys.} {\bf B744} (2006) 59
  [\href{http://xxx.lanl.gov/abs/hep-ph/0602188}{{\tt hep-ph/0602188}}].

\bibitem{Weber:1991wd}
A.~Weber, {\it Soft gluon resummations for polarized Drell-Yan dimuon
  production},  {\em Nucl. Phys.} {\bf B382} (1992) 63.

\bibitem{Nadolsky:2003fz}
P.~M. Nadolsky and C.~P. Yuan, {\it Soft parton radiation in polarized vector
  boson production: {T}heoretical issues},  {\em Nucl. Phys.} {\bf B666} (2003)
  3 [\href{http://xxx.lanl.gov/abs/hep-ph/0304001}{{\tt hep-ph/0304001}}].

\bibitem{Collins:1992xw}
J.~C. Collins, {\it Hard scattering in QCD with polarized beams},  {\em Nucl.
  Phys.} {\bf B394} (1993) 169
  [\href{http://xxx.lanl.gov/abs/hep-ph/9207265}{{\tt hep-ph/9207265}}].

\bibitem{Ji:2004xq}
X.~Ji, J.-P. Ma, and F.~Yuan, {\it QCD factorization for spin-dependent cross
  sections in DIS and Drell-Yan processes at low transverse momentum},  {\em
  Phys. Lett.} {\bf B597} (2004) 299
  [\href{http://xxx.lanl.gov/abs/hep-ph/0405085}{{\tt hep-ph/0405085}}].

\bibitem{Sivers:1990cc}
D.~W. Sivers, {\it Single spin production asymmetries from the hard scattering
  of pointlike constituents},  {\em Phys. Rev.} {\bf D41} (1990) 83.

\bibitem{Qiu:1991pp}
J.-W.~Qiu and G.~Sterman, {\it Single transverse spin asymmetries},
  {\em Phys. Rev. Lett.} {\bf 67} (1991) 2264.

\bibitem{Idilbi:2004vb}
A.~Idilbi, X.~Ji, J.-P. Ma, and F.~Yuan, {\it Collins-Soper equation for the
  energy evolution of transverse-momentum and spin dependent parton
  distributions},  {\em Phys. Rev.} {\bf D70} (2004) 074021
  [\href{http://xxx.lanl.gov/abs/hep-ph/0406302}{{\tt hep-ph/0406302}}].

\bibitem{Mendez:1978zx}
A.~M{\'e}ndez, {\it QCD predictions for semi-inclusive and inclusive
  leptoproduction},  {\em Nucl. Phys.} {\bf B145} (1978) 199.

\bibitem{Meng:1995yn}
R.~Meng, F.~I. Olness, and D.~E. Soper, {\it Semi-inclusive deeply inelastic
  scattering at small $q_T$},  {\em Phys. Rev.} {\bf D54} (1996) 1919
  [\href{http://xxx.lanl.gov/abs/hep-ph/9511311}{{\tt hep-ph/9511311}}].

\bibitem{Diehl:2005pc}
M.~Diehl and S.~Sapeta, {\it On the analysis of lepton scattering on
  longitudinally or transversely polarized protons},  {\em Eur. Phys. J.} {\bf
  C41} (2005) 515 [\href{http://xxx.lanl.gov/abs/hep-ph/0503023}{{\tt
  hep-ph/0503023}}].

\bibitem{Eguchi:2006qz}
H.~Eguchi, Y.~Koike, and K.~Tanaka, {\it Single transverse spin asymmetry for
  {large-$p_T$} pion production in semi-inclusive deep inelastic scattering},
  {\em Nucl. Phys.} {\bf B752} (2006) 1
  [\href{http://xxx.lanl.gov/abs/hep-ph/0604003}{{\tt hep-ph/0604003}}].

\bibitem{Eguchi:2006mc}
H.~Eguchi, Y.~Koike, and K.~Tanaka, {\it Twist-3 formalism for single
  transverse spin asymmetry reexamined: Semi-inclusive deep inelastic
  scattering},  {\em Nucl. Phys.} {\bf B763} (2007) 198
  [\href{http://xxx.lanl.gov/abs/hep-ph/0610314}{{\tt hep-ph/0610314}}].

\bibitem{Hagiwara:1983cq}
K.~Hagiwara, K.~Hikasa, and N.~Kai, {\it Time reversal odd asymmetry in
  semiinclusive leptoproduction in quantum chromodynamics},  {\em Phys. Rev.}
  {\bf D27} (1983) 84.

\bibitem{Gehrmann:1995be}
T.~Gehrmann, {\it {Time-reversal-odd asymmetries at HERA}},
  \href{http://xxx.lanl.gov/abs/hep-ph/9608469}{{\tt hep-ph/9608469}};\\
%
M.~Ahmed and T.~Gehrmann, {\it Azimuthal asymmetries in hadronic final states
  at HERA},  {\em Phys. Lett.} {\bf B465} (1999) 297
  [\href{http://xxx.lanl.gov/abs/hep-ph/9906503}{{\tt hep-ph/9906503}}].

\bibitem{Belitsky:2002sm}
A.~V. Belitsky, X.~Ji, and F.~Yuan, {\it Final state interactions and gauge
  invariant parton distributions},  {\em Nucl. Phys.} {\bf B656} (2003)
  165 [\href{http://xxx.lanl.gov/abs/hep-ph/0208038}{{\tt
  hep-ph/0208038}}].

\bibitem{Bomhof:2004aw}
C.~J. Bomhof, P.~J. Mulders, and F.~Pijlman, {\it Gauge link structure in quark
  quark correlators in hard processes},  {\em Phys. Lett.} {\bf B596} (2004)
  277 [\href{http://xxx.lanl.gov/abs/hep-ph/0406099}{{\tt
  hep-ph/0406099}}].

\bibitem{Bomhof:2006dp}
C.~J. Bomhof, P.~J. Mulders, and F.~Pijlman, {\it The construction of
  gauge-links in arbitrary hard processes},  {\em Eur. Phys. J.} {\bf C47}
  (2006) 147 [\href{http://xxx.lanl.gov/abs/hep-ph/0601171}{{\tt
  hep-ph/0601171}}].

\bibitem{Collins:2004nx}
J.~C. Collins and A.~Metz, {\it Universality of soft and collinear factors in
  hard-scattering factorization},  {\em Phys. Rev. Lett.} {\bf 93} (2004)
  252001 [\href{http://xxx.lanl.gov/abs/hep-ph/0408249}{{\tt
  hep-ph/0408249}}].

\bibitem{Hautmann:2007uw}
F.~Hautmann, {\it Endpoint singularities in unintegrated parton distributions},
   {\em Phys. Lett.} {\bf B655} (2007) 26
  [\href{http://xxx.lanl.gov/abs/hep-ph/0702196}{{\tt hep-ph/0702196}}].

\bibitem{Cherednikov:2007tw}
I.~O. Cherednikov and N.~G. Stefanis, {\it Renormalization, {W}ilson lines, and
  transverse-momentum dependent parton distribution functions},
  {\em Phys.\ Rev.}  {\bf D77} (2008) 094001
  [\href{http://xxx.lanl.gov/abs/arXiv:0710.1955}{{\tt arXiv:0710.1955
  [hep-ph]}}].

\bibitem{Goeke:2005hb}
K.~Goeke, A.~Metz, and M.~Schlegel, {\it Parameterization of the quark-quark
  correlator of a spin-1/2 hadron},  {\em Phys. Lett.} {\bf B618} (2005)
  90 [\href{http://xxx.lanl.gov/abs/hep-ph/0504130}{{\tt
  hep-ph/0504130}}].

\bibitem{Collins:2002kn}
J.~C. Collins, {\it Leading-twist single-transverse-spin asymmetries: Drell-Yan
  and deep-inelastic scattering},  {\em Phys. Lett.} {\bf B536} (2002) 43
  [\href{http://xxx.lanl.gov/abs/hep-ph/0204004}{{\tt hep-ph/0204004}}].

\bibitem{Collins:1982uw}
J.~C. Collins and D.~E. Soper, {\it Parton distribution and decay functions},
  {\em Nucl. Phys.} {\bf B194} (1982) 445.

\bibitem{Boer:2003xz}
D.~Boer, {\it Theoretical aspects of spin physics},
  \href{http://xxx.lanl.gov/abs/hep-ph/0312149}{{\tt hep-ph/0312149}}.

\bibitem{Henneman:2005th}
A.~A. Henneman, {\it Scale dependence of correlations on the light-front},
  PhD thesis, Vrije Universiteit Amsterdam, 2005,
  \href{http://www.nikhef.nl/pub/services/newbiblio/theses.php}{{\tt
      http://www.nikhef.nl/pub/services/newbiblio/theses.php}}.

\bibitem{Boer:2001he}
D.~Boer, {\it Sudakov suppression in azimuthal spin asymmetries},  {\em Nucl.
  Phys.} {\bf B603} (2001) 195
  [\href{http://xxx.lanl.gov/abs/hep-ph/0102071}{{\tt hep-ph/0102071}}].

\bibitem{Anselmino:2005nn}
M.~Anselmino, M.~Boglione, U.~D'Alesio, A.~Kotzinian, F.~Murgia, and
  A.~Prokudin, {\it The role of Cahn and Sivers effects in deep inelastic
  scattering},  {\em Phys. Rev.} {\bf D71} (2005) 074006
  [\href{http://xxx.lanl.gov/abs/hep-ph/0501196}{{\tt hep-ph/0501196}}].

\bibitem{Mendez:1978vr}
A.~M{\'e}ndez, A.~Raychaudhuri, and V.~J.~Stenger,
  {\it QCD Effects In Semiinclusive Neutrino Processes},
  {\em Nucl. Phys.} {\bf B148} (1979) 499.

\bibitem{Konig:1982uk}
A.~K\"onig and P.~Kroll,
  {\it A Realistic Calculation Of The Azimuthal Asymmetry In
    Semiinclusive Deep Inelastic Scattering},
  {\em Z. Phys.} {\bf C16} (1982) 89.

\bibitem{Chay:1991nh}
J.~Chay, S.~D. Ellis, and W.~J. Stirling, {\it Azimuthal asymmetry in
  lepton-photon scattering at high energies},  {\em Phys. Rev.} {\bf D45}
  (1992) 46.

\bibitem{Oganessyan:1997jq}
K.~A. Oganessyan, H.~R. Avakian, N.~Bianchi, and P.~Di~Nezza, {\it
  Investigations of azimuthal asymmetry in semi-inclusive leptoproduction},
  {\em Eur. Phys. J.} {\bf C5} (1998) 681
  [\href{http://xxx.lanl.gov/abs/hep-ph/9709342}{{\tt hep-ph/9709342}}].

\bibitem{Anselmino:2006rv}
M.~Anselmino, M.~Boglione, A.~Prokudin, and C.~T\"urk,
  {\it Semi-inclusive deep inelastic scattering processes from small
    to large $P_T$},
  {\em Eur. Phys. J.} {\bf A31} (2007) 373
  [\href{http://xxx.lanl.gov/abs/hep-ph/0606286}{{\tt hep-ph/0606286}}].

\bibitem{Kotzinian:1996cz}
A.~M. Kotzinian and P.~J. Mulders, {\it Longitudinal quark polarization in
  transversely polarized nucleons},  {\em Phys. Rev.} {\bf D54} (1996)
  1229 [\href{http://xxx.lanl.gov/abs/hep-ph/9511420}{{\tt
  hep-ph/9511420}}].

\bibitem{Chay:1997qy}
J.~Chay and S.~M. Kim, {\it Azimuthal correlation in lepton hadron scattering
  via charged weak-current processes},  {\em Phys. Rev.} {\bf D57} (1998)
  224 [\href{http://xxx.lanl.gov/abs/hep-ph/9705284}{{\tt
  hep-ph/9705284}}].

\bibitem{Collins:1993kk}
J.~C. Collins, {\it Fragmentation of transversely polarized quarks probed in
  transverse momentum distributions},  {\em Nucl. Phys.} {\bf B396} (1993)
  161 [\href{http://xxx.lanl.gov/abs/hep-ph/9208213}{{\tt
  hep-ph/9208213}}].

\bibitem{Meissner:2007rx}
S.~Meissner, A.~Metz, and K.~Goeke, {\it Relations between generalized and
  transverse momentum dependent parton distributions}, {\em
  Phys. Rev.} {\bf D76} (2007) 034002
  [\href{http://xxx.lanl.gov/abs/hep-ph/0703176}{{\tt hep-ph/0703176}}].

\bibitem{Artru:1990zv}
X.~Artru and M.~Mekhfi, {\it Transversely polarized parton densities, their
  evolution and their measurement},  {\em Z. Phys.} {\bf C45} (1990) 669.

\bibitem{Falciano:1986wk}
{NA10} Collaboration, S.~Falciano {\em et~al.}, {\it {Angular distributions
  of muon pairs produced by 194 GeV/$c$ negative pions}},  {\em Z. Phys.} {\bf
  C31} (1986) 513;\\
%
{NA10} Collaboration, M.~Guanziroli {\em et~al.}, {\it Angular
  distributions of muon pairs produced by negative pions on deuterium and
  tungsten},  {\em Z. Phys.} {\bf C37} (1988) 545;\\
%
J.~S. Conway {\em et~al.}, {\it Experimental study of muon pairs produced by
  252-{GeV} pions on tungsten},  {\em Phys. Rev.} {\bf D39} (1989) 92;\\
%
{FNAL-E866/NuSea} Collaboration, L.~Y. Zhu {\em et~al.}, {\it {Measurement
  of angular distributions of Drell-Yan dimuons in $p + d$ interaction at 800
  GeV/$c$}},  {\em Phys. Rev. Lett.} {\bf 99} (2007) 082301
  [\href{http://xxx.lanl.gov/abs/hep-ex/0609005}{{\tt hep-ex/0609005}}].

\bibitem{Abe:2005zx}
{BELLE} Collaboration, K.~Abe {\em et~al.}, {\it Measurement of azimuthal
  asymmetries in inclusive production of hadron pairs in $e^+ e^-$ annihilation
  at BELLE},  {\em Phys. Rev. Lett.} {\bf 96} (2006) 232002
  [\href{http://xxx.lanl.gov/abs/hep-ex/0507063}{{\tt hep-ex/0507063}}].

\bibitem{Boer:1997mf}
D.~Boer, R.~Jakob, and P.~J. Mulders, {\it Asymmetries in polarized hadron
  production in $e^+ e^-$ annihilation up to order 1/Q},  {\em Nucl. Phys.}
  {\bf B504} (1997) 345
  [\href{http://xxx.lanl.gov/abs/hep-ph/9702281}{{\tt hep-ph/9702281}}].

\bibitem{Binosi:2003yf}
D.~Binosi and L.~Theussl, {\it JaxoDraw: A graphical user interface for drawing
  Feynman diagrams},  {\em Comput. Phys. Commun.} {\bf 161} (2004) 76
  [\href{http://xxx.lanl.gov/abs/hep-ph/0309015}{{\tt hep-ph/0309015}}].

\end{thebibliography}
\end{document}